\documentclass[pre, aps,twocolumn,superscriptaddress,longbibliography,floatfix,notitlepage]
{revtex4-1}

\usepackage{dcolumn}
\usepackage{bm}
\usepackage[utf8x]{inputenc}
\usepackage[english]{babel}
\usepackage[T1]{fontenc}
\usepackage{lmodern}
\usepackage{amsmath,amsfonts,amssymb,amsthm,bm,times,dcolumn}
\usepackage{mwe}
\usepackage{microtype}
\usepackage{braket}
\usepackage{mathrsfs}
\usepackage{array}
\usepackage{tikz}
\usepackage{blkarray}
\usepackage[colorlinks={true}, citecolor={blue}, filecolor={blue}, linkcolor={blue}, urlcolor={blue}]{hyperref}
\usepackage{graphicx,epstopdf,color}
\usepackage{subfigure}

\usepackage{graphicx}
\usepackage{epstopdf}

\usepackage[mathscr]{euscript}

\usepackage{setspace}
\usepackage{physics}

\begin{document}

\title{Quantum Advantage of Thermal Machines with Bose and Fermi Gases}

\author{Saikat Sur}
\email{saikat.sur@weizmann.ac.il}
\affiliation{Department of Chemical and Biological Physics, Weizmann Institute of Science, Rehovot 7610001, Israel}
\author{Arnab Ghosh}
\affiliation{Department of Chemistry, Indian Institute of Technology, Kanpur 208016, India}

\date{\today}

\begin{abstract}
In this article, we show that a quantum gas, a collection of massive, non-interacting, indistinguishable quantum particles can be realized as a thermodynamic machine as an artifact of energy quantization and hence bears no classical analog. Such a thermodynamic machine depends on the statistics of the particles, the chemical potential, and the spatial dimension of the system. Our detailed analysis demonstrates the fundamental features of quantum Stirling cycles from the viewpoint of particle statistics and system dimensions that helps us to realize desired quantum heat engines and refrigerators by exploiting the role of quantum statistical mechanics. In particular, a clear distinction between the behavior of a Fermi gas and a Bose gas is observed in one dimension than in higher dimensions, solely due to the innate differences in their particle statistics indicating the conspicuous role of a quantum thermodynamic signature in lower dimensions.
\end{abstract}

\maketitle

\section{\label{introduction}Introduction}
The study of quantum thermodynamics comprises the basis for analyzing heat engines and refrigerators at the microscopic level~\cite{scully_2002,kosloff2013quantum,binder2019thermodynamics,deffner2019quantum, mukherjee_2021_review,vinjanampathy2016,bhattacharjee2021, opatrny_2021_prl,misra_2022}. Although thermodynamics is quite successful in the classical regime, its applications towards quantum systems must be reconciled in view of energy quantization, degeneracies, and most importantly particle statistics~\cite{zheng_et_al_2014,david_et_al_2018, niedenzu_2019,deffner_2019, thomas_2019,myers2020bosons,myers2021quantum,gupt2021statistical,myers_2021, bouton_2021,yadin_2022}. A fundamental departure from classical mechanics is therefore imperative for analyzing the thermodynamic behavior of quantum systems. As a result, one might expect new thermodynamic effects would emerge in the quantum domain without having any classical correspondence. Most of the cases that have been analyzed for understanding the generic thermodynamic features in quantum thermal machines, do not exhibit any proper evidence of quantum advantage over their classical counterparts~\cite{scully2003extracting,rossnagel2014nanoscale,klaers2017squeezed,niedenzu2018quantum,ghosh2019are}. In recent years, several proposals for constructing quantum mechanical versions of thermal machines such as Otto, Carnot, Stirling and Diesel~\cite{salamon1980minimum,chen1991effect,geva1992classical,geva1992quantum,chen1994maximam,chen1998effect,chen1999comprehensive,arnaud2002carnot,bhattacharyya2001comment,abah2012single, das2020, mukherjee2020} with different working media like a particle in a box, harmonic oscillator and spin systems~\cite{geva1992quantum,feldmann_2000,chen2002performance,lin2003optimal,henrich2007quantum,thomas2014,gelbwaser2018single,levy2018,kosloff2017quantum,huang_2013,huang_2014,abah_2016,stefanatos_2014,stefanatos_2017} have been presented. Among them only a few analyzes the quantum-enhanced machines specifically from the viewpoint of a single particle within a potential well~\cite{david_et_al_2018,thomas_2019,chatterjee_2020,Chattopadhyay_2019}.

\par The role of quantum statistics comes into play if the system comprises an ensemble of identical particles in contact with a pair of low-temperature baths. In this limit, the dynamics of the system are governed by the principle of how a single particle state is occupied. This distinguishes the behavior of two different types of particles, viz. fermions, and bosons. The chemical potential that depends on the dimensionality of the system, in addition to its statistics, can play a huge role in determining its behavior. Therefore, the \textcolor{black}{collective behavior of quantum particles is interlinked with the dimensionality and hence the degeneracy of the energy levels~\cite{chowdhury_2000,PATHRIA2011179,PATHRIA2011231}.}

\par There have been several proposals to achieve \textcolor{black}{quantum supremacy~\cite{jaramillo2016quantum,chen2019,watanabe2020}} by using many-particle thermal machines over single particle machine. Many-particle quantum thermal machines are shown to outperform the classical counterparts using interacting Bose gases in a time-dependent harmonic trap and  a tight waveguide~\cite{jaramillo2016quantum,chen2019}. In these studies, the volume equivalent of the heat machines is taken to be the frequency of the harmonic oscillator and the interparticle interaction strength, respectively. Both these studies convey the quantum supremacy arising out of the interplay between many-particle quantum effects. Moreover, multi-component engines have been proposed, where quantum enhancement is shown exploiting the indistinguishability of quantum components~\cite{watanabe2020}.

\par In this article, we elaborately, rigorously, and methodically study the dependence of particle statistics and dimensionality of the system in the context of a quantum Stirling cycle based on an infinite potential box. In the present scenario, when the potential is distorted by introducing an infinite barrier in the middle of the box, the odd levels shift upwards and overlap with the immediately next even energy levels, creating a new energy level structure with degeneracies \textcolor{black}{(see e.g., Fig.~\ref{fig:1}(b))} \cite{griffithsbookquantummechanics,belloni_2014}. In quantum Stirling cycles, this distortion of the potential may constitute a certain amount of work and heat exchange between the baths exclusively based on quantized energy levels without having any classical analog of volume~\cite{gurtin_fried_anand_2010,paolucci_2016}. Depending on the behavior of the cycle, whether work is extracted or not, one can construct a desired Stirling heat engine or refrigerator just by exploiting the quantum nature of the identical particles. \textcolor{black}{The approach to a many-particle quantum system with symmetrizing/anti-symmetrizing wavefunctions is ineffective, especially in the $N \rightarrow \infty$ limit, where distinct statistical behavior is observed. Here, we show that one can bypass the problem by using the tools of quantum statistical mechanics.}

\par Our approach is interesting  both from theoretical and practical viewpoints as the recent advances in realizing heat machines with quantum components pave the way for the realization of multi-particle energy-quantization assisted machines in future~\cite{rossnagel2014nanoscale,abah2012single,rossnagel2016single,lindenfels2019spin, levy2018, deng2018, batalho_2014, josefsson_2018}. Particularly, quantum dots (electrons or holes confined in three  spatial dimensions),  wires (electrons or holes confined in two spatial dimensions and free in the other direction) and wells (electrons or holes confined in one  spatial dimension but free in the other two directions)~\cite{sattler_2010,liu_2013,du_2020,stern_2014} and ion-trap systems~\cite{ gangloff2015, bylinskii_2015, karpa_2013} are potential candidates for realizing quantum particles entrapped in a potential well.

\par This paper is organized as follows: In Sec.~\ref{Sec.II}, we present the basic formulation with a brief overview of the quantum Stirling cycle, followed by detailed characteristics of the working medium, and the fundamental principles behind quantum statistics of identical particles. In Sec.~\ref{Sec.III}, we summarize the main results and their interpretations  together with the analytical approach in the \textcolor{black} {limiting case of a large number of particles and extremely low temperature}. Finally, we conclude in Sec.~\ref{Sec.IV}.

\section{\label{Sec.II} Formulation}
\subsection{\label{Sec.IIA}Quantum Stirling cycle}

\begin{figure}[ht]
\centering
\includegraphics[width=0.43\textwidth]{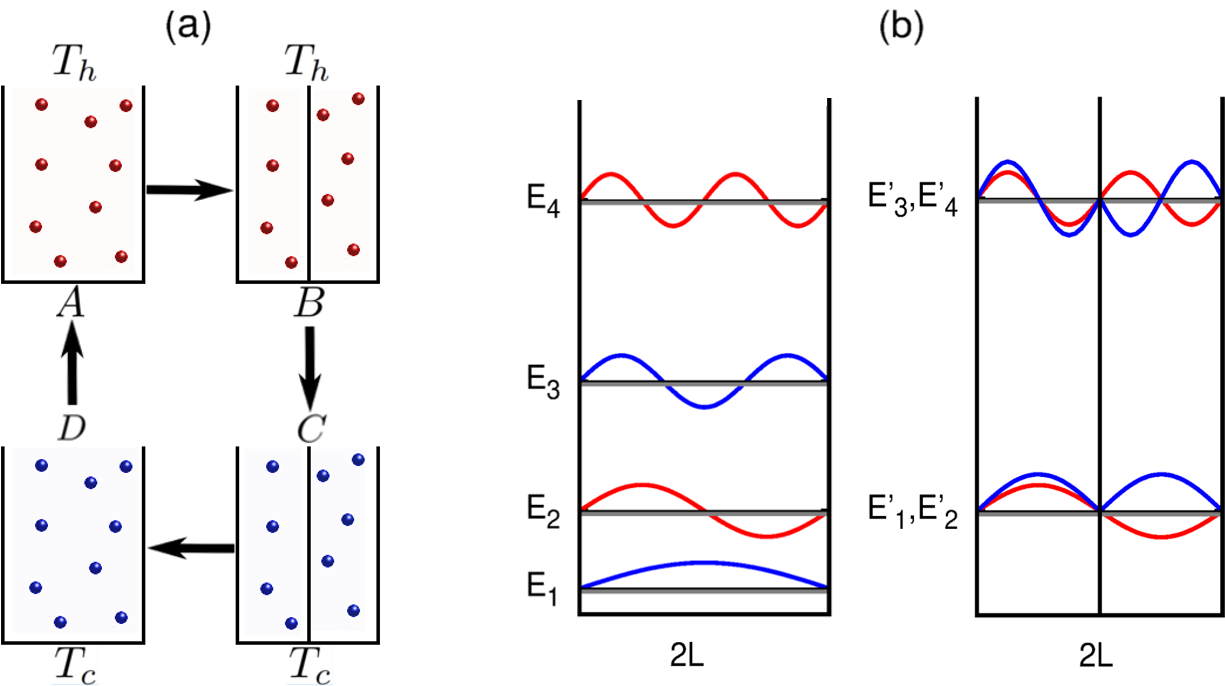}\hfill
\caption{(a) A schematic representation of a Stirling engine with $N$ massive quantum particles confined in an infinite potential box. (b) \textcolor{black}{Single particle energy levels and wavefunctions of a particle in an infinite potential well with and without the barrier.}}
\label{fig:1}
\end{figure}

\par A quantum Stirling cycle~\cite{thomas_2019} consists of two configurations, governed by two different  Hamiltonians $H$ and $H'$ respectively, and each configuration is kept in equilibrium with two thermal baths. Hence, the cycle has four stages, each connected to its preceding and succeeding stages through two isothermal and two isochoric processes as shown schematically in Fig~\ref{fig:1}(a). Initially, at stage $A$, the system is in equilibrium  with a hot bath at a temperature $T_h$. In the first step $(A \rightarrow B)$, the potential is distorted quasi-statically and isothermally to change  the configurations of the energy levels and the wavefunctions, while the system is still coupled to the hot bath [Fig~\ref{fig:1}(b)].   In the next step $(B \rightarrow C)$, the system is detached from the hot bath and connected with a cold bath at a temperature $T_c$ via an isochoric process. Consequently, the system fully thermalizes with the cold bath and  the temperature of the system falls from $T_h$ to $T_c$, without performing any mechanical work. In the next step $(C \rightarrow D)$, the system is brought back to its original configuration quasi-statically and isothermally while keeping the system connected to the cold  bath. In the final step $(D \rightarrow A)$, the system is detached from the cold bath and connected to the hot bath through an isochoric process without performing any mechanical work. The system again fully thermalizes with the hot bath and the temperature rises from $T_c$ to $T_h$.\\ 

The total heat transferred to the system from the bath in the isothermal processes $(A \rightarrow B)$ and $(C \rightarrow D)$, while keeping the system in equilibrium with the bath at temperature $T_h$ and $T_c$ respectively, are  combinations of the mechanical works performed due to deformation of the potential and the change of the internal energies. Thus the heat transferred to the bath  from the system during two isothermal processes is given by
\begin{eqnarray}
Q_{AB} = -(U_B-U_A+k_B T_h \ln Z_{B} - k_B T_h \ln Z_{A}), 
\end{eqnarray}
and
\begin{eqnarray}
Q_{CD} = -(U_D-U_C+k_B T_c \ln Z_{D} - k_B T_c \ln Z_{C}),
\end{eqnarray}
respectively. By definition, the internal energy at a temperature $T$ in terms of the thermal partition function $Z_T$ is given as 
\begin{equation}
 U_T = k_B T^2\frac{\partial}{\partial T} \ln Z_T.
\end{equation}
In what follows in Sec.~\ref{Sec.IIC}, we explicitly evaluate the thermal partition functions of the Stirling cycle with noninteracting quantum gas of fermions and bosons. On the contrary, the heat transferred to the system during the isochoric processes $(B \rightarrow C)$ and $(D \rightarrow A)$ are only the differences between the average energies of the initial and the final configurations
\begin{eqnarray}
Q_{BC} = -(U_C-U_B),\;\; 
\text{and}\;\; Q_{DA} = -(U_A-U_D).
\end{eqnarray}

\par Provided all the processes involved in the cycle are reversible and no leakage takes place, the expressions for the net work done on the system $W$ and the heat transferred to the system with the hot and cold baths, $Q_h$ and $Q_c$ after completion of one cycle are given as follow:
\begin{eqnarray}
  W &=& (Q_{AB} + Q_{BC} +Q_{CD} + Q_{DA})  =  -k_B \left(T_h \ln \frac{Z_{B}}{Z_{A}} - T_c \ln \frac{Z_{C}}{Z_{D}}\right),\nonumber\\
  Q_h &=& -(Q_{AB} + Q_{DA}) = k_B T_h \ln \frac{Z_{B}}{Z_{A}}+U_B -U_D, \nonumber\\
  Q_c &=& -(Q_{BC} + Q_{CD}) = -k_B T_c \ln \frac{Z_{C}}{Z_{D}}-U_B +U_D. \nonumber\\
  \label{eq:w}
\end{eqnarray}
By the principle of conservation of energy or the first law of thermodynamics, one can check that $Q_h +Q_c+W = 0$. In our convention, if the quantities $Q_h$ or $Q_c$ are positive,  the heat  is flowing into the system, similarly, if work $W$ is positive, work is done on the system. In conformity with the second law of thermodynamics, only four modes of operation are possible. The four possible modes can be identified by the signs of $W$, $Q_h$ and $Q_c$, as depicted in Table~\ref{tab:1}.   
\begin{table}[ht]
\centering
\begin{tabular}{ |p{4cm}||p{1cm}|p{1cm}|p{1cm}|  }
 \hline
Modes of operation & $Q_h$ & $Q_c$ & $W$ \\
 \hline
{ Engine}  & $>0$ & $<0$ &  $<0$\\
 { Refrigerator}&   $<0$  & $>0$   & $>0$\\
{ Accelerator }   & $>0$    & $<0$&   $>0$\\
 { Heater}  & $<0$ & $<0$&  $>0$\\
  \hline
\end{tabular}
\caption{Different modes of operation of a quantum thermodynamic cycle.}
\label{tab:1}
\end{table}

Now, if heat is absorbed from the hot bath at a higher temperature by the system, converts a fraction of it into work, and rejects the rest to the bath at a lower temperature, we term it as a heat engine mode. The system works as a heat engine if $W$ is negative, i.e., work can be extracted from the system. The efficiency $\eta_E$ of the engine, defined as the ratio of the work extracted to the total heat absorbed by the system is
\begin{eqnarray}
\hspace{-0.2 cm}\eta_E = -\frac{W}{Q_h}.
\label{eq:eta}
\end{eqnarray}

In contrast, if heat is absorbed from the cold bath and rejected into the hot bath with external work done on the system, we term it refrigerator mode. Here, $W$ is positive, i.e., work is done on the system. The coefficient of performance of a refrigerator, the ratio of the  heat extracted from the cold bath to the work done on the system is given as, 

\begin{eqnarray}
 \eta_R = \frac{Q_c}{W}.
  \label{eq:eta_R}
\end{eqnarray}

Apart from the two modes discussed, there can be two other modes. In the accelerator mode, there is a free flow of heat from the hot bath to the cold bath, and work is done on the system. On the other hand in the heater mode,  a fraction of the work done on the system is rejected to the hot bath and the rest to the cold bath.

     \textcolor{black} {The Carnot cycle,  a theoretical proposition of Carnot, consists of only reversible processes and therefore conserves the total entropy while the system completes one cycle. However, it provides the maximum possible efficiency for a given pair of thermal baths, which is solely dependent on the bath temperatures~\cite{carnot1897}. In order to provide an estimate of its performance, we will present the efficiency and coefficient of performance of the quantum cycle scaled by the same to its Carnot equivalent. The efficiency of a Carnot engine and the coefficient of performance of the Carnot refrigerator are given respectively as}

\begin{eqnarray}
 \eta^{max}_E = 1 - \frac{T_c}{T_h}, \mbox{~~and ~~}
\eta^{max}_R = \frac{T_c}{T_h - T_c}.
\end{eqnarray}

The work done during a cycle is primarily dependent on the potential deformation that changes the wavefunction of the working substance and shifts the energy eigenvalues along with the bath temperatures. As discussed, quasi-static  deformation of the potential can cause extraction of work from the system and hence serve as a quantum heat engine or a refrigerator depending on whether heat is transferred from a hot bath to a cold bath and vice versa. As a practical example of the aforementioned situations, we consider a quantum thermal machine that uses a gas of quantum particles in a hard potential box as its working medium~\cite{david_et_al_2018,thomas_2019} and discusses the details of its properties.

\textcolor{black}{Although in this article, we demonstrate our results for a Stirling cycle, one can study other thermodynamics cycles with Bose and Fermi gases case by case. For example, an Otto  cycle consists of a pair of isochoric and isentropic (adiabatic) processes. In our context, the barriers are to be introduced adiabatically and the rest of the steps are identical.  With the approximation of adiabaticity, that ensures that the level populations do not change upon removal of the bath, one can have  results  similar to that of a Stirling engine~\cite{david_et_al_2018}. One can generalize the results replacing a single particle working medium with a quantum gas with fermions and bosons. Whereas, an exact equivalence with a Carnot cycle is difficult to construct in this setup since there should be a pair of expansion steps, an isothermal and an adiabatic. }


\subsection{\label{Sec.IIB} The working medium}
We consider an ensemble of non-interacting, massive, indistinguishable particles of mass $M$ confined in a $d$ dimensional potential box with length $2L$ in each dimension as our working medium of the thermal machine.  As the particles are non-interacting, the full Hamiltonian of the system is the sum of identical single-particle Hamiltonians. The single particle Hamiltonian is given as
\begin{equation}
H = \sum^d_{i =1}\frac{p^2_{x_i}}{2M}, {~~~} \mbox{for~} |x_i|  < L.
\end{equation}
The potential is distorted by introducing $d$ impenetrable barriers at $x_i = 0$, one in each dimension, that changes the energy levels and distorts the wavefunctions. In this context, to highlight the quantum advantage over its classical counterpart,  we note that introducing delta function barriers does not change the volume of the classical system and \textcolor{black}{thereby has zero contribution to work}. In addition to that, if the box becomes very large, gaps between the energy levels go to zero forming free particles with a continuum of energy levels. 

The single-particle Hamiltonian of the system in the distorted configuration, with $d$ impenetrable barriers, is given by
\begin{equation}
H' = \sum^d_{i =1}\frac{p^2_{x_i}}{2M} + \lambda \delta(x_i), {~~} \mbox{for~} |x_i|  < L,
\end{equation}
where $\lambda$ is the strength of the delta function barrier. The  single particle eigen energies of the Hamiltonians $H$ and $H'$, labeled by  a set of integer quantum numbers $n_{x_i}$ are given by the following expressions
\begin{eqnarray}
 &&E(n_{x_1}, n_{x_2},...,n_{x_d})  =  \frac{ \pi ^2\hbar^2 }{8ML^2} \sum^d_{i =1}  n^2_{x_i}; \quad n_{x_i} = 1, 2, 3,  ... \nonumber\\
&&E' (n_{x_1}, n_{x_2},...,n_{x_d}) =  \frac{ \pi ^2\hbar^2 }{8ML^2}\sum^d_{i =1} \left[n_{x_i}+\frac{\varepsilon(n_{x_i})}{2}({1-(-1)^{n_{x_i}}}) \right]^2; \quad \nonumber\\ &&{~~~~~~~~~~~~~~~~~~~~~~~~~~~~~~~~~~~~~~~~~~~~~~~~~~~~~~~~~~~~~~~~}n_{x_i} = 1, 2, 3,  ...
  \label{eq: energy_level}
\end{eqnarray}
where $0 \le \varepsilon(n_{x_i}) \le 1$. The value of $\varepsilon(n_{x_i})$ depends on the barrier strength $\lambda$ and can be obtained from the graphical solution of the transcendental equation (see \cite{belloni_2014,griffithsbookquantummechanics} for details). Since we consider here an infinitely strong barrier, i.e., $\lambda \rightarrow \infty$, the quantity $\varepsilon(n_{x_i}) = 1$. Inserting an impenetrable barrier shifts each odd energy eigenstate towards its next even one and thus creates twice degenerate energy levels. \textcolor{black}{The single-particle energy  levels and eigenfunctions are schematically shown in  Fig.~\ref{fig:1}(b). In a system with $d$ dimensions, one can show that there will be $2^d$ degenerate energy levels~\cite{griffithsbookquantummechanics}. For example, in two dimensions, the energy levels given by Eq.~\eqref{eq: energy_level} are denoted by a pair of quantum numbers $(n_{x_1},n_{x_2})$. Now, without the barriers the pair $(n_{x_1},n_{x_2})$ takes values $(1,1), (1,2), (2,1), (2,2), (1,3), (2,3), (1,4), (2,4), ....$ and so on. After introducing two impenetrable barriers at $x_1=0$ and $x_2=0$, the energies of the same set change to the energy levels corresponding to the quantum numbers $(2,2), (2,2), (2,2), (2,2), (2,4), (2,4), (2,4), (2,4),....$ of the undeformed box and thus introduces a degeneracy of $4$.} 

The approach of dealing with  a large number of quantum particles in the system by symmetrizing/anti-symmetrizing their states is inefficient and intractable even in computers, particularly in the limit where an average number of particles $N \rightarrow \infty$. But the collective effects of bosons or fermions are expected to be prominent in the said limit. At this juncture, the problem can be bypassed using the tools of quantum statistical mechanics.

\subsection{Statistics of quantum particles}
\label{Sec.IIC}

Quantum statistical mechanics is the foundation of understanding the low-temperature behavior of multi-particle physical systems consisting of identical particles.  In quantum mechanics, all identical particles are classified into two categories; the class of particles with half-integer spin, known as fermions, is described by Fermi-Dirac statistics, and the other class with integer spins,  known as bosons, is described by Bose-Einstein statistics. The fundamental difference between these two categories arises entirely from the principle of how a single particle state is occupied. Two fermions can never occupy a single particle state; on the other hand, multiple bosons can occupy the same state.

\par Quantum statistics dominate only when the interparticle spacing becomes smaller than the thermal de Broglie wavelength under low temperature, which can be termed as ``\textit{quantum regime}''. On the contrary, both statistics can be approximated by Maxwell-Boltzmann distribution in the classical regime, where quantum effects are negligible. The thermal grand partition functions 
$Z^+_T$ and $Z^-_T$  for $N$  non-interacting fermions and bosons at a temperature $T$ will respectively be~\cite{schroeder,arovas}
\begin{eqnarray}
Z^+_T &=&\prod^\infty_{n=1} \left(\sum^1_{j = 0} e^{-j(\tilde{E}_n-\tilde{\mu})/T}\right)^{g_n} 
= \prod^\infty_{n=1}\left(1+e^{-(\tilde{E}_n -\tilde{\mu})/T}\right)^{g_n},\nonumber\\
Z^-_T &=& \prod^\infty_{n=1} \left(\sum^\infty_{j = 0}  e^{-j(\tilde{E}_n-\tilde{\mu})/T}\right)^{g_n}
= \prod^\infty_{n=1}
\left(\frac{1}{1-e^{-(\tilde{E}_n -\tilde{\mu})/T}}\right)^{g_n},\nonumber\\
\label{partition_function}
\end{eqnarray}
where $g_n$ is the degeneracy of the single particle $n$-th energy level and $\tilde \mu$ is the scaled chemical potential. \textcolor{black}{Note that in the above equation, the dependence on the particle number is hidden in the chemical potential.} For bosons, it is to be noted that  the $N$ dependence on $Z^-_T$ vanishes as the constraint on the particle number is relaxed in the grand canonical ensemble.  For the sake of simplicity, we express all the energies and the chemical potentials in the units of $k_B$, i.e. $\tilde E = E/k_B$ and $\tilde{\mu} = \mu/k_B$ in Eq.~\eqref{partition_function}.


As we will see in the next section, the chemical potential of the system plays an essential role in determining the properties of the thermal machine. Now, to estimate the value of the chemical potential for fermions in terms of an average number of particles $N$, it is important to define the Fermi energy $E_F$, the energy of the highest filled state at $T=0$, which is given as $ E_F = \frac{\hbar^2k_F^2}{2M}$ where $\vec{k}_F$ is the Fermi wave vector. In $d$-dimension, the relation \cite{chowdhury_2000,PATHRIA2011231,PATHRIA2011179,ma_2010} $\frac{C_d k^d_F}{(2\pi/2L)^d} = N$ yields $ k_F = 2\pi(\frac{\sigma}{C_d})^{1/d}$. Hence the dimensionless Fermi energy in terms of the volume of a $d$-dimensional unit sphere \cite{blumenson,chowdhury_2000} $C_d = \frac{\pi^{d/2}}{\Gamma(d/2+1)}$ and the particle density $\sigma = \frac{N}{(2L)^d}$, can be expressed as
\begin{eqnarray}
\tilde{E}_F = \frac{E_F}{k_B} = 2 \alpha L^2\left(\frac{\sigma}{C_d}\right)^{2/d} = \frac{\alpha N^{2/d}}{2C^{2/d}_d}.
\label{eq:fermi_energy}
\end{eqnarray}
For example,
\begin{eqnarray}
 &&\tilde{E}_F^{(d = 1)}  = \frac{\alpha L^2 \sigma^2}{2} = \frac{\alpha N^2}{8},\nonumber\\
  &&\tilde{E}_F^{(d = 2)}  = \frac{2\alpha L^2 \sigma}{\pi} = \frac{\alpha N}{2\pi},\nonumber\\
   &&\tilde{E}_F^{(d = 3)}  = \Big(\frac{6}{\pi}\Big)^{2/3}\frac{\alpha L^2 \sigma^{2/3}}{2} = \Big(\frac{6}{\pi}\Big)^{2/3} \frac{\alpha N^{2/3}}{8}. 
 \label{eq:fermi_energy_d}  
\end{eqnarray}
 At extremely low temperatures, for fermions, all the energy levels below the Fermi energy are occupied, introducing barrier(s) in the system does not change the Fermi energy appreciably. We, therefore, can consider the chemical potentials in the two configurations, with and without the barrier(s) to be equal, i.e., $\mu = \mu'$.

For the convenience of our calculations, here, we define a parameter $\alpha = \frac{\hbar^2 \pi^2}{k_BML^2}$ with the dimension of temperature~\footnote{$[\alpha] = [\hbar]^2  [k_B]^{-1} M^{-1} L^{-2}\\ = (ML^2T^{-1})^2(ML^2T^{-2}\Theta^{-1})^{-1} M^{-1} L^{-2} = \Theta$ }, such that the low-temperature condition is scaled w.r.t $\alpha$ as  $T_{h,c}/\alpha \lesssim 1$. We re-emphasize here that as the bath temperatures $T_h,T_c $ become greater than $ \alpha$, the energy levels become continuous and the quantum advantage is gradually lost; we term this limit as ``classical regime''. In order to get an estimate of the different parameters related to the system, let us consider a system of electrons entrapped in a potential well. In terms of the fundamental constants, $\alpha = 1$ corresponds to a box length $L \sim 100~$nm. Since $\alpha$ is the characteristic temperature associated with the system, the low-temperature limit, in this case, reduces to $T_h,T_c \lesssim 1 K$. From the expression of $\alpha$, it is evident with the decrease of the size of the box, i.e., the more confined the quantum system becomes, the constraint on the two bath temperatures becomes more relaxed.

Finally, we note that in the low-temperature limit, the chemical potential for fermions can be taken to be equal to the Fermi energy; as the difference is extremely small even at room temperature~\cite{cowan_2019,chowdhury_2000}.

In the case of bosons, one can easily observe from the Bose-Einstein distribution that the chemical potential must be smaller than the ground state of the system in order to ensure a positive value for the occupation number. As the ground state energies of a system with $d$ dimensions are $E_1 = \alpha d/8$ and $E'_1 = \alpha d/2$ for the configurations without and with the barriers respectively, the corresponding chemical potentials must satisfy the conditions  $\mu  \le  \alpha d/8$ and $\mu' \le \alpha d/2$. Hence, the relation between the chemical potential and the average particle number becomes involved in this case. Starting from the boundary condition that the  sum over the occupation numbers $N_n$ of all energy levels equals the total number of particles, i.e., $N = \sum_n N_n$, one can write 

\begin{eqnarray}
N &&= \sum^{\infty}_{n=1}\frac{g_n}{e^{(\tilde{E}_n -\tilde{\mu})/T}-1},
\end{eqnarray}

 In the extremely low-temperature limit, all the terms except the one corresponding to the ground state are negligible. With this approximation, the $N$ dependence of the  chemical potential for the bosons is given by the following form,

\begin{eqnarray}
 \tilde{\mu} \approx  \tilde{E}_1 - T \ln (1+ \frac{g_1}{N}). 
\end{eqnarray}

For a system of $d$ dimensions, the values of the chemical potential will be given as
\begin{eqnarray}
&&\tilde{\mu} \approx \frac{\alpha d}{8} -  T\ln (1+ \frac{1}{N}), \nonumber\\
&&\tilde{\mu}' \approx \frac{\alpha d}{2} -  T\ln (1+ \frac{2^d}{N}). \nonumber\\
\label{eq: mu_boson}
\end{eqnarray}
Further note that, in the limit $N \rightarrow \infty$, the values of  $\tilde{\mu}$ limit the  respective ground state energies from below. Unlike fermions, the ground state occupation of bosons contributes to the chemical potential predominantly in the low-temperature limit. Since inserting the barrier(s) significantly changes the ground state, the chemical potentials with and without the barrier(s) are different.

\textcolor{black} {The chemical potential  for a Bose gas is slightly less than the ground state energy at a very low temperature and the difference between them is of the order of inverse volume of the system~\cite{cowan_2019,chowdhury_2000}. It goes to zero at a finite temperature only if the volume of the system becomes large enough so that the energy level spacing becomes continuous, i.e., $\alpha \rightarrow 0$, forming free particles and hence Bose-Einstein condensate (BEC) is formed at a finite temperature for $d \ge 3$. In the limit $\alpha \rightarrow 0$, as explained before, the quantum advantage is lost implying that  quantum thermal machines based on quantized energy levels and BEC cannot be achieved in the same limit. A few proposals for quantum heat engines or refrigerators that use the BEC phase of the working medium or the bath have been made~\cite{myers_bec_2022,niedenzu_bec_2019}, but their thermodynamic advantage is not related to quantized energy levels as they are in this case. }

  \section{Results}
  
  \label{Sec.III}

It is often useful to express the work done in terms of relative partition functions, i.e., the ratios of the partition functions for the adjacent stages, i.e., $\zeta(T_h) = {Z_B}/{Z_A}$ and $\zeta(T_c) = {Z_C}/{Z_D}$. Then the work done given by Eq.~\eqref{eq:w} can be rewritten as
\begin{eqnarray}
W = - (T_h  \ln  \zeta(T_h) - T_c  \ln  \zeta(T_c)).
\label{eq:work}
\end{eqnarray}
The relative partition function is the ratio between the partition functions of the system with and without the barrier(s) at the same temperature and thus implies the  work done due to inserting the barrier(s) to deform the wavefunctions to create a new set of degenerate energy levels~\cite{thomas_2019}. This insertion of the barrier(s) accounts for an extra amount of work $-T \ln \zeta(T)$. Given a pair of bath temperatures $(T_h, T_c)$, one can compute the total work done in a cycle using Eq.~\eqref{eq:work} in terms of the relative partition functions. For $\zeta(T) = 1$, the insertion of the barrier(s) does not contribute to any extra work and therefore no advantage is obtained  as we expect in the classical limit. In what follows, before we discuss the quantum regime, we briefly review the well-known results of the classical limit for the sake of completeness.

 In the classical regime, i.e., $T \gg \alpha$, the thermodynamics of the particles are governed by Maxwell-Boltzmann statistics. The single particle thermal partition function in this regime, in $d$ dimensions, is given by

\begin{eqnarray}
 Z_T = \sum^{\infty}_{n_{x_1}=1} ... \sum^{\infty}_{n_{x_d}=1} e^{-\tilde{E}(n_{x_1},...,n_{x_d})/T}.
\end{eqnarray}
In terms of the eigenenergies given in Eq.~\eqref{eq: energy_level}, one readily finds the expression for the relative partition function
\begin{eqnarray}
\zeta(T) =  &&\frac{\sum^{\infty}_{n_{x_1}=1} ... \sum^{\infty}_{n_{x_d}=1} e^{-{\alpha (n^2_{x_1} +...+n^2_{x_d})}/8T} }{2^d\sum^{\infty}_{n_{x_1}=1}... \sum^{\infty}_{n_{x_d}=1} e^{-{\alpha (n_{x_1}^2+...+n_{x_d}^2)}/2T}} \nonumber\\ &&\approx \textcolor{black}{ \frac{\left(\frac{1}{2}\sqrt{\frac{8\pi T}{\alpha}}\right)^d}{2^d \times \left(\frac{1}{2}\sqrt{\frac{2\pi T}{\alpha}}\right)^d}} = 1.
\label{zeta_classical}
\end{eqnarray}

This implies that introducing barrier(s) at $x=0$ does not cost any extra work in the classical limit. In view of Eqs.~\eqref{eq:work} and~\eqref{zeta_classical}, one concludes that  the work done (extracted) by (from) the system is identically zero in the classical limit. Therefore, one can easily see that the counterpart of such a quantum cycle operating between two thermal baths, is an  incompressible classical engine/refrigerator with zero efficiency~\cite{david_et_al_2018,paolucci_2016,gurtin_fried_anand_2010}. This paradigm breaks down in the quantum domain, where the discreteness in the energy levels and the inhomogeneous shift of the population distribution can lead to efficient quantum thermal machines with no classical analogue~\cite{david_et_al_2018, thomas_2019}.

First, let us consider the case with particles in a one-dimensional potential well. The relative partition functions from Eq.~\eqref{partition_function} for fermions and bosons are found to be
\begin{eqnarray}
 \zeta^+(T) = \prod^\infty_{n =1} \frac{\left(1+e^{-(\alpha(2n)^2/8 -\tilde{\mu}')/T}\right)}{\left(1+e^{-(\alpha (2n-1)^2/8 -\tilde{\mu})/T}\right)},\label{xi-f-1d}
\end{eqnarray}
and
\begin{eqnarray} 
\zeta^-(T) = \prod^\infty_{n =1} \frac{\left(1-e^{-(\alpha (2n-1)^2/8 -\tilde{\mu})/T}\right)}{\textcolor{black}{\left(1-e^{-(\alpha (2n)^2/8 -\tilde{\mu}')/T}\right)}}.
\label{xi-b-1d}
\end{eqnarray}
The expressions for the internal energies at $B$ and $D$ for fermions and bosons are respectively given by
\begin{eqnarray}
&&U^{\pm}_B = \sum^\infty_{n =1}2\frac{\alpha n^2 /2 -\tilde{\mu}'}{ e^{(\alpha n^2/2 -\tilde{\mu}')/T_h} \pm 1};\nonumber\\ 
&&U^{\pm}_D = \sum^\infty_{n =1}\frac{\alpha n^2 /8-\tilde{\mu}}{ e^{(\alpha n^2/8 -\tilde{\mu})/T_c} \pm 1}. 
\end{eqnarray}

We now generalize the result given in the previous section for a quantum gas in a $d$-dimensional box, with a barrier in each dimension. As discussed in Sec.~\ref{Sec.IIB}, this will render an energy level structure with a degeneracy of $2^d$. The expressions for the relative partition functions of the particles are modified according to Eq.~\eqref{eq: energy_level} as
\begin{eqnarray}
 \zeta^+(T) = \prod^\infty_{n_{x_i}=1} \prod^1_{j_{x_i}=0}\frac{\left(1+e^{-(\alpha \sum_{i}((2n_{x_i})^2/8 -\tilde{\mu}')/T}\right)}{\left(1+e^{-(\alpha \sum_{i}((2n_{x_i}-j_{x_i})^2/8 -\tilde{\mu})/T}\right)}, \nonumber\\
 \label{eq: zeta_fermion_d}
\end{eqnarray}
for fermions and
\begin{eqnarray}
 \zeta^-(T) = \prod^\infty_{n_{x_i}=1} \prod^1_{j_{x_i}=0}  \frac{\left(1-e^{-(\alpha \sum_{i}(2n_{x_i}-j_{x_i})^2/8 -\tilde{\mu})/T}\right)}{\left(1-e^{-(\alpha \sum_{i}(2n_{x_i})^2/8 -\tilde{\mu}')/T}\right)}, \nonumber\\
  \label{eq: zeta_boson_d}
\end{eqnarray}
for bosons, respectively. 

The extension for the $d$-dimensional internal energies at $B$ and $D$, for a system of fermions and bosons are respectively given by
\begin{eqnarray}
&&U^{\pm}_B =  \sum^\infty_{n_{x_i} =1}2^d\frac{\alpha (\sum_i (2n_{x_i})^2) /8 -\tilde{\mu}'}{ e^{(\alpha  (\sum_i (2n_{x_i})^2)/8 -\tilde{\mu}')/T_h} \pm 1},\nonumber\\
&&U^{\pm}_D = \sum^\infty_{n_{x_i} =1} \frac{\alpha (\sum_i n^2_{x_i}) /8-\tilde{\mu}}{ e^{(\alpha (\sum_i n^2_{x_i})/8 -\tilde{\mu})/T_c} \pm 1}.
\end{eqnarray}

\begin{figure}
 \begin{center}
\includegraphics[width=0.49\textwidth]{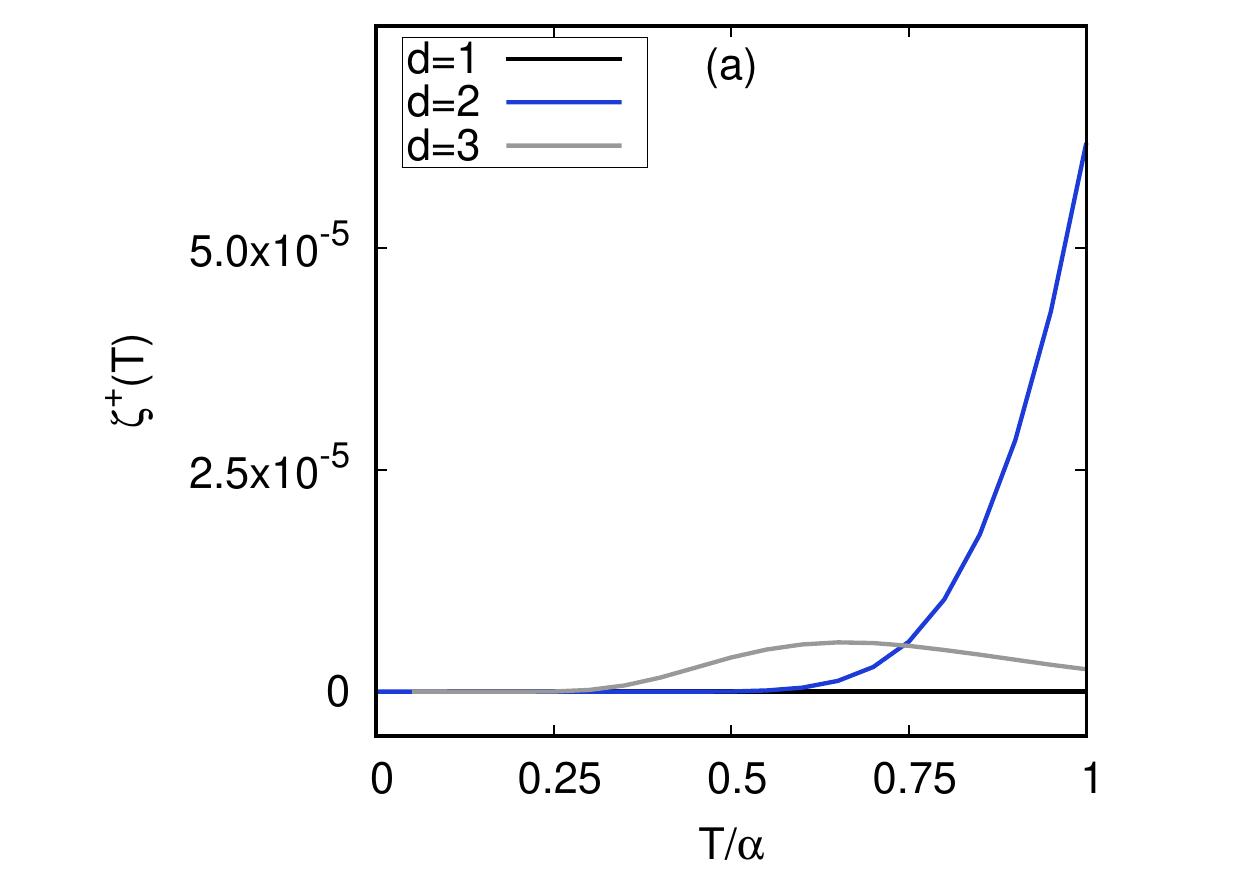}
\includegraphics[width=0.49\textwidth]{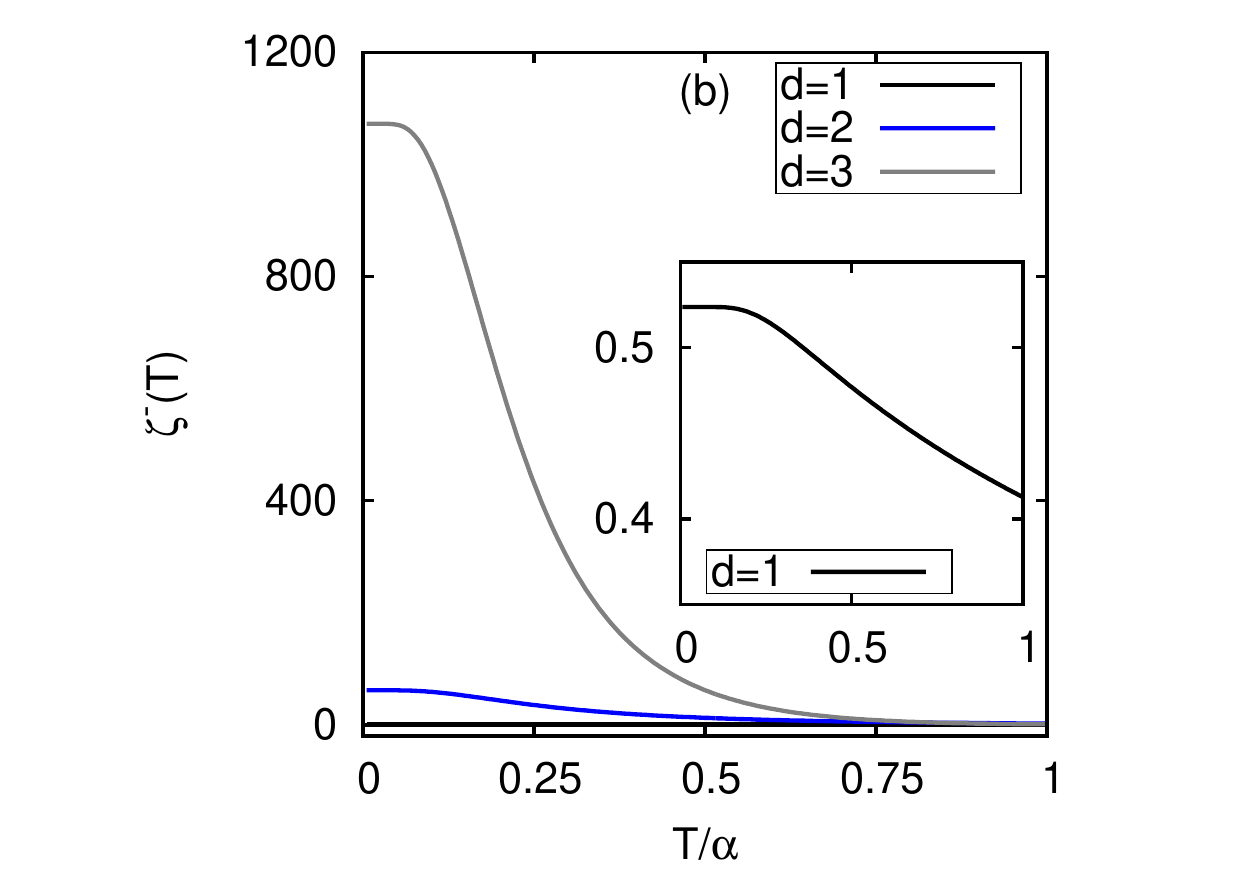}
  \end{center}
  \caption{{Relative Partition function $\zeta^{\pm}(T)$ [Cf. Eqs.~\eqref{eq: zeta_fermion_d} and~\eqref{eq: zeta_boson_d}] as a function of $T$ for (a)  fermions  and (b)  bosons with $N = 20$. }}
  \label{fig:2}
  \end{figure}

To see the contrasting role of the chemical potential on the relative partition functions for fermions and bosons, the temperature dependence of $\zeta^{\pm}(T)$ have been plotted in Fig.~\ref{fig:2} following Eqs.~\eqref{eq: zeta_fermion_d} and~\eqref{eq: zeta_boson_d} for $d=1, 2$ and $3$. It is evident from Fig.~\ref{fig:2}(a) that the dimension of the system of fermions decides the qualitative behavior of $\zeta^+(T)$. The value of  $\zeta^+(T)$ is zero at $T$ tends to zero irrespective of $d$ but shows different behavior depending on the $d$ owing to the particle number dependence of $\tilde{\mu}$ ($\sim \tilde{E}_F$). In $T/\alpha \lesssim 1$ regime, the value of $\zeta^+(T)$ does not change appreciably from zero for $d=1$ ($\zeta^+(T) \sim 10^{-10}$  at $T = 1$), but increases for $d=2$ and $d=3$. However,  the value of $\zeta^+(T)$ for $d=3$ is much smaller compared to $d=2$. As we have explained earlier, the relative partition function is related to the amount of work $-T \ln \zeta(T)$ to change the wavefunctions while inserting the barrier. This work done is always non-zero irrespective of $d$.  On the other hand, for bosons, the qualitative behavior of $\zeta^-(T)$ is similar irrespective of the dimension as seen in Fig.~\ref{fig:2}(b). The value of $\zeta^-(T)$ is the maximum at $T=0$ and then decays to zero quickly irrespective of the dimension as the lowest energy level is filled with all the particles  but the value at $T=0$ increases with $d$ mainly owing to the values of dimension dependent chemical potentials given in Eq.~\ref{eq: mu_boson}.

It is further evident  from Fig.~\ref{fig:2} that low-temperature behavior of fermions and bosons would be significantly contrasting as the relative partition functions for fermions and bosons tend to zero and finite nonzero values respectively as $T \rightarrow 0$ and their slopes are positive and negative respectively.  It is also expected that the behavior of both fermions and bosons will strongly depend on the dimension of the system. These corollaries solely follow particle statistics and are independent of the specific form of the cycle. Therefore, it captures the generic nature of the low-temperature behavior of any quantum thermal machines with quantum gases. We show in the next subsection in detail that our results with the Stirling cycle are consistent with these observations.

Given the two baths with two different temperatures and a system with arbitrary dimensions and a number of particles, it is not straightforward to simplify the above expressions to predict the behavior of the cycle. One can easily plot the expressions  to observe the characteristics of the system for a given pair of baths set at two arbitrary temperatures.  However, in the next subsection, we elaborate that it is actually possible to look at the extremely low temperature of the system in $N \rightarrow \infty$ limit analytically and show the contrasting behavior of the system with fermions and bosons in different dimensions.

\subsection{ Analytical approach for extremely low temperature \& $N \rightarrow \infty$ limit of the cycle}

We are specifically interested in the behavior of the Stirling cycle in the quantum regime, i.e., in the extremely low temperatures, given by $T_h \rightarrow 0, T_c \rightarrow 0$ with $\Delta T= (T_h-T_c)>0$. In this limit the expression for work [Eq.~\eqref{eq:work}] reduces to
\begin{equation}
W = \lim_{T \rightarrow 0, \Delta T \rightarrow 0} -((T+\Delta T) \ln \zeta ((T+\Delta T)) + T \ln \zeta(T).   
\end{equation}
Now, let us define the following function of temperature,

\begin{equation}
 \omega(T) = T \ln \zeta(T).
\end{equation}

The slope of this function solely determines, whether, at low temperatures, work can be extracted from the system or not. Work can be extracted from the system only if the slope of the function, i.e., $\lim_{T \rightarrow 0} \frac{d}{dT} \omega(T)$ as $T \rightarrow 0$ is positive and work is done on the system if the same is negative.  Now, in the thermodynamic limit, i.e., $N \rightarrow \infty$, one can analytically evaluate the sign of the above quantity and thereby decide the nature of the system. With reference to Fig. 2, one can find that the slope of the relative function for fermions and bosons are opposite in the low-temperature regime and predict their distinctive behaviors.

\begin{figure*}
\begin{center}             
\includegraphics[width=0.49\textwidth]{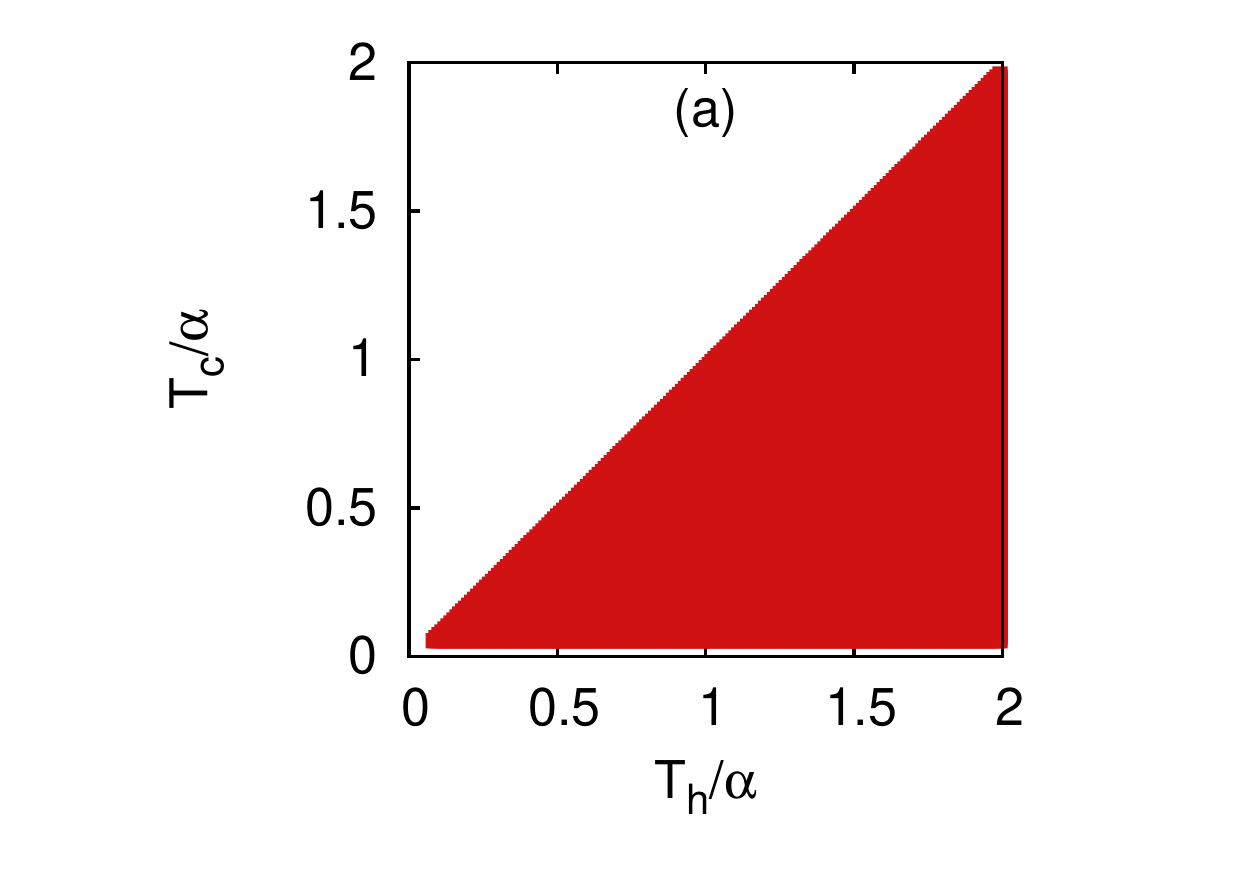}
\includegraphics[width=0.49\textwidth]{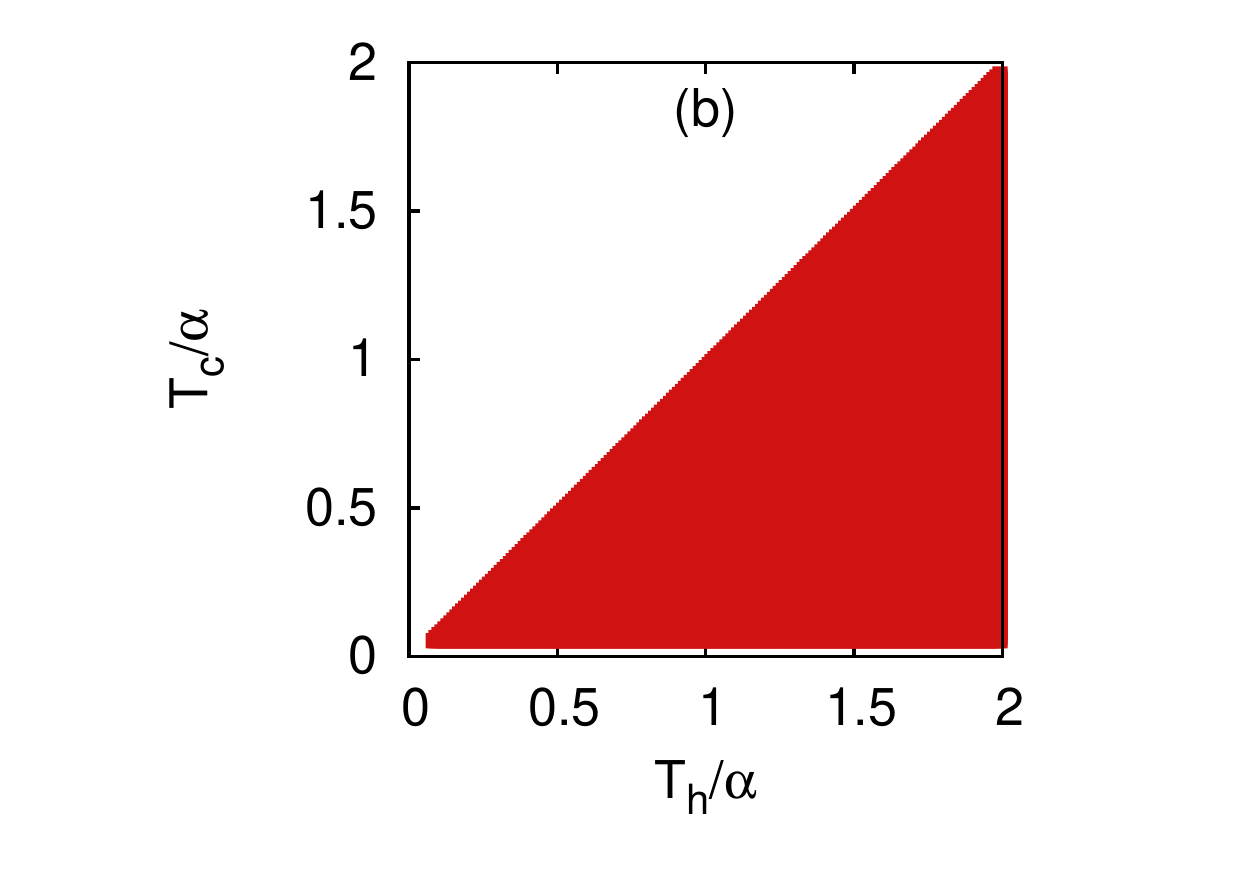}
\includegraphics[width=0.49\textwidth]{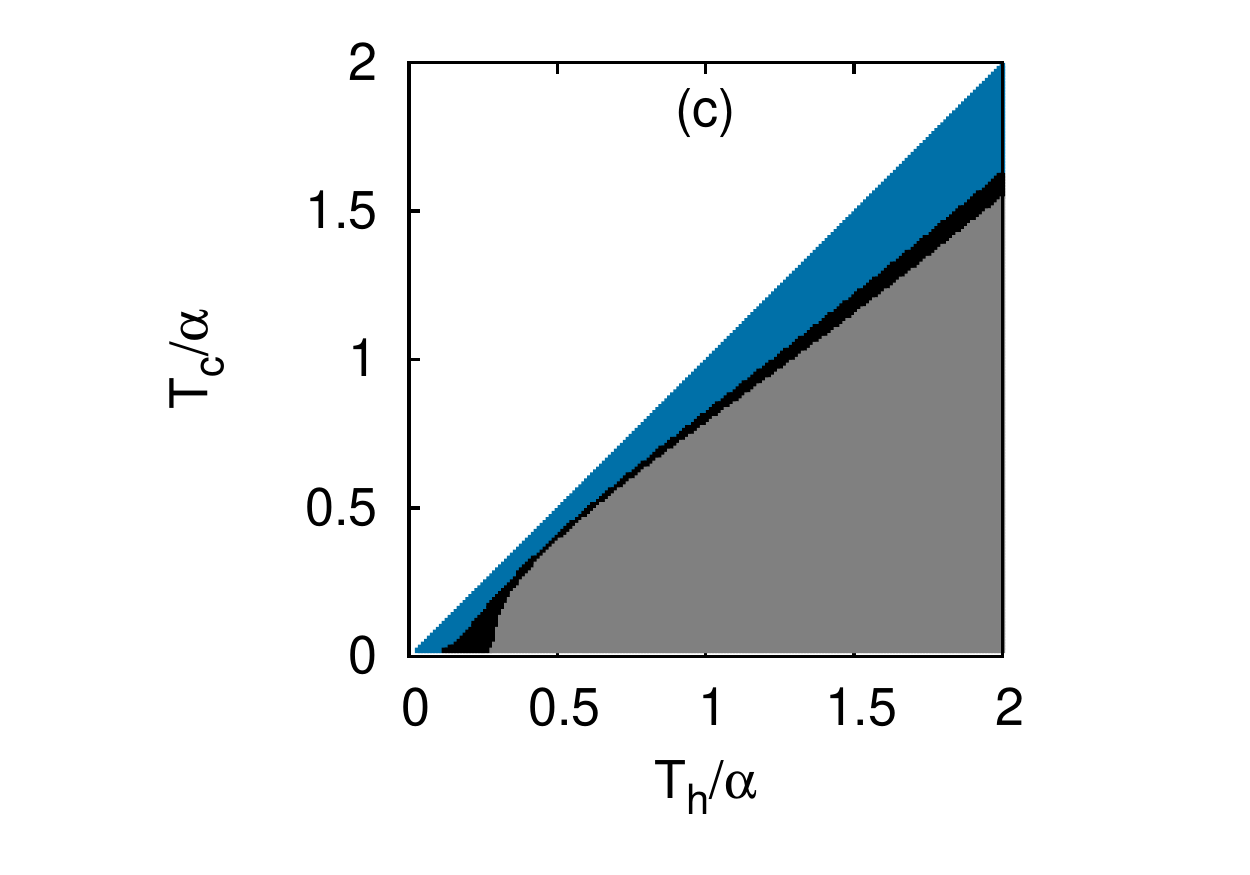}
\includegraphics[width=0.49\textwidth]{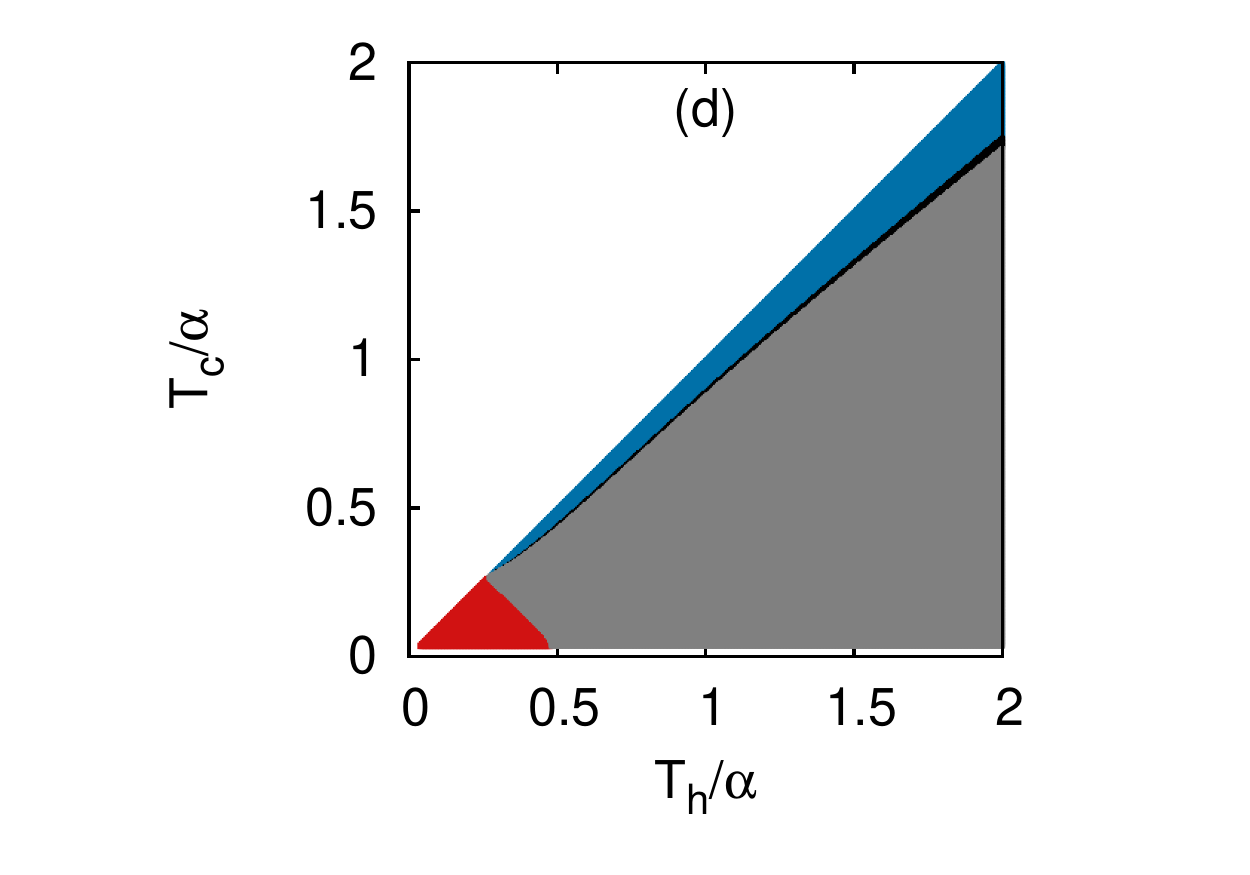}
\includegraphics[width=0.49\textwidth]{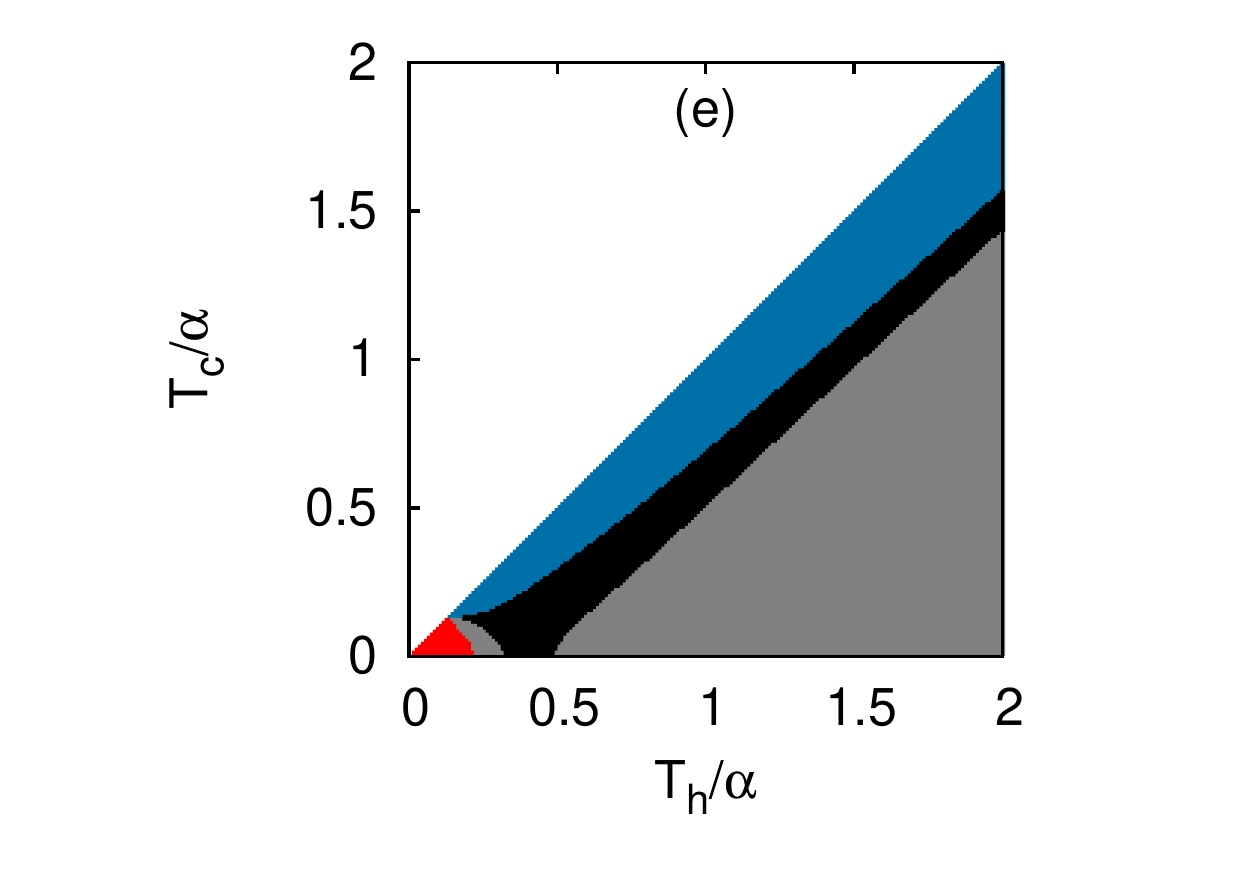}
\includegraphics[width=0.49\textwidth]{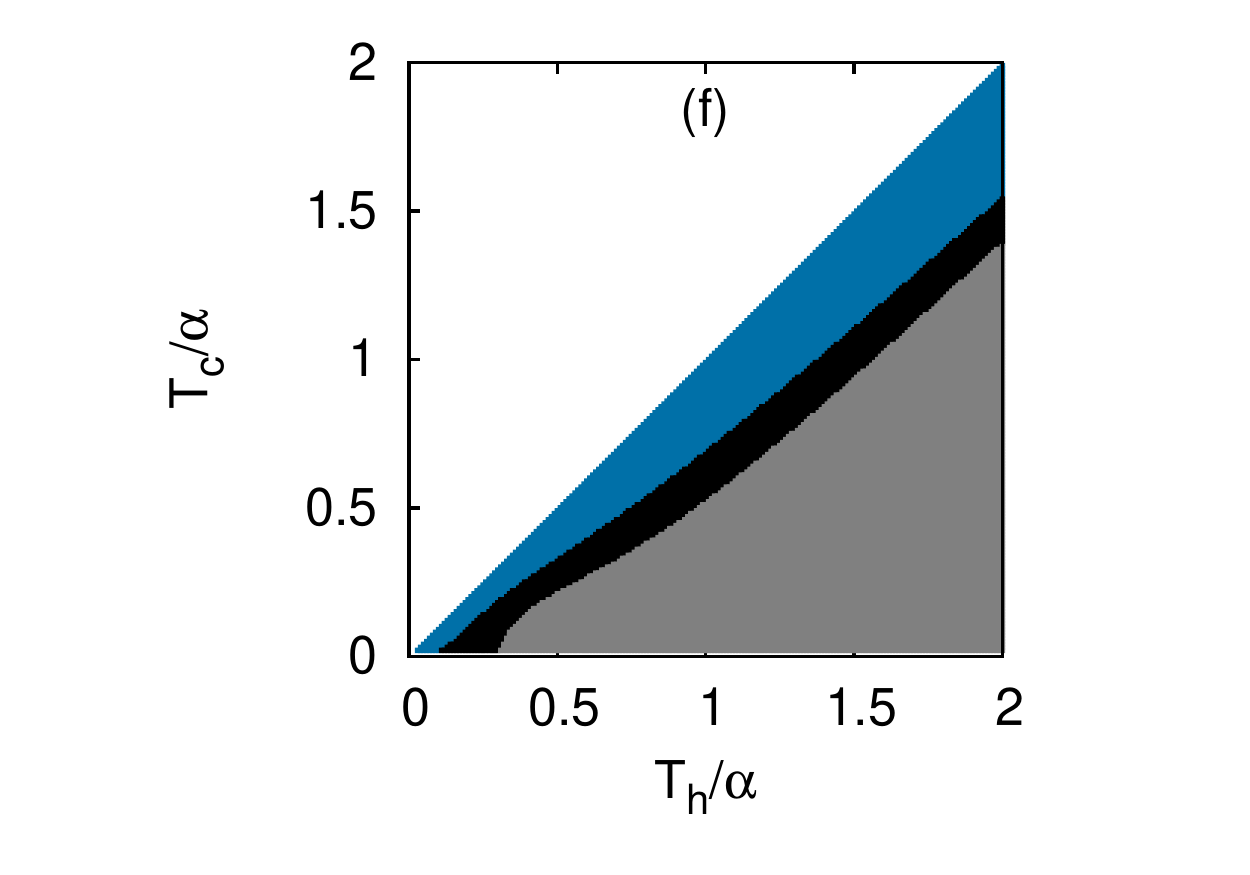}   
  \end{center}
  \caption{{Different modes of operation of a Stirling engine with a gas of fermions.  The modes of operation as functions of $T_h$ and $T_c$ for: (a)  $d = 1$, $N = 20$; (b) $d = 1$, $N = 40$; (c) $d = 2$, $N = 20$; (d) $d = 2$, $N = 40$; (e) $d = 3$, $N = 20$; (f) $d = 3$, $N = 40$. The regimes corresponding to Engine, Refrigerator, Accelerator, and Heater are marked with red, blue, grey, and black respectively.}}
  \label{fig:3_1}
  \end{figure*}

\begin{figure*}
\begin{center}             
\includegraphics[width=0.49\textwidth]{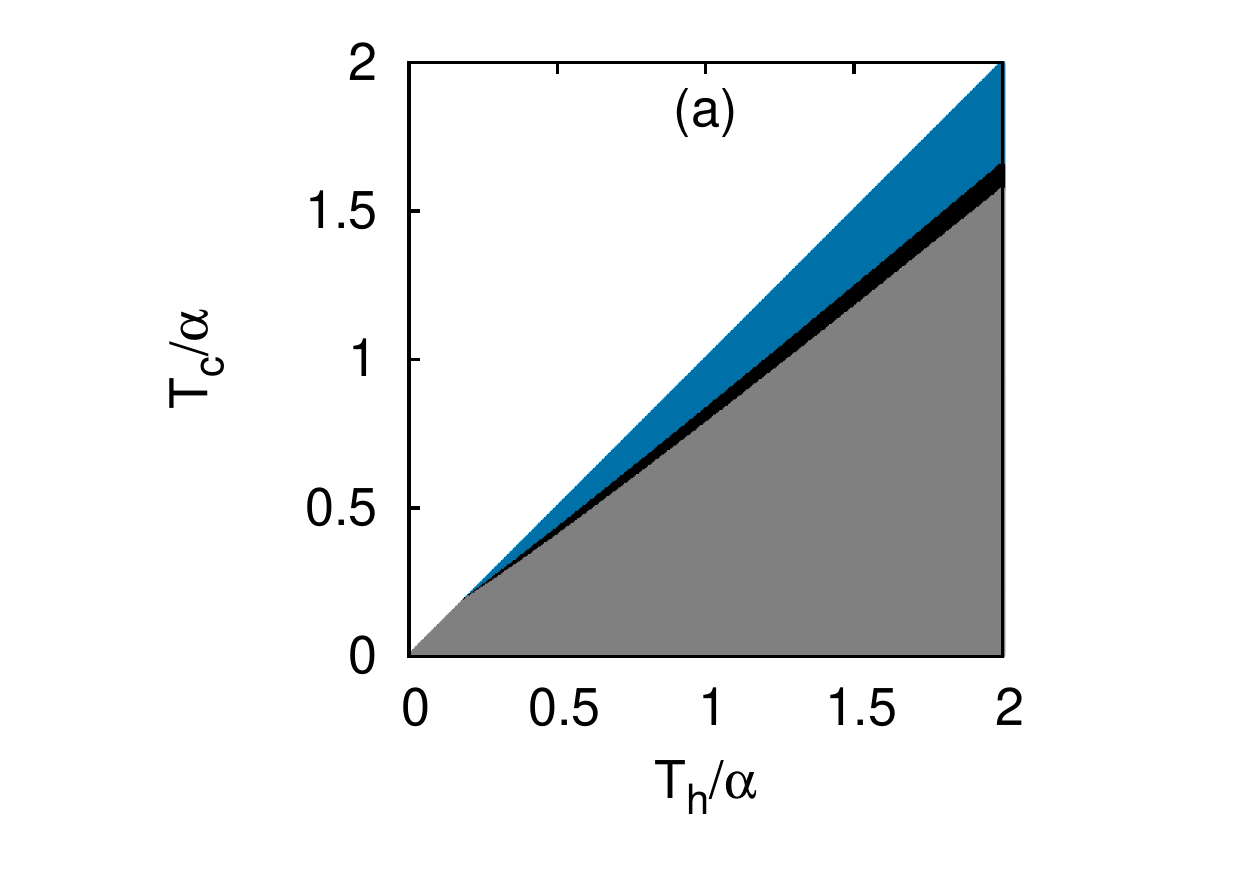}
\includegraphics[width=0.49\textwidth]{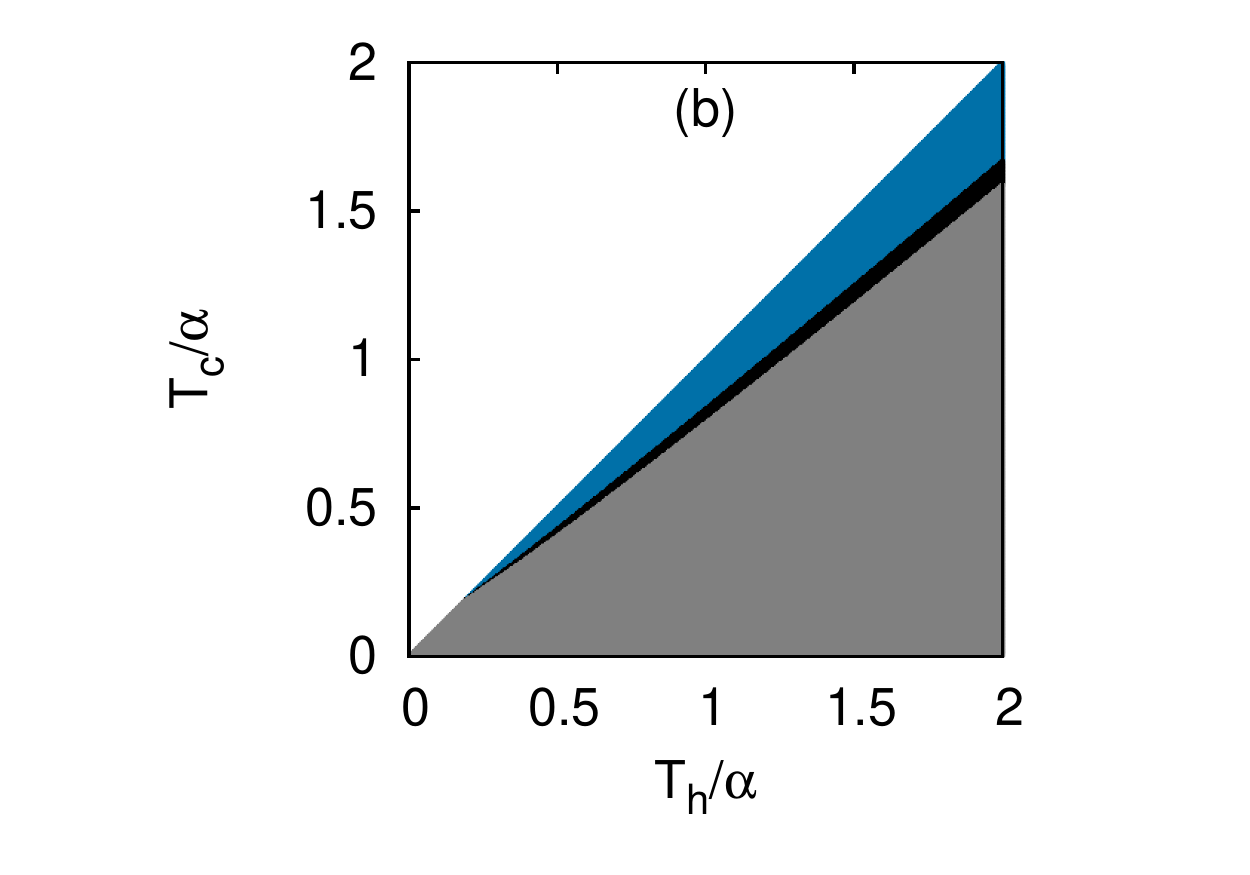}
\includegraphics[width=0.49\textwidth]{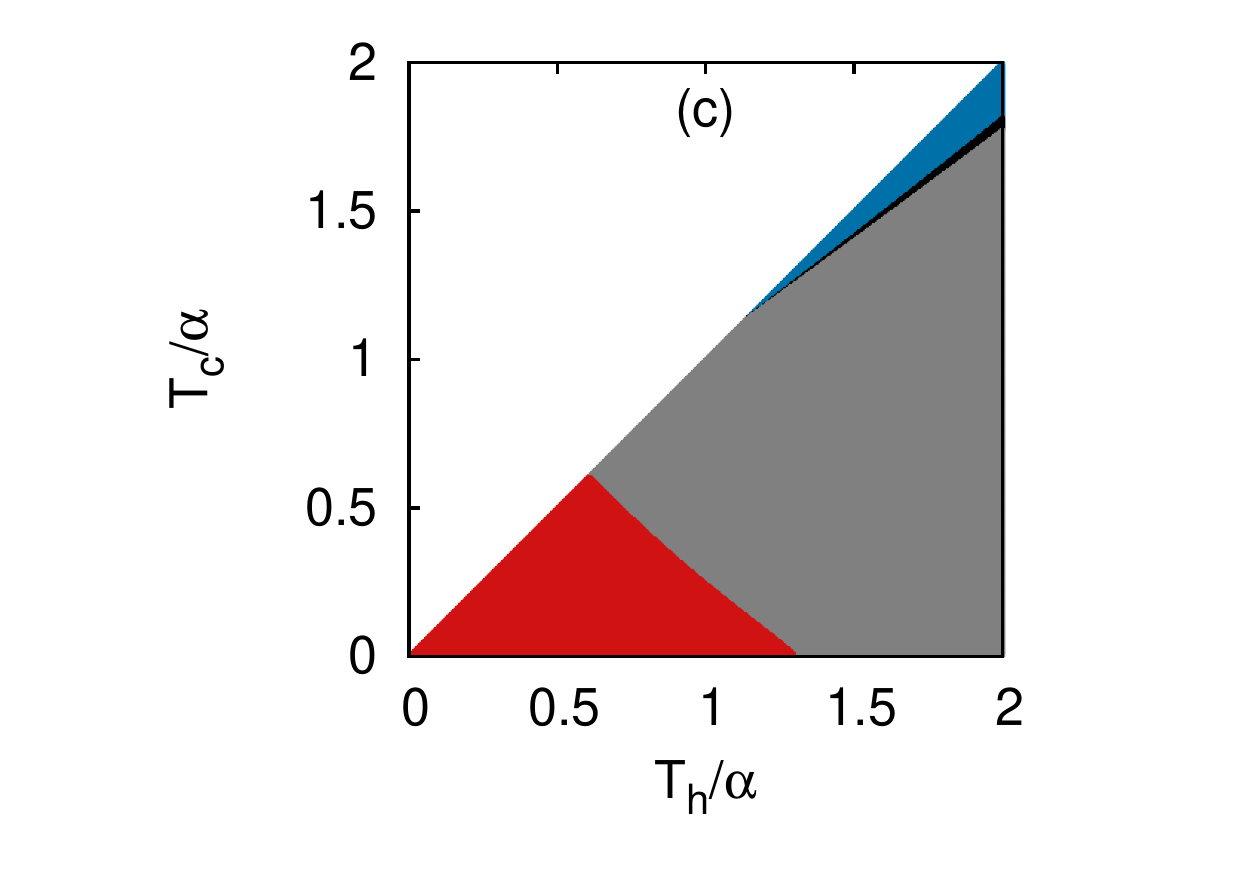}
\includegraphics[width=0.49\textwidth]{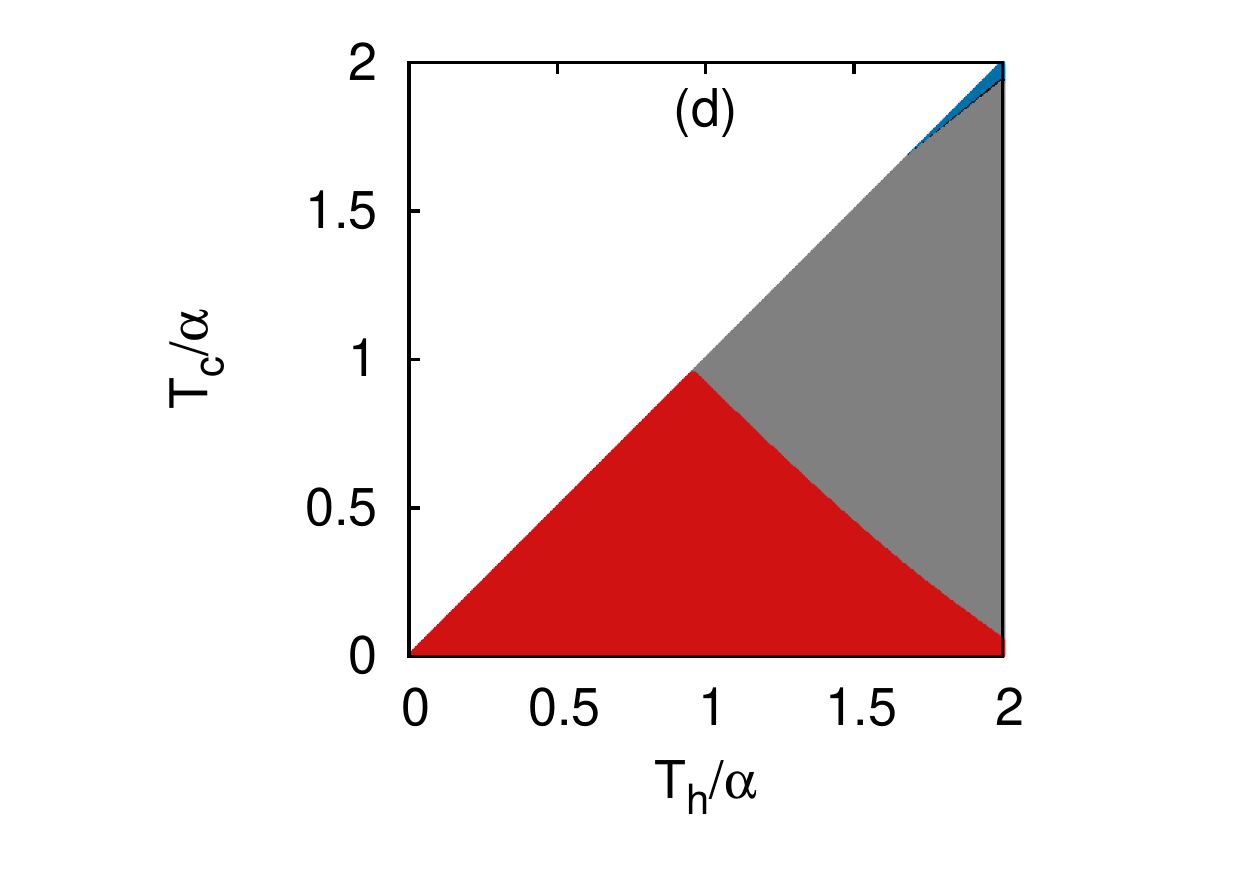}
\includegraphics[width=0.49\textwidth]{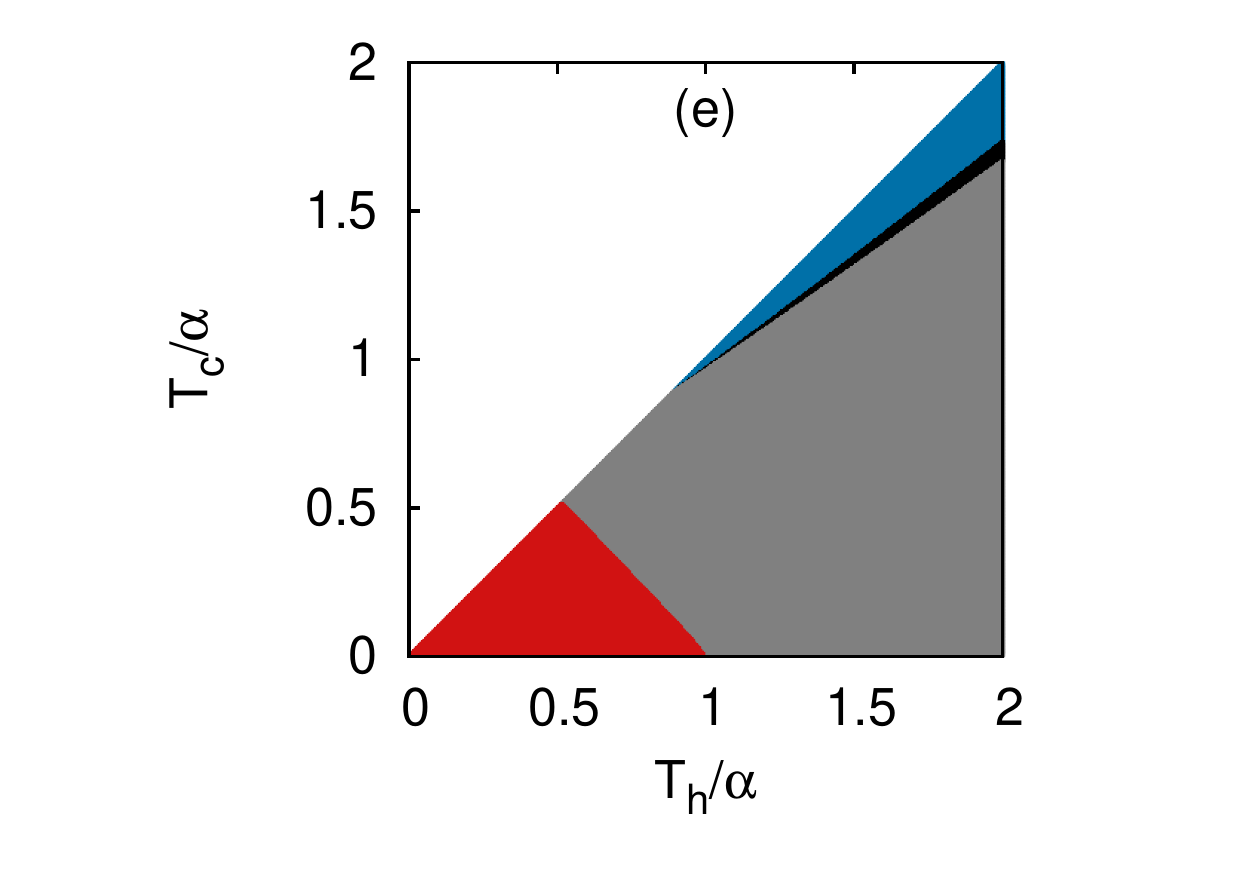}
\includegraphics[width=0.49\textwidth]{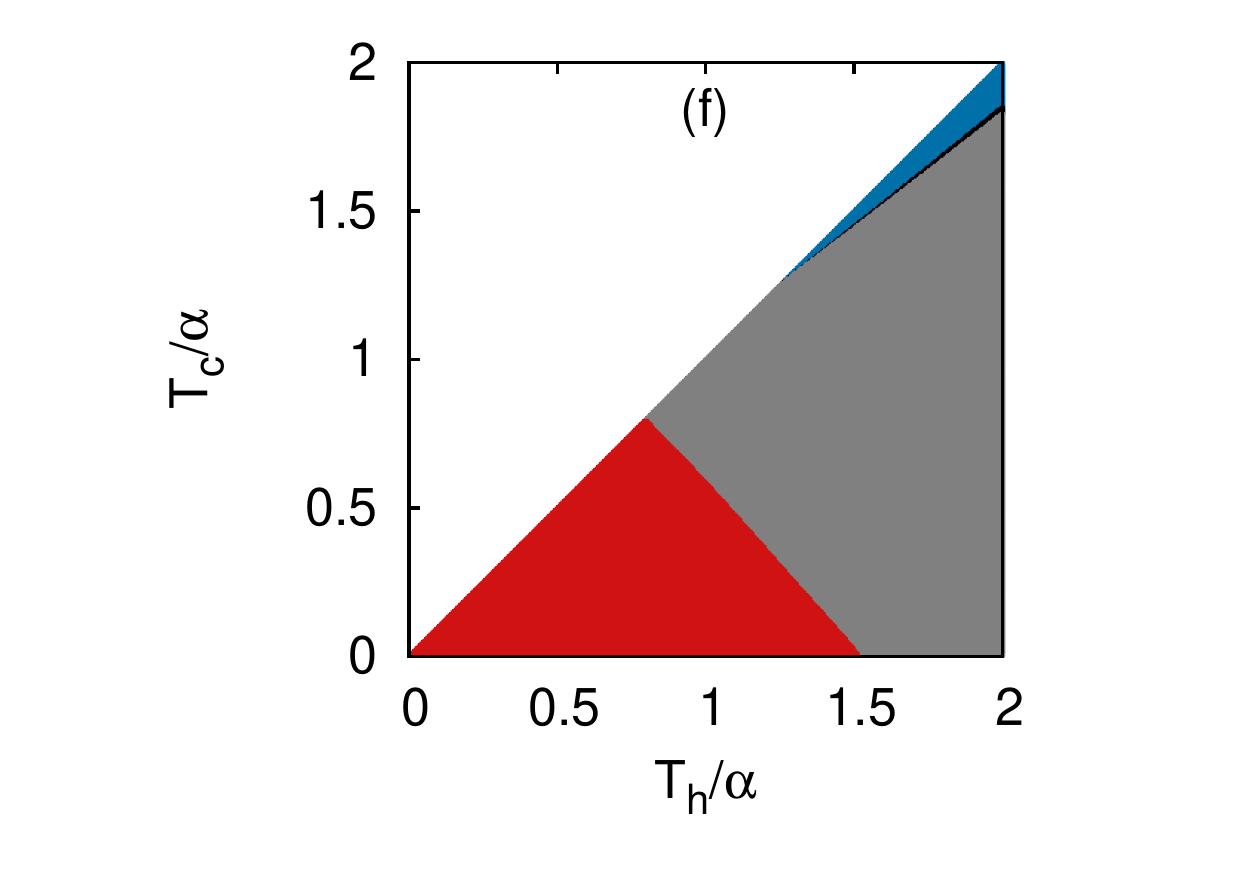}   
  \end{center}
  \caption{{Different modes of operation of a Stirling engine with a gas of bosons. The modes of operation as functions of $T_h$ and $T_c$ for: (a)  $d = 1$, $N = 20$; (b) $d = 1$, $N = 40$; (c) $d = 2$, $N = 20$; (d) $d = 2$, $N = 40$; (e) $d = 3$, $N = 20$; (f) $d = 3$, $N = 40$. The regimes corresponding to Engine, Refrigerator, Accelerator, and Heater are marked with red, blue, grey, and black respectively.}}
  \label{fig:3_2}
  \end{figure*}
  
  \begin{figure*}
\begin{center}
\includegraphics[width=0.49\textwidth]{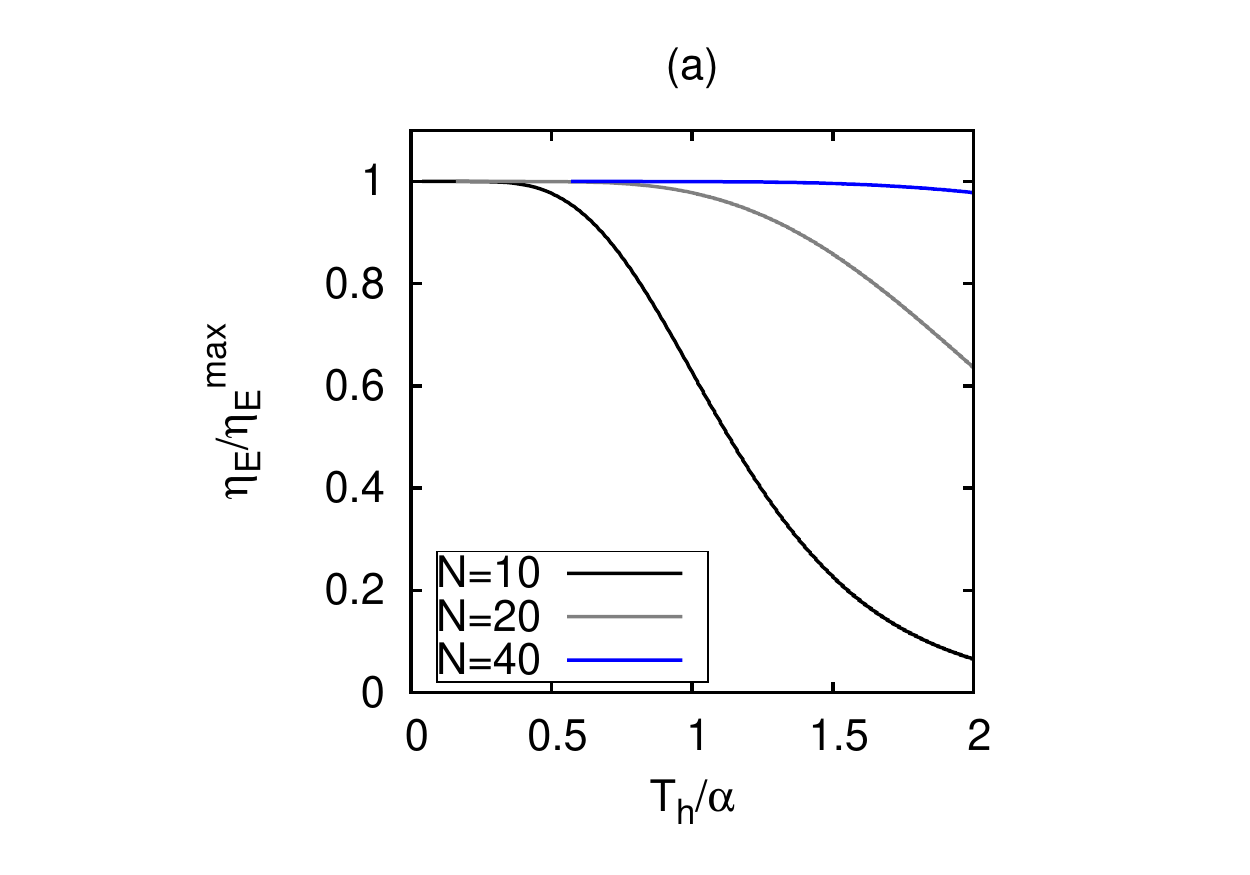}
\includegraphics[width=0.49\textwidth]{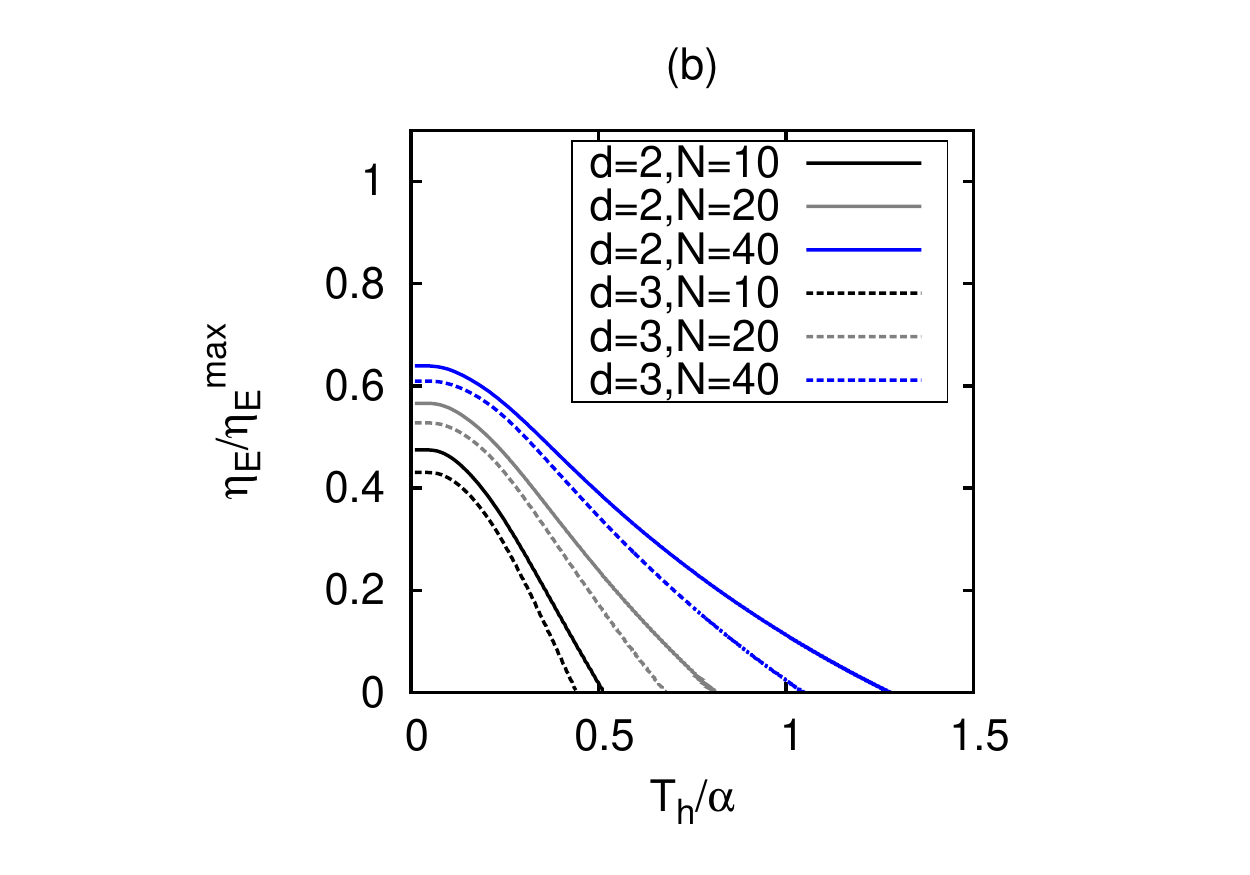}
\end{center}
\caption{{Dependence on the particle number $N$ on engine efficiency. Engine efficiency scaled w.r.t. Carnot efficiency $\eta_E/\eta^{max}_E$ for a fixed $T_c/T_h = 0.50$ for (a) fermions in $d=1$ and (b) bosons in $d=2, 3$.  }}
\label{fig:4}
\end{figure*}

\subsubsection{Fermions}
Let us first consider the case of non-interacting fermions in a one-dimensional potential well. The function $\omega(T)$ takes the form of [Cf. Eq.~\eqref{xi-f-1d}]

\begin{eqnarray}
 \omega(T) = T \ln  \prod^\infty_{n =1} \frac{\left(1+e^{-\alpha((2n)^2 - N^2/)/8T}\right)}{\left(1+e^{-\alpha( (2n-1)^2 -N^2)/8T}\right)}.
\end{eqnarray}

In the aforementioned limiting cases, the quantities $\lim_{T \rightarrow 0} \lim_{N \rightarrow \infty} e^{\frac{\alpha}{8T}(4n^2-N^2)}$ and $\lim_{T \rightarrow 0} \lim_{N \rightarrow \infty} e^{\frac{\alpha}{8T}((2n-1)^2-N^2)}$ both are very small positive numbers and $$ e^{\frac{\alpha}{8T}(4n^2-N^2)} > e^{\frac{\alpha}{8T}((2n-1)^2-N^2)} {~} \quad \quad \forall {~} n.$$ Hence, expanding $\frac{d}{dT} \omega(T)$ series of $n$  we get
\begin{eqnarray}
 \frac{d}{dT} \omega(T) = \frac{\alpha}{8T} \sum^\infty_{n=1} \left(N^2-4n^2+1\right)\left[e^{\frac{\alpha}{8T}(4n^2-N^2)} \right.\nonumber\\\left.-  e^{\frac{\alpha}{8T}((2n-1)^2-N^2)}\right] + (1-4n) e^{\frac{\alpha}{8T}((2n-1)^2-N^2)}.
 \label{eq:fermion_w} 
\end{eqnarray}
 Clearly, for $N \rightarrow \infty$, the quantity $\frac{d}{dT} \omega(T)$ is positive. This implies that work can be extracted from a system with a large number of non-interacting fermions in low-temperature limits. Therefore from Table~\ref{tab:1} one concludes that the system behaves exclusively as a heat engine. 

The efficiency of the heat engine is then given as 
\begin{eqnarray}
 \eta_E = -\frac{W}{Q_h} = \frac{1- \frac{T_c \ln \zeta^+(T_c)}{T_h \ln \zeta^+(T_h)}} {1+\frac{U^+_B- U^+_D}{T_h\ln \zeta^+(T_h)}}.
\end{eqnarray}
In the aforementioned limit, the quantities $T_h\ln \zeta^+(T_h) \rightarrow -\infty$ and $\frac{T_c \ln \zeta^+(T_c)}{T_h \ln \zeta^+(T_h)} \rightarrow \frac{T_c}{T_h}$. Therefore, efficiency $\eta_E \rightarrow 1-\frac{T_c}{T_h}$, i.e., the Carnot limit. The above equation connotes that a large number of non-interacting fermions entrapped in a one-dimensional potential well behave like a heat engine and its efficiency tends to Carnot limit as the bath temperatures approach absolute zero. Here we restate that it is purely a quantum effect with no classical correspondence, yet the machine still works with the highest possible efficiency and abides by the second law of thermodynamics. This is the first important result of our analysis exhibiting true quantum signature at a macroscopic scale.  

But for fermions at higher dimensions, the behavior of the system is entirely different because of the different number dependence of the chemical potential. The function $\omega(T)$, in this case, is given by [Cf. Eq.~\eqref{eq: zeta_fermion_d}]
\begin{eqnarray}
 \omega(T) = T \ln \prod^\infty_{n_{x_i} =1}\prod^1_{j_{x_i}=0}\frac{\left(1+e^{-(\alpha \sum_{i}(2n_{x_i})^2/8 -\tilde{\mu})/T}\right)}{\left(1+e^{-(\alpha \sum_{i}(2n_{x_i}-j_{x_i})^2/8 -\tilde{\mu})/T}\right)}.\nonumber\\
\end{eqnarray} 
 We know that in the low-temperature limit, the chemical potential $\tilde{\mu}$ can be taken equal to be the Fermi energy $\tilde{E}_F$. It can be seen from Eq.~\eqref{eq:fermi_energy} that the chemical potential in  higher dimensions varies in a sub-quadratic fashion with $N$ as $\tilde{\mu}  \propto N^\gamma$, $\gamma < 2$, whereas the energy dispersion is still quadratic as given in Eq.~\eqref{eq: energy_level}. As a result, in the macroscopic thermodynamic limit $N \rightarrow \infty$, the energy always outgrows the chemical potential. Now, in the limit $N \rightarrow \infty$ and $T \rightarrow 0$, the expression $\frac{1}{T} (\alpha\sum_{i}(2n_{x_i})^2/8 -\tilde{\mu})$ tends to $\infty$ or $-\infty$  depending on the values of $\tilde{\mu}$.  If $\frac{1}{T} (\alpha\sum_{i}(2n_{x_i})^2/8 -\tilde{\mu})$ tends to $\infty$, then $e^{-\frac{1}{T} (\alpha \sum_{i}(2n_{x_i})^2/8 -\tilde{\mu})} \rightarrow 0$, on the other hand if $\frac{1}{T} (\alpha\sum_{i}(2n_{x_i})^2/8 -\tilde{\mu})$ tends to $-\infty$, then $e^{\frac{1}{T} (\alpha\sum_{i}(2n_{x_i})^2/8 -\tilde{\mu})} \rightarrow 0$. So, we split up the sum over $n_{x_i}, j_{x_i}$ in two parts and denote them by $\sum_1$ and $\sum_2$ for these two cases respectively. Hence,

  \begin{eqnarray}
  &&\frac{d}{dT} \omega(T) = \sum_{1} \left[e^{-F_1(n_{x_i}, j_{x_i})} -e^{-F_2(n_{x_i}, j_{x_i})}\right] \nonumber\\
  && + \sum_1 \left[F_1(n_{x_i}, j_{x_i})e^{-F_1(n_{x_i}, j_{x_i})} -F_2(n_{x_i}, j_{x_i})e^{-F_2(n_{x_i}, j_{x_i})}\right] \nonumber\\ 
  &&-\sum_2  \left[F_1(n_{x_i}, j_{x_i})e^{F_1(n_{x_i}, j_{x_i})} -F_2(n_{x_i}, j_{x_i})e^{F_2(n_{x_i}, j_{x_i})}\right],\nonumber\\
  \label{eq: fermion_d}
 \end{eqnarray}

{where, $F_1(n_{x_i}, j_{x_i}) = \frac{1}{T}(\frac{\alpha}{8}\sum_{i}(2n_{x_i})^2 -\tilde{\mu})$ and $F_2(n_{x_i}, j_{x_i}) = \frac{1}{T}(\frac{\alpha}{8}\sum_{i}(2n_{x_i} - j_{x_i})^2 -\tilde{\mu})$.

Using the following inequalities for each set of integers $(n_{x_1}, n_{x_2}, ...)$ and $j_{x_i} = 0,1$,
\begin{eqnarray}
&& e^{\frac{1}{T}(\frac{\alpha}{8}\sum_{i}(2n_{x_i})^2 -\tilde{\mu})} \ge e^{\frac{1}{T}(\frac{\alpha}{8}\sum_{i}(2n_{x_i}-j_{x_i})^2 -\tilde{\mu})},  \nonumber\\
&& e^{-\frac{1}{T}(\frac{\alpha}{8}\sum_{i}(2n_{x_i} -j_{x_i})^2 -\tilde{\mu})} \ge e^{-\frac{1}{T}(\frac{\alpha}{8}\sum_{i}(2n_{x_i})^2 -\tilde{\mu})},
\end{eqnarray}
one finds that the first and the third terms are negative and positive respectively and the second term can be both positive and negative. Therefore, $ \frac{d}{dT} \omega(T)$ can be both positive and negative depending on the actual functional forms of the chemical potential and dimension. Therefore, the non-interacting fermions at low temperatures in more than one dimension can behave as a heat engine or a refrigerator, accelerator, and heater depending on the sign of $Q_h$ and $Q_c$. The interplay between the number dependence on the chemical potential and the energy levels in $d>2$ dictates the behavior of fermions, which we explicitly show in numerical results later. This is the second important observation of our analysis.

\subsubsection{Bosons}

Now, we consider the case of non-interacting bosons in a one-dimensional box. The function $\omega(T)$ takes the form of [Eq.~\eqref{xi-b-1d}]
\begin{eqnarray}
 \omega(T) = T \ln  \prod^\infty_{n =1} \frac{\left(1-e^{-(\alpha(2n-1)^2/8 -\tilde{\mu})/T}\right)}{\left(1-e^{-(\alpha(2n)^2/8 -\tilde{\mu}')/T}\right)},
\end{eqnarray}

Expanding $\frac{d}{dT} \omega(T)$ in the $T \rightarrow 0$ limit we get 

\begin{eqnarray}
 &&\frac{d}{dT} \omega(T) = \sum^\infty_{n=1}  \ln \frac{ (1-e^{-(\alpha(2n-1)^2/8 -\tilde{\mu})/T})}{(1-e^{-(\alpha(2n)^2/8 -\tilde{\mu}')/T})}\nonumber\\
 && + \frac{1}{T} \sum^\infty_{n=1} ( (\frac{\alpha}{8}(2n)^2 -\tilde{\mu}')e^{-(\alpha(2n)^2/8 -\tilde{\mu}')/T} \nonumber\\ &&- (\frac{\alpha}{8}(2n-1)^2 -\tilde{\mu})e^{-(\alpha(2n-1)^2/8 -\tilde{\mu})/T}).\nonumber\\
  \end{eqnarray}

In the $N \rightarrow \infty$ limit, using  the inequality
\begin{eqnarray}
\frac{\alpha}{8}(2n)^2 -\tilde{\mu}' > \frac{\alpha}{8}(2n-1)^2 -\tilde{\mu} {~~} \forall n \ge 1
\label{eq: inequality_boson}
\end{eqnarray}
with $\mu = \alpha/8$ and $\mu' = \alpha/2$ (Cf. Eq.~\ref{eq: mu_boson}), one readily shows the quantity $d\omega(T)/dT$ to be negative and therefore, implying  that a system with large number of non-interacting bosons in $d=1$, behaves as a refrigerator or an accelerator or a heater depending on the signs of $Q_h$ and $Q_c$.

But for bosons at higher dimensions, the function $\omega(T)$, in this case, is given by [Cf. Eq.~\eqref{eq: zeta_boson_d}]
\begin{eqnarray}
 \omega(T) = T \ln \prod^\infty_{n_{x_i} =1}\prod^1_{j_{x_i}=0}\frac{\left(1-e^{-(\alpha \sum_{i}(2n_{x_i}-j_{x_i})^2/8 -\tilde{\mu})/T}\right)}{\left(1-e^{-(\alpha \sum_{i}(2n_{x_i})^2/8 -\tilde{\mu}')/T}\right)}.\nonumber\\
\end{eqnarray} 

In the limit $T \rightarrow 0$, unlike fermions, only the ground state is occupied by the bosons and consequently, the occupation at the excited states is negligible. In the $N \rightarrow \infty$ limit, we use the following inequality for the set of integers $n_{x_i} = 1$ and $j_{x_i} = 0,1$,

\begin{eqnarray}
 \frac{\alpha}{8}(\sum_i(2n_{x_i})^2) - \tilde{\mu}' \le \frac{\alpha}{8}(\sum_i(2n_{x_i}-j_{x_i})^2) - \tilde{\mu},{~~} \forall d\ge 2, \nonumber\\
 \label{eq: inequality_boson_2}
\end{eqnarray}
with $\tilde{\mu} = \alpha d/8$ and $\tilde{\mu}' = \alpha d/2$, one finds that $d\omega/dT$ is positive (note that the inequalities in Eqs. \eqref{eq: inequality_boson} and \ref{eq: inequality_boson_2} have reversed signs). This implies that the system with non-interacting bosons in $d\ge 2$ will behave as a heat engine in the low-temperature limit.
The efficiency of the heat engine is then given as}
\begin{eqnarray}
 \eta_E = -\frac{W}{Q_h} = \frac{1- \frac{T_c \ln \zeta^-(T_c)}{T_h \ln \zeta^-(T_h)}} {1+\frac{U^-_B- U^-_D}{T_h\ln \zeta^-(T_h)}}.
\end{eqnarray}
In the limit $T_h,T_c \rightarrow 0$,  both the quantities $ \zeta^-(T_h)$ and $\zeta^-(T_c)$ are nonzero,  $\ln \zeta^-(T_c) > \ln \zeta^-(T_h)$ [Fig.~\ref{fig:2}(b)] and $U^-_B > U^-_D$ ensure an efficiency less than the Carnot bound, unlike the fermions in $d=1$. Later, we plot the efficiency as a function of $N$ in Fig.~\ref{fig:4}(b) for $d=2$ and $d=3$. 

In Fig.~\ref{fig:3_1} we present different modes of operations for fermions in one, two, and three dimensions for $N = 20$ and $40$ from numerical calculations. Fermions in a one-dimensional box behave strictly as a heat engine in the $T_h-T_c$ parameter space [Fig.~\ref{fig:3_1}(a),(b)] as predicted from the analytical approach. The efficiency is almost constant and equal to the Carnot bound in the low-temperature regime $T_h \lesssim \alpha$. On the other hand, in $d=2$ and $d=3$ we predicted from Eq.~\eqref{eq: fermion_d} that all the four modes can  coexist in  the $T_h-T_c$ plane in the low-temperature limit. In $d=2$, from numerical calculations also we find that the three modes (refrigerator, accelerator, and heater) with $W>0$ for $N=20$ and all the four modes (engine, refrigerator, accelerator, and heater) with $N=40$ coexist in the $T_h-T_c$ plane [Fig.~\ref{fig:3_1}(c), (d)].  In $d=3$,  we find that all the four  modes  (refrigerator, engine, heater, and accelerator) for $N=20$ and the three modes (refrigerator, heater, and accelerator) with $W>0$ for $N=40$ coexist ~[Fig.~\ref{fig:3_1}(e), (f)]. In all the cases above, when $T_c \lesssim T_h$, the system works as a refrigerator but as one increases $T_h$ while keeping $T_c$ fixed, the region of the accelerator is reached. A narrow region of the heater demarcates the boundary between the regions of the refrigerator and accelerator where $Q_h$ flips its sign. The numerical results conclude that the low-temperature behavior of fermions for $d>1$ depends both on the dimensions and the number of particles in the system owing to the interplay between the number dependence of the chemical potential and energy levels, as explained through our analytical approach.   

In contrast to fermions, as predicted from our analytical approach, the system with bosons in $d=1$ predominantly behaves as an accelerator with  small regions of refrigerator and heater [Figs.~\ref{fig:3_2}(a)-(b)]. But it shows qualitatively different behavior in $d=2,3$ [Figs.~\ref{fig:3_2}(c)-(f)], where there exist all four distinct regions but near $T_h, T_c \rightarrow 0$, the system behaves as a heat engine as predicted from the analysis. We have shown the results with two different values of $N$; Figs.~\ref{fig:3_2}(a)-(e) for $N=20$ and Figs.~\ref{fig:3_2}(b)-(f) for $N=40$, but no qualitative difference depending on $N$ is observed.

Some common remarks can be made from the results shown in Fig.~\ref{fig:3_1} and Fig.~\ref{fig:3_2}. (a) The fermi gas in one dimension is the most desired system to construct a heat engine with Carnot efficiency. However, a boson gas is also a candidate for heat engines provided the dimension of the system is greater than one.  (b) The region of refrigerator mode is located near the boundary $T_h = T_c$ and the region of the accelerator on the other boundary $T_c \sim 0$ for bosons and fermions in $d=2,3$. A small region of the heater demarcates the boundary of transition between the refrigerator and the accelerator. Therefore, the refrigerator is not a suitable mode for quantum gases, in general.

\subsection{Dependence on average particle number $N$} 

We have already shown that a Stirling cycle with fermions in one dimension exclusively works as an engine whereas with bosons it both behaves as a refrigerator and a heater.  To explore the $N$ dependence for fermions, we plot the heat engine efficiency scaled w.r.t the Carnot efficiency, $\eta_E/ \eta^{max}_E$ in Fig.~\ref{fig:4}(a), with the number of particles $N$ and $T_h$, in one-dimensional box keeping the ratio $T_c/T_h =0.5$ fixed. The efficiency reaches the Carnot bound as the bath temperatures tend to zero. It is interesting to see that for a given pair of baths with non-zero temperatures, a system with larger $N$ yields better efficiency. Owing to a prominent number of dependence on fermions, the engine efficiency can be boosted by adding more fermions to the system. 

In Fig.~\ref{fig:4}(b), we plot the engine efficiency for bosons scaled w.r.t the Carnot efficiency,  with the number of particles $N$, and $T_h$, for $d=2,3$ keeping the ratio $T_c/T_h = 0.5$ fixed. The scaled efficiency tends to values much less than 1 as the bath temperatures tend to zero irrespective of $N$ and $d$ as predicted from our analysis and it also goes to zero when the system switches to the accelerator mode as shown already in Figs.~\ref{fig:3_2}(c)-(f). The maximum coefficient of performance increases slightly with $N$, as seen from Fig.~\ref{fig:4}(b), reveals that bosons do not show any prominent dependence on $N$, unlike fermions.


\section{Conclusions}\label{Sec.IV}

To conclude, we have demonstrated some fundamental features of quantum particles in the context of the quantum Stirling cycle. \textcolor{black} {Given a thermodynamics cycle, one can anticipate that the  behaviour of the working medium in the quantum regime is fundamentally determined only by the quantum statistics of the particles.}  We analyze the role of quantum statistics and system dimensions in determining the collective thermodynamic behaviour of particles.  It is worth mentioning that the efficiency considered here exhibits a pure quantum signature with no classical analogue and manifests only in the quantum regime. Though our analysis is focused on a specific working medium, viz., quantum gas trapped in an infinite potential well or a square well, one can trivially generalize our analysis for any system. A similar analysis with other trapping potentials would connote a qualitatively similar result as a consequence of the statistical nature of identical quantum particles.

We find that bosons behave completely opposite to fermions as a manifestation of the fundamental difference between the particle statistics in the quantum domain. Therefore, fermions and bosons are useful in a different way from the viewpoint of constructing a desirable thermodynamic machine. \textcolor{black}{A Stirling cycle with fermions confined in a  one-dimensional potential well, when connected to two low-temperature baths, behaves exclusively like a heat engine ($W<0$) while a Stirling cycle with bosons behaves like an accelerator, refrigerator and a heater ($W>0$) depending on the bath temperatures. On the other hand, the behaviour of bosons and fermions is qualitatively similar in two and three-dimensional systems. In both cases, unlike in one dimension, a mixture of four different modes can be observed depending on the bath temperatures and the number of particles.}

We have also shown that the particle number or the dimension does not appreciably affect the performance of bosons. Unlike bosons, the number dependence of the chemical potential for fermions decides the behaviour of the system. As the dimension of the system decides the number dependence of the chemical potential, it determines the overall thermodynamic behaviour of the system. The behaviour of fermions and bosons in a one-dimensional well is completely different compared to that of two or higher-dimensional boxes.  We have found that increasing the number of particles in a system  can boost its performance in spite of the engine efficiency/coefficient of performance being bounded by the Carnot bound.



\acknowledgments{S.S. thanks Dr. Amit Kumar Jash (WIS) for fruitful discussions regarding recent experiments on quantum wires and wells. A.G. acknowledges ICTS for participating in the program - Physics with Trapped Atoms, Molecules and Ions (ICTS/TAMIONs-2022/5).}


\bibliography{manuscript.bib}

\begin{thebibliography}{83}%
\makeatletter
\providecommand \@ifxundefined [1]{%
 \@ifx{#1\undefined}
}%
\providecommand \@ifnum [1]{%
 \ifnum #1\expandafter \@firstoftwo
 \else \expandafter \@secondoftwo
 \fi
}%
\providecommand \@ifx [1]{%
 \ifx #1\expandafter \@firstoftwo
 \else \expandafter \@secondoftwo
 \fi
}%
\providecommand \natexlab [1]{#1}%
\providecommand \enquote  [1]{``#1''}%
\providecommand \bibnamefont  [1]{#1}%
\providecommand \bibfnamefont [1]{#1}%
\providecommand \citenamefont [1]{#1}%
\providecommand \href@noop [0]{\@secondoftwo}%
\providecommand \href [0]{\begingroup \@sanitize@url \@href}%
\providecommand \@href[1]{\@@startlink{#1}\@@href}%
\providecommand \@@href[1]{\endgroup#1\@@endlink}%
\providecommand \@sanitize@url [0]{\catcode `\\12\catcode `\$12\catcode
  `\&12\catcode `\#12\catcode `\^12\catcode `\_12\catcode `\%12\relax}%
\providecommand \@@startlink[1]{}%
\providecommand \@@endlink[0]{}%
\providecommand \url  [0]{\begingroup\@sanitize@url \@url }%
\providecommand \@url [1]{\endgroup\@href {#1}{\urlprefix }}%
\providecommand \urlprefix  [0]{URL }%
\providecommand \Eprint [0]{\href }%
\providecommand \doibase [0]{http://dx.doi.org/}%
\providecommand \selectlanguage [0]{\@gobble}%
\providecommand \bibinfo  [0]{\@secondoftwo}%
\providecommand \bibfield  [0]{\@secondoftwo}%
\providecommand \translation [1]{[#1]}%
\providecommand \BibitemOpen [0]{}%
\providecommand \bibitemStop [0]{}%
\providecommand \bibitemNoStop [0]{.\EOS\space}%
\providecommand \EOS [0]{\spacefactor3000\relax}%
\providecommand \BibitemShut  [1]{\csname bibitem#1\endcsname}%
\let\auto@bib@innerbib\@empty
\bibitem [{\citenamefont {Scully}(2002)}]{scully_2002}%
  \BibitemOpen
  \bibfield  {author} {\bibinfo {author} {\bibfnamefont {Marlan~O.}\
  \bibnamefont {Scully}},\ }\bibfield  {title} {\enquote {\bibinfo {title}
  {Quantum afterburner: Improving the efficiency of an ideal heat engine},}\
  }\href {\doibase 10.1103/PhysRevLett.88.050602} {\bibfield  {journal}
  {\bibinfo  {journal} {Phys. Rev. Lett.}\ }\textbf {\bibinfo {volume} {88}},\
  \bibinfo {pages} {050602} (\bibinfo {year} {2002})}\BibitemShut {NoStop}%
\bibitem [{\citenamefont {Kosloff}(2013)}]{kosloff2013quantum}%
  \BibitemOpen
  \bibfield  {author} {\bibinfo {author} {\bibfnamefont {Ronnie}\ \bibnamefont
  {Kosloff}},\ }\bibfield  {title} {\enquote {\bibinfo {title} {Quantum
  thermodynamics: A dynamical viewpoint},}\ }\href {\doibase 10.3390/e15062100}
  {\bibfield  {journal} {\bibinfo  {journal} {Entropy}\ }\textbf {\bibinfo
  {volume} {15}},\ \bibinfo {pages} {2100--2128} (\bibinfo {year}
  {2013})}\BibitemShut {NoStop}%
\bibitem [{\citenamefont {Binder}\ \emph {et~al.}(2019)\citenamefont {Binder},
  \citenamefont {Correa}, \citenamefont {Gogolin}, \citenamefont {Anders},\
  and\ \citenamefont {Adesso}}]{binder2019thermodynamics}%
  \BibitemOpen
  \bibfield  {author} {\bibinfo {author} {\bibfnamefont {F.}~\bibnamefont
  {Binder}}, \bibinfo {author} {\bibfnamefont {L.A.}\ \bibnamefont {Correa}},
  \bibinfo {author} {\bibfnamefont {C.}~\bibnamefont {Gogolin}}, \bibinfo
  {author} {\bibfnamefont {J.}~\bibnamefont {Anders}}, \ and\ \bibinfo {author}
  {\bibfnamefont {G.}~\bibnamefont {Adesso}},\ }\href
  {https://books.google.co.in/books?id=IQlouQEACAAJ} {\emph {\bibinfo {title}
  {Thermodynamics in the Quantum Regime: Fundamental Aspects and New
  Directions}}},\ Fundamental Theories of Physics\ (\bibinfo  {publisher}
  {Springer International Publishing},\ \bibinfo {year} {2019})\BibitemShut
  {NoStop}%
\bibitem [{\citenamefont {Deffner}\ and\ \citenamefont
  {Campbell}(2019)}]{deffner2019quantum}%
  \BibitemOpen
  \bibfield  {author} {\bibinfo {author} {\bibfnamefont {S.}~\bibnamefont
  {Deffner}}\ and\ \bibinfo {author} {\bibfnamefont {S.E.}\ \bibnamefont
  {Campbell}},\ }\href {https://books.google.co.in/books?id=-IwbyAEACAAJ}
  {\emph {\bibinfo {title} {Quantum Thermodynamics: An Introduction to the
  Thermodynamics of Quantum Information}}}\ (\bibinfo  {publisher} {Morgan \&
  Claypool Publishers},\ \bibinfo {year} {2019})\BibitemShut {NoStop}%
\bibitem [{\citenamefont {Mukherjee}\ and\ \citenamefont
  {Divakaran}(2021)}]{mukherjee_2021_review}%
  \BibitemOpen
  \bibfield  {author} {\bibinfo {author} {\bibfnamefont {Victor}\ \bibnamefont
  {Mukherjee}}\ and\ \bibinfo {author} {\bibfnamefont {Uma}\ \bibnamefont
  {Divakaran}},\ }\bibfield  {title} {\enquote {\bibinfo {title} {Many-body
  quantum thermal machines},}\ }\href {\doibase 10.1088/1361-648X/ac1b60}
  {\bibfield  {journal} {\bibinfo  {journal} {Journal of Physics: Condensed
  Matter}\ }\textbf {\bibinfo {volume} {33}},\ \bibinfo {pages} {454001}
  (\bibinfo {year} {2021})}\BibitemShut {NoStop}%
\bibitem [{\citenamefont {Vinjanampathy}\ and\ \citenamefont
  {Anders}(2016)}]{vinjanampathy2016}%
  \BibitemOpen
  \bibfield  {author} {\bibinfo {author} {\bibfnamefont {Sai}\ \bibnamefont
  {Vinjanampathy}}\ and\ \bibinfo {author} {\bibfnamefont {Janet}\ \bibnamefont
  {Anders}},\ }\bibfield  {title} {\enquote {\bibinfo {title} {Quantum
  thermodynamics},}\ }\href {\doibase 10.1080/00107514.2016.1201896} {\bibfield
   {journal} {\bibinfo  {journal} {Contemporary Physics}\ }\textbf {\bibinfo
  {volume} {57}},\ \bibinfo {pages} {545--579} (\bibinfo {year} {2016})},\
  \Eprint {http://arxiv.org/abs/https://doi.org/10.1080/00107514.2016.1201896}
  {https://doi.org/10.1080/00107514.2016.1201896} \BibitemShut {NoStop}%
\bibitem [{\citenamefont {Bhattacharjee}\ and\ \citenamefont
  {Dutta}(2021)}]{bhattacharjee2021}%
  \BibitemOpen
  \bibfield  {author} {\bibinfo {author} {\bibfnamefont {Sourav}\ \bibnamefont
  {Bhattacharjee}}\ and\ \bibinfo {author} {\bibfnamefont {Amit}\ \bibnamefont
  {Dutta}},\ }\bibfield  {title} {\enquote {\bibinfo {title} {Quantum thermal
  machines and batteries},}\ }\href {\doibase 10.1140/epjb/s10051-021-00235-3}
  {\bibfield  {journal} {\bibinfo  {journal} {The European Physical Journal B}\
  }\textbf {\bibinfo {volume} {94}},\ \bibinfo {pages} {239} (\bibinfo {year}
  {2021})}\BibitemShut {NoStop}%
\bibitem [{\citenamefont {Opatrn\'y}\ \emph {et~al.}(2021)\citenamefont
  {Opatrn\'y}, \citenamefont {Misra},\ and\ \citenamefont
  {Kurizki}}]{opatrny_2021_prl}%
  \BibitemOpen
  \bibfield  {author} {\bibinfo {author} {\bibfnamefont {Tomas}\ \bibnamefont
  {Opatrn\'y}}, \bibinfo {author} {\bibfnamefont {Avijit}\ \bibnamefont
  {Misra}}, \ and\ \bibinfo {author} {\bibfnamefont {Gershon}\ \bibnamefont
  {Kurizki}},\ }\bibfield  {title} {\enquote {\bibinfo {title} {Work generation
  from thermal noise by quantum phase-sensitive observation},}\ }\href
  {\doibase 10.1103/PhysRevLett.127.040602} {\bibfield  {journal} {\bibinfo
  {journal} {Phys. Rev. Lett.}\ }\textbf {\bibinfo {volume} {127}},\ \bibinfo
  {pages} {040602} (\bibinfo {year} {2021})}\BibitemShut {NoStop}%
\bibitem [{\citenamefont {Misra}\ \emph {et~al.}(2022)\citenamefont {Misra},
  \citenamefont {Opatrn\'y},\ and\ \citenamefont {Kurizki}}]{misra_2022}%
  \BibitemOpen
  \bibfield  {author} {\bibinfo {author} {\bibfnamefont {Avijit}\ \bibnamefont
  {Misra}}, \bibinfo {author} {\bibfnamefont {Tomas}\ \bibnamefont
  {Opatrn\'y}}, \ and\ \bibinfo {author} {\bibfnamefont {Gershon}\ \bibnamefont
  {Kurizki}},\ }\bibfield  {title} {\enquote {\bibinfo {title} {Work extraction
  from single-mode thermal noise by measurements: How important is
  information?}}\ }\href {\doibase 10.1103/PhysRevE.106.054131} {\bibfield
  {journal} {\bibinfo  {journal} {Phys. Rev. E}\ }\textbf {\bibinfo {volume}
  {106}},\ \bibinfo {pages} {054131} (\bibinfo {year} {2022})}\BibitemShut
  {NoStop}%
\bibitem [{\citenamefont {Zheng}\ and\ \citenamefont
  {Poletti}(2014)}]{zheng_et_al_2014}%
  \BibitemOpen
  \bibfield  {author} {\bibinfo {author} {\bibfnamefont {Yuanjian}\
  \bibnamefont {Zheng}}\ and\ \bibinfo {author} {\bibfnamefont {Dario}\
  \bibnamefont {Poletti}},\ }\bibfield  {title} {\enquote {\bibinfo {title}
  {Work and efficiency of quantum otto cycles in power-law trapping
  potentials},}\ }\href {\doibase 10.1103/PhysRevE.90.012145} {\bibfield
  {journal} {\bibinfo  {journal} {Phys. Rev. E}\ }\textbf {\bibinfo {volume}
  {90}},\ \bibinfo {pages} {012145} (\bibinfo {year} {2014})}\BibitemShut
  {NoStop}%
\bibitem [{\citenamefont {Gelbwaser-Klimovsky}\ \emph
  {et~al.}(2018{\natexlab{a}})\citenamefont {Gelbwaser-Klimovsky},
  \citenamefont {Bylinskii}, \citenamefont {Gangloff}, \citenamefont {Islam},
  \citenamefont {Aspuru-Guzik},\ and\ \citenamefont
  {Vuletic}}]{david_et_al_2018}%
  \BibitemOpen
  \bibfield  {author} {\bibinfo {author} {\bibfnamefont {David}\ \bibnamefont
  {Gelbwaser-Klimovsky}}, \bibinfo {author} {\bibfnamefont {Alexei}\
  \bibnamefont {Bylinskii}}, \bibinfo {author} {\bibfnamefont {Dorian}\
  \bibnamefont {Gangloff}}, \bibinfo {author} {\bibfnamefont {Rajibul}\
  \bibnamefont {Islam}}, \bibinfo {author} {\bibfnamefont {Al\'an}\
  \bibnamefont {Aspuru-Guzik}}, \ and\ \bibinfo {author} {\bibfnamefont
  {Vladan}\ \bibnamefont {Vuletic}},\ }\bibfield  {title} {\enquote {\bibinfo
  {title} {Single-atom heat machines enabled by energy quantization},}\ }\href
  {\doibase 10.1103/PhysRevLett.120.170601} {\bibfield  {journal} {\bibinfo
  {journal} {Phys. Rev. Lett.}\ }\textbf {\bibinfo {volume} {120}},\ \bibinfo
  {pages} {170601} (\bibinfo {year} {2018}{\natexlab{a}})}\BibitemShut
  {NoStop}%
\bibitem [{\citenamefont {Niedenzu}\ \emph
  {et~al.}(2019{\natexlab{a}})\citenamefont {Niedenzu}, \citenamefont {Mazets},
  \citenamefont {Kurizki},\ and\ \citenamefont
  {Jendrzejewski}}]{niedenzu_2019}%
  \BibitemOpen
  \bibfield  {author} {\bibinfo {author} {\bibfnamefont {Wolfgang}\
  \bibnamefont {Niedenzu}}, \bibinfo {author} {\bibfnamefont {Igor}\
  \bibnamefont {Mazets}}, \bibinfo {author} {\bibfnamefont {Gershon}\
  \bibnamefont {Kurizki}}, \ and\ \bibinfo {author} {\bibfnamefont {Fred}\
  \bibnamefont {Jendrzejewski}},\ }\bibfield  {title} {\enquote {\bibinfo
  {title} {Quantized refrigerator for an atomic cloud},}\ }\href
  {https://doi.org/10.22331/q-2019-06-28-155} {\bibfield  {journal} {\bibinfo
  {journal} {Quantum}\ }\textbf {\bibinfo {volume} {3}},\ \bibinfo {pages}
  {155} (\bibinfo {year} {2019}{\natexlab{a}})}\BibitemShut {NoStop}%
\bibitem [{\citenamefont {Deffner}(2019)}]{deffner_2019}%
  \BibitemOpen
  \bibfield  {author} {\bibinfo {author} {\bibfnamefont {Sebastian}\
  \bibnamefont {Deffner}},\ }\bibfield  {title} {\enquote {\bibinfo {title}
  {Quantum refrigerators – the quantum thermodynamics of cooling bose
  gases},}\ }\href {https://doi.org/10.22331/qv-2019-08-13-20} {\bibfield
  {journal} {\bibinfo  {journal} {Quantum views}\ }\textbf {\bibinfo {volume}
  {3}},\ \bibinfo {pages} {20} (\bibinfo {year} {2019})}\BibitemShut {NoStop}%
\bibitem [{\citenamefont {Thomas}\ \emph {et~al.}(2019)\citenamefont {Thomas},
  \citenamefont {Das},\ and\ \citenamefont {Ghosh}}]{thomas_2019}%
  \BibitemOpen
  \bibfield  {author} {\bibinfo {author} {\bibfnamefont {George}\ \bibnamefont
  {Thomas}}, \bibinfo {author} {\bibfnamefont {Debmalya}\ \bibnamefont {Das}},
  \ and\ \bibinfo {author} {\bibfnamefont {Sibasish}\ \bibnamefont {Ghosh}},\
  }\bibfield  {title} {\enquote {\bibinfo {title} {Quantum heat engine based on
  level degeneracy},}\ }\href {\doibase 10.1103/PhysRevE.100.012123} {\bibfield
   {journal} {\bibinfo  {journal} {Phys. Rev. E}\ }\textbf {\bibinfo {volume}
  {100}},\ \bibinfo {pages} {012123} (\bibinfo {year} {2019})}\BibitemShut
  {NoStop}%
\bibitem [{\citenamefont {Myers}\ and\ \citenamefont
  {Deffner}(2020)}]{myers2020bosons}%
  \BibitemOpen
  \bibfield  {author} {\bibinfo {author} {\bibfnamefont {Nathan~M.}\
  \bibnamefont {Myers}}\ and\ \bibinfo {author} {\bibfnamefont {Sebastian}\
  \bibnamefont {Deffner}},\ }\bibfield  {title} {\enquote {\bibinfo {title}
  {Bosons outperform fermions: The thermodynamic advantage of symmetry},}\
  }\href {\doibase 10.1103/PhysRevE.101.012110} {\bibfield  {journal} {\bibinfo
   {journal} {Phys. Rev. E}\ }\textbf {\bibinfo {volume} {101}},\ \bibinfo
  {pages} {012110} (\bibinfo {year} {2020})}\BibitemShut {NoStop}%
\bibitem [{\citenamefont {Myers}\ \emph {et~al.}(2021)\citenamefont {Myers},
  \citenamefont {McCready},\ and\ \citenamefont {Deffner}}]{myers2021quantum}%
  \BibitemOpen
  \bibfield  {author} {\bibinfo {author} {\bibfnamefont {Nathan~M.}\
  \bibnamefont {Myers}}, \bibinfo {author} {\bibfnamefont {Jacob}\ \bibnamefont
  {McCready}}, \ and\ \bibinfo {author} {\bibfnamefont {Sebastian}\
  \bibnamefont {Deffner}},\ }\href {https://www.mdpi.com/2073-8994/13/6/978}
  {\bibfield  {journal} {\bibinfo  {journal} {Symmetry}\ }\textbf {\bibinfo
  {volume} {13}} (\bibinfo {year} {2021})}\BibitemShut {NoStop}%
\bibitem [{\citenamefont {Gupt}\ \emph {et~al.}(2021)\citenamefont {Gupt},
  \citenamefont {Bhattacharyya},\ and\ \citenamefont
  {Ghosh}}]{gupt2021statistical}%
  \BibitemOpen
  \bibfield  {author} {\bibinfo {author} {\bibfnamefont {Nikhil}\ \bibnamefont
  {Gupt}}, \bibinfo {author} {\bibfnamefont {Srijan}\ \bibnamefont
  {Bhattacharyya}}, \ and\ \bibinfo {author} {\bibfnamefont {Arnab}\
  \bibnamefont {Ghosh}},\ }\bibfield  {title} {\enquote {\bibinfo {title}
  {Statistical generalization of regenerative bosonic and fermionic stirling
  cycles},}\ }\href {\doibase 10.1103/PhysRevE.104.054130} {\bibfield
  {journal} {\bibinfo  {journal} {Phys. Rev. E}\ }\textbf {\bibinfo {volume}
  {104}},\ \bibinfo {pages} {054130} (\bibinfo {year} {2021})}\BibitemShut
  {NoStop}%
\bibitem [{\citenamefont {Myers}\ and\ \citenamefont
  {Deffner}(2021)}]{myers_2021}%
  \BibitemOpen
  \bibfield  {author} {\bibinfo {author} {\bibfnamefont {Nathan~M.}\
  \bibnamefont {Myers}}\ and\ \bibinfo {author} {\bibfnamefont {Sebastian}\
  \bibnamefont {Deffner}},\ }\bibfield  {title} {\enquote {\bibinfo {title}
  {Thermodynamics of statistical anyons},}\ }\href {\doibase
  10.1103/PRXQuantum.2.040312} {\bibfield  {journal} {\bibinfo  {journal} {PRX
  Quantum}\ }\textbf {\bibinfo {volume} {2}},\ \bibinfo {pages} {040312}
  (\bibinfo {year} {2021})}\BibitemShut {NoStop}%
\bibitem [{\citenamefont {Bouton}\ \emph {et~al.}(2021)\citenamefont {Bouton},
  \citenamefont {Nettersheim}, \citenamefont {Burgardt}, \citenamefont {Adam},
  \citenamefont {Lutz},\ and\ \citenamefont {Widera}}]{bouton_2021}%
  \BibitemOpen
  \bibfield  {author} {\bibinfo {author} {\bibfnamefont {Quentin}\ \bibnamefont
  {Bouton}}, \bibinfo {author} {\bibfnamefont {Jens}\ \bibnamefont
  {Nettersheim}}, \bibinfo {author} {\bibfnamefont {Sabrina}\ \bibnamefont
  {Burgardt}}, \bibinfo {author} {\bibfnamefont {Daniel}\ \bibnamefont {Adam}},
  \bibinfo {author} {\bibfnamefont {Eric}\ \bibnamefont {Lutz}}, \ and\
  \bibinfo {author} {\bibfnamefont {Artur}\ \bibnamefont {Widera}},\ }\bibfield
   {title} {\enquote {\bibinfo {title} {A quantum heat engine driven by atomic
  collisions},}\ }\href {\doibase 10.1038/s41467-021-22222-z} {\bibfield
  {journal} {\bibinfo  {journal} {Nature Communications}\ }\textbf {\bibinfo
  {volume} {12}},\ \bibinfo {pages} {2063} (\bibinfo {year}
  {2021})}\BibitemShut {NoStop}%
\bibitem [{\citenamefont {Benjamin~Yadin}(2017)}]{yadin_2022}%
  \BibitemOpen
  \bibfield  {author} {\bibinfo {author} {\bibfnamefont {Kay~Brandner}\
  \bibnamefont {Benjamin~Yadin}, \bibfnamefont {Benjamin~Morris}},\ }\bibfield
  {title} {\enquote {\bibinfo {title} {Thermodynamics of permutation-invariant
  quantum many-body systems: A group-theoretical framework},}\ }\href
  {https://arxiv.org/abs/2206.12639} {\bibfield  {journal} {\bibinfo  {journal}
  {arXiv preprint arXiv:2206.12639}\ } (\bibinfo {year} {2017})}\BibitemShut
  {NoStop}%
\bibitem [{\citenamefont {Scully}\ \emph {et~al.}(2003)\citenamefont {Scully},
  \citenamefont {Zubairy}, \citenamefont {Agarwal},\ and\ \citenamefont
  {Walther}}]{scully2003extracting}%
  \BibitemOpen
  \bibfield  {author} {\bibinfo {author} {\bibfnamefont {Marlan~O.}\
  \bibnamefont {Scully}}, \bibinfo {author} {\bibfnamefont {M.~Suhail}\
  \bibnamefont {Zubairy}}, \bibinfo {author} {\bibfnamefont {Girish~S.}\
  \bibnamefont {Agarwal}}, \ and\ \bibinfo {author} {\bibfnamefont {Herbert}\
  \bibnamefont {Walther}},\ }\bibfield  {title} {\enquote {\bibinfo {title}
  {Extracting work from a single heat bath via vanishing quantum coherence},}\
  }\href {\doibase 10.1126/science.1078955} {\bibfield  {journal} {\bibinfo
  {journal} {Science}\ }\textbf {\bibinfo {volume} {299}},\ \bibinfo {pages}
  {862--864} (\bibinfo {year} {2003})}\BibitemShut {NoStop}%
\bibitem [{\citenamefont {Ro\ss{}nagel}\ \emph {et~al.}(2014)\citenamefont
  {Ro\ss{}nagel}, \citenamefont {Abah}, \citenamefont {Schmidt-Kaler},
  \citenamefont {Singer},\ and\ \citenamefont {Lutz}}]{rossnagel2014nanoscale}%
  \BibitemOpen
  \bibfield  {author} {\bibinfo {author} {\bibfnamefont {J.}~\bibnamefont
  {Ro\ss{}nagel}}, \bibinfo {author} {\bibfnamefont {O.}~\bibnamefont {Abah}},
  \bibinfo {author} {\bibfnamefont {F.}~\bibnamefont {Schmidt-Kaler}}, \bibinfo
  {author} {\bibfnamefont {K.}~\bibnamefont {Singer}}, \ and\ \bibinfo {author}
  {\bibfnamefont {E.}~\bibnamefont {Lutz}},\ }\bibfield  {title} {\enquote
  {\bibinfo {title} {Nanoscale heat engine beyond the carnot limit},}\ }\href
  {\doibase 10.1103/PhysRevLett.112.030602} {\bibfield  {journal} {\bibinfo
  {journal} {Phys. Rev. Lett.}\ }\textbf {\bibinfo {volume} {112}},\ \bibinfo
  {pages} {030602} (\bibinfo {year} {2014})}\BibitemShut {NoStop}%
\bibitem [{\citenamefont {Klaers}\ \emph {et~al.}(2017)\citenamefont {Klaers},
  \citenamefont {Faelt}, \citenamefont {Imamoglu},\ and\ \citenamefont
  {Togan}}]{klaers2017squeezed}%
  \BibitemOpen
  \bibfield  {author} {\bibinfo {author} {\bibfnamefont {Jan}\ \bibnamefont
  {Klaers}}, \bibinfo {author} {\bibfnamefont {Stefan}\ \bibnamefont {Faelt}},
  \bibinfo {author} {\bibfnamefont {Atac}\ \bibnamefont {Imamoglu}}, \ and\
  \bibinfo {author} {\bibfnamefont {Emre}\ \bibnamefont {Togan}},\ }\bibfield
  {title} {\enquote {\bibinfo {title} {Squeezed thermal reservoirs as a
  resource for a nanomechanical engine beyond the carnot limit},}\ }\href
  {\doibase 10.1103/PhysRevX.7.031044} {\bibfield  {journal} {\bibinfo
  {journal} {Phys. Rev. X}\ }\textbf {\bibinfo {volume} {7}},\ \bibinfo {pages}
  {031044} (\bibinfo {year} {2017})}\BibitemShut {NoStop}%
\bibitem [{\citenamefont {Niedenzu}\ \emph {et~al.}(2018)\citenamefont
  {Niedenzu}, \citenamefont {Mukherjee}, \citenamefont {Ghosh}, \citenamefont
  {Kofman},\ and\ \citenamefont {Kurizki}}]{niedenzu2018quantum}%
  \BibitemOpen
  \bibfield  {author} {\bibinfo {author} {\bibfnamefont {Wolfgang}\
  \bibnamefont {Niedenzu}}, \bibinfo {author} {\bibfnamefont {Victor}\
  \bibnamefont {Mukherjee}}, \bibinfo {author} {\bibfnamefont {Arnab}\
  \bibnamefont {Ghosh}}, \bibinfo {author} {\bibfnamefont {Abraham~G.}\
  \bibnamefont {Kofman}}, \ and\ \bibinfo {author} {\bibfnamefont {Gershon}\
  \bibnamefont {Kurizki}},\ }\bibfield  {title} {\enquote {\bibinfo {title}
  {Quantum engine efficiency bound beyond the second law of thermodynamics},}\
  }\href {\doibase 10.1038/s41467-017-01991-6} {\bibfield  {journal} {\bibinfo
  {journal} {Nature Communications}\ }\textbf {\bibinfo {volume} {9}},\
  \bibinfo {pages} {165} (\bibinfo {year} {2018})}\BibitemShut {NoStop}%
\bibitem [{\citenamefont {Ghosh}\ \emph {et~al.}(2019)\citenamefont {Ghosh},
  \citenamefont {Mukherjee}, \citenamefont {Niedenzu},\ and\ \citenamefont
  {Kurizki}}]{ghosh2019are}%
  \BibitemOpen
  \bibfield  {author} {\bibinfo {author} {\bibfnamefont {Arnab}\ \bibnamefont
  {Ghosh}}, \bibinfo {author} {\bibfnamefont {Victor}\ \bibnamefont
  {Mukherjee}}, \bibinfo {author} {\bibfnamefont {Wolfgang}\ \bibnamefont
  {Niedenzu}}, \ and\ \bibinfo {author} {\bibfnamefont {Gershon}\ \bibnamefont
  {Kurizki}},\ }\bibfield  {title} {\enquote {\bibinfo {title} {Are quantum
  thermodynamic machines better than their classical counterparts?}}\ }\href
  {\doibase 10.1140/epjst/e2019-800060-7} {\bibfield  {journal} {\bibinfo
  {journal} {The European Physical Journal Special Topics}\ }\textbf {\bibinfo
  {volume} {227}},\ \bibinfo {pages} {2043--2051} (\bibinfo {year}
  {2019})}\BibitemShut {NoStop}%
\bibitem [{\citenamefont {Salamon}\ \emph {et~al.}(1980)\citenamefont
  {Salamon}, \citenamefont {Nitzan}, \citenamefont {Andresen},\ and\
  \citenamefont {Berry}}]{salamon1980minimum}%
  \BibitemOpen
  \bibfield  {author} {\bibinfo {author} {\bibfnamefont {Peter}\ \bibnamefont
  {Salamon}}, \bibinfo {author} {\bibfnamefont {Abraham}\ \bibnamefont
  {Nitzan}}, \bibinfo {author} {\bibfnamefont {Bjarne}\ \bibnamefont
  {Andresen}}, \ and\ \bibinfo {author} {\bibfnamefont {R.~Stephen}\
  \bibnamefont {Berry}},\ }\bibfield  {title} {\enquote {\bibinfo {title}
  {Minimum entropy production and the optimization of heat engines},}\ }\href
  {\doibase 10.1103/PhysRevA.21.2115} {\bibfield  {journal} {\bibinfo
  {journal} {Phys. Rev. A}\ }\textbf {\bibinfo {volume} {21}},\ \bibinfo
  {pages} {2115--2129} (\bibinfo {year} {1980})}\BibitemShut {NoStop}%
\bibitem [{\citenamefont {Chen}\ and\ \citenamefont
  {Yan}(1991)}]{chen1991effect}%
  \BibitemOpen
  \bibfield  {author} {\bibinfo {author} {\bibfnamefont {Jincan}\ \bibnamefont
  {Chen}}\ and\ \bibinfo {author} {\bibfnamefont {Zijun}\ \bibnamefont {Yan}},\
  }\bibfield  {title} {\enquote {\bibinfo {title} {The effect of field
  dependent heat capacity on regeneration in magnetic ericsson cycles},}\
  }\href {\doibase 10.1063/1.348821} {\bibfield  {journal} {\bibinfo  {journal}
  {Journal of Applied Physics}\ }\textbf {\bibinfo {volume} {69}},\ \bibinfo
  {pages} {6245--6247} (\bibinfo {year} {1991})}\BibitemShut {NoStop}%
\bibitem [{\citenamefont {Geva}\ and\ \citenamefont
  {Kosloff}(1992{\natexlab{a}})}]{geva1992classical}%
  \BibitemOpen
  \bibfield  {author} {\bibinfo {author} {\bibfnamefont {Eitan}\ \bibnamefont
  {Geva}}\ and\ \bibinfo {author} {\bibfnamefont {Ronnie}\ \bibnamefont
  {Kosloff}},\ }\bibfield  {title} {\enquote {\bibinfo {title} {On the
  classical limit of quantum thermodynamics in finite time},}\ }\href {\doibase
  10.1063/1.463909} {\bibfield  {journal} {\bibinfo  {journal} {J. Chem.
  Phys.}\ }\textbf {\bibinfo {volume} {97}},\ \bibinfo {pages} {4398--4412}
  (\bibinfo {year} {1992}{\natexlab{a}})}\BibitemShut {NoStop}%
\bibitem [{\citenamefont {Geva}\ and\ \citenamefont
  {Kosloff}(1992{\natexlab{b}})}]{geva1992quantum}%
  \BibitemOpen
  \bibfield  {author} {\bibinfo {author} {\bibfnamefont {Eitan}\ \bibnamefont
  {Geva}}\ and\ \bibinfo {author} {\bibfnamefont {Ronnie}\ \bibnamefont
  {Kosloff}},\ }\bibfield  {title} {\enquote {\bibinfo {title} {A
  quantum-mechanical heat engine operating in finite time. a model consisting
  of spin-1/2 systems as the working fluid},}\ }\href {\doibase
  10.1063/1.461951} {\bibfield  {journal} {\bibinfo  {journal} {J. Chem.
  Phys.}\ }\textbf {\bibinfo {volume} {96}},\ \bibinfo {pages} {3054--3067}
  (\bibinfo {year} {1992}{\natexlab{b}})}\BibitemShut {NoStop}%
\bibitem [{\citenamefont {Chen}(1994)}]{chen1994maximam}%
  \BibitemOpen
  \bibfield  {author} {\bibinfo {author} {\bibfnamefont {Jincan}\ \bibnamefont
  {Chen}},\ }\bibfield  {title} {\enquote {\bibinfo {title} {The maximum power
  output and maximum efficiency of an irreversible carnot heat engine},}\
  }\href {\doibase 10.1088/0022-3727/27/6/011} {\bibfield  {journal} {\bibinfo
  {journal} {Journal of Physics D: Applied Physics}\ }\textbf {\bibinfo
  {volume} {27}},\ \bibinfo {pages} {1144--1149} (\bibinfo {year}
  {1994})}\BibitemShut {NoStop}%
\bibitem [{\citenamefont {Chen}\ and\ \citenamefont
  {Yan}(1998)}]{chen1998effect}%
  \BibitemOpen
  \bibfield  {author} {\bibinfo {author} {\bibfnamefont {Jincan}\ \bibnamefont
  {Chen}}\ and\ \bibinfo {author} {\bibfnamefont {Zijun}\ \bibnamefont {Yan}},\
  }\bibfield  {title} {\enquote {\bibinfo {title} {The effect of thermal
  resistances and regenerative losses on the performance characteristics of a
  magnetic ericsson refrigeration cycle},}\ }\href {\doibase 10.1063/1.368349}
  {\bibfield  {journal} {\bibinfo  {journal} {Journal of Applied Physics}\
  }\textbf {\bibinfo {volume} {84}},\ \bibinfo {pages} {1791--1795} (\bibinfo
  {year} {1998})}\BibitemShut {NoStop}%
\bibitem [{\citenamefont {Chen}\ and\ \citenamefont
  {Schouten}(1999)}]{chen1999comprehensive}%
  \BibitemOpen
  \bibfield  {author} {\bibinfo {author} {\bibfnamefont {Jincan}\ \bibnamefont
  {Chen}}\ and\ \bibinfo {author} {\bibfnamefont {Jan~A.}\ \bibnamefont
  {Schouten}},\ }\bibfield  {title} {\enquote {\bibinfo {title} {The
  comprehensive influence of several major irreversibilities on the performance
  of an ericsson heat engine},}\ }\href {\doibase
  https://doi.org/10.1016/S1359-4311(98)00059-3} {\bibfield  {journal}
  {\bibinfo  {journal} {Applied Thermal Engineering}\ }\textbf {\bibinfo
  {volume} {19}},\ \bibinfo {pages} {555--564} (\bibinfo {year}
  {1999})}\BibitemShut {NoStop}%
\bibitem [{\citenamefont {Arnaud}\ \emph {et~al.}(2002)\citenamefont {Arnaud},
  \citenamefont {Chusseau},\ and\ \citenamefont {Philippe}}]{arnaud2002carnot}%
  \BibitemOpen
  \bibfield  {author} {\bibinfo {author} {\bibfnamefont {Jacques}\ \bibnamefont
  {Arnaud}}, \bibinfo {author} {\bibfnamefont {Laurent}\ \bibnamefont
  {Chusseau}}, \ and\ \bibinfo {author} {\bibfnamefont {Fabrice}\ \bibnamefont
  {Philippe}},\ }\bibfield  {title} {\enquote {\bibinfo {title} {Carnot cycle
  for an oscillator},}\ }\href {\doibase 10.1088/0143-0807/23/5/306} {\bibfield
   {journal} {\bibinfo  {journal} {Eur. J. Phys.}\ }\textbf {\bibinfo {volume}
  {23}},\ \bibinfo {pages} {489} (\bibinfo {year} {2002})}\BibitemShut
  {NoStop}%
\bibitem [{\citenamefont {Bhattacharyya}\ and\ \citenamefont
  {Mukhopadhyay}(2001)}]{bhattacharyya2001comment}%
  \BibitemOpen
  \bibfield  {author} {\bibinfo {author} {\bibfnamefont {K}~\bibnamefont
  {Bhattacharyya}}\ and\ \bibinfo {author} {\bibfnamefont {S}~\bibnamefont
  {Mukhopadhyay}},\ }\bibfield  {title} {\enquote {\bibinfo {title} {Comment on
  {\textasciigrave}quantum-mechanical carnot engine{\textquotesingle}},}\
  }\href {\doibase 10.1088/0305-4470/34/7/401} {\bibfield  {journal} {\bibinfo
  {journal} {Journal of Physics A: Mathematical and General}\ }\textbf
  {\bibinfo {volume} {34}},\ \bibinfo {pages} {1529--1533} (\bibinfo {year}
  {2001})}\BibitemShut {NoStop}%
\bibitem [{\citenamefont {Abah}\ \emph {et~al.}(2012)\citenamefont {Abah},
  \citenamefont {Ro\ss{}nagel}, \citenamefont {Jacob}, \citenamefont {Deffner},
  \citenamefont {Schmidt-Kaler}, \citenamefont {Singer},\ and\ \citenamefont
  {Lutz}}]{abah2012single}%
  \BibitemOpen
  \bibfield  {author} {\bibinfo {author} {\bibfnamefont {O.}~\bibnamefont
  {Abah}}, \bibinfo {author} {\bibfnamefont {J.}~\bibnamefont {Ro\ss{}nagel}},
  \bibinfo {author} {\bibfnamefont {G.}~\bibnamefont {Jacob}}, \bibinfo
  {author} {\bibfnamefont {S.}~\bibnamefont {Deffner}}, \bibinfo {author}
  {\bibfnamefont {F.}~\bibnamefont {Schmidt-Kaler}}, \bibinfo {author}
  {\bibfnamefont {K.}~\bibnamefont {Singer}}, \ and\ \bibinfo {author}
  {\bibfnamefont {E.}~\bibnamefont {Lutz}},\ }\bibfield  {title} {\enquote
  {\bibinfo {title} {Single-ion heat engine at maximum power},}\ }\href
  {\doibase 10.1103/PhysRevLett.109.203006} {\bibfield  {journal} {\bibinfo
  {journal} {Phys. Rev. Lett.}\ }\textbf {\bibinfo {volume} {109}},\ \bibinfo
  {pages} {203006} (\bibinfo {year} {2012})}\BibitemShut {NoStop}%
\bibitem [{\citenamefont {Das}\ and\ \citenamefont
  {Mukherjee}(2020)}]{das2020}%
  \BibitemOpen
  \bibfield  {author} {\bibinfo {author} {\bibfnamefont {Arpan}\ \bibnamefont
  {Das}}\ and\ \bibinfo {author} {\bibfnamefont {Victor}\ \bibnamefont
  {Mukherjee}},\ }\bibfield  {title} {\enquote {\bibinfo {title}
  {Quantum-enhanced finite-time otto cycle},}\ }\href {\doibase
  10.1103/PhysRevResearch.2.033083} {\bibfield  {journal} {\bibinfo  {journal}
  {Phys. Rev. Research}\ }\textbf {\bibinfo {volume} {2}},\ \bibinfo {pages}
  {033083} (\bibinfo {year} {2020})}\BibitemShut {NoStop}%
\bibitem [{\citenamefont {Mukherjee}\ \emph {et~al.}(2020)\citenamefont
  {Mukherjee}, \citenamefont {Kofman},\ and\ \citenamefont
  {Kurizki}}]{mukherjee2020}%
  \BibitemOpen
  \bibfield  {author} {\bibinfo {author} {\bibfnamefont {Victor}\ \bibnamefont
  {Mukherjee}}, \bibinfo {author} {\bibfnamefont {Abraham~G.}\ \bibnamefont
  {Kofman}}, \ and\ \bibinfo {author} {\bibfnamefont {Gershon}\ \bibnamefont
  {Kurizki}},\ }\bibfield  {title} {\enquote {\bibinfo {title} {Anti-zeno
  quantum advantage in fast-driven heat machines},}\ }\href {\doibase
  10.1038/s42005-019-0272-z} {\bibfield  {journal} {\bibinfo  {journal}
  {Communications Physics}\ }\textbf {\bibinfo {volume} {3}},\ \bibinfo {pages}
  {8} (\bibinfo {year} {2020})}\BibitemShut {NoStop}%
\bibitem [{\citenamefont {Feldmann}\ and\ \citenamefont
  {Kosloff}(2000)}]{feldmann_2000}%
  \BibitemOpen
  \bibfield  {author} {\bibinfo {author} {\bibfnamefont {Tova}\ \bibnamefont
  {Feldmann}}\ and\ \bibinfo {author} {\bibfnamefont {Ronnie}\ \bibnamefont
  {Kosloff}},\ }\bibfield  {title} {\enquote {\bibinfo {title} {Performance of
  discrete heat engines and heat pumps in finite time},}\ }\href {\doibase
  10.1103/PhysRevE.61.4774} {\bibfield  {journal} {\bibinfo  {journal} {Phys.
  Rev. E}\ }\textbf {\bibinfo {volume} {61}},\ \bibinfo {pages} {4774--4790}
  (\bibinfo {year} {2000})}\BibitemShut {NoStop}%
\bibitem [{\citenamefont {Chen}\ \emph {et~al.}(2002)\citenamefont {Chen},
  \citenamefont {Lin},\ and\ \citenamefont {Hua}}]{chen2002performance}%
  \BibitemOpen
  \bibfield  {author} {\bibinfo {author} {\bibfnamefont {Jincan}\ \bibnamefont
  {Chen}}, \bibinfo {author} {\bibfnamefont {Bihong}\ \bibnamefont {Lin}}, \
  and\ \bibinfo {author} {\bibfnamefont {Ben}\ \bibnamefont {Hua}},\ }\bibfield
   {title} {\enquote {\bibinfo {title} {The performance of a quantum heat
  engine working with spin systems},}\ }\href {\doibase
  10.1088/0022-3727/35/16/322} {\bibfield  {journal} {\bibinfo  {journal}
  {Journal of Physics D: Applied Physics}\ }\textbf {\bibinfo {volume} {35}},\
  \bibinfo {pages} {2051--2057} (\bibinfo {year} {2002})}\BibitemShut {NoStop}%
\bibitem [{\citenamefont {Lin}\ \emph {et~al.}(2003)\citenamefont {Lin},
  \citenamefont {Chen},\ and\ \citenamefont {Hua}}]{lin2003optimal}%
  \BibitemOpen
  \bibfield  {author} {\bibinfo {author} {\bibfnamefont {Bihong}\ \bibnamefont
  {Lin}}, \bibinfo {author} {\bibfnamefont {Jincan}\ \bibnamefont {Chen}}, \
  and\ \bibinfo {author} {\bibfnamefont {Ben}\ \bibnamefont {Hua}},\ }\bibfield
   {title} {\enquote {\bibinfo {title} {The optimal performance of a quantum
  refrigeration cycle working with harmonic oscillators},}\ }\href {\doibase
  10.1088/0022-3727/36/4/313} {\bibfield  {journal} {\bibinfo  {journal}
  {Journal of Physics D: Applied Physics}\ }\textbf {\bibinfo {volume} {36}},\
  \bibinfo {pages} {406--413} (\bibinfo {year} {2003})}\BibitemShut {NoStop}%
\bibitem [{\citenamefont {Henrich}\ \emph {et~al.}(2007)\citenamefont
  {Henrich}, \citenamefont {Rempp},\ and\ \citenamefont
  {Mahler}}]{henrich2007quantum}%
  \BibitemOpen
  \bibfield  {author} {\bibinfo {author} {\bibfnamefont {M.~J.}\ \bibnamefont
  {Henrich}}, \bibinfo {author} {\bibfnamefont {F.}~\bibnamefont {Rempp}}, \
  and\ \bibinfo {author} {\bibfnamefont {G.}~\bibnamefont {Mahler}},\
  }\bibfield  {title} {\enquote {\bibinfo {title} {Quantum thermodynamic otto
  machines: A spin-system approach},}\ }\href {\doibase
  10.1140/epjst/e2007-00371-8} {\bibfield  {journal} {\bibinfo  {journal} {Eur.
  Phys. J. Spec. Top.}\ }\textbf {\bibinfo {volume} {151}},\ \bibinfo {pages}
  {157--165} (\bibinfo {year} {2007})}\BibitemShut {NoStop}%
\bibitem [{\citenamefont {Thomas}\ and\ \citenamefont
  {Johal}(2014)}]{thomas2014}%
  \BibitemOpen
  \bibfield  {author} {\bibinfo {author} {\bibfnamefont {George}\ \bibnamefont
  {Thomas}}\ and\ \bibinfo {author} {\bibfnamefont {Ramandeep~S.}\ \bibnamefont
  {Johal}},\ }\bibfield  {title} {\enquote {\bibinfo {title} {Friction due to
  inhomogeneous driving of coupled spins in a quantum heat engine},}\ }\href
  {\doibase 10.1140/epjb/e2014-50231-1} {\bibfield  {journal} {\bibinfo
  {journal} {The European Physical Journal B}\ }\textbf {\bibinfo {volume}
  {87}},\ \bibinfo {pages} {166} (\bibinfo {year} {2014})}\BibitemShut
  {NoStop}%
\bibitem [{\citenamefont {Gelbwaser-Klimovsky}\ \emph
  {et~al.}(2018{\natexlab{b}})\citenamefont {Gelbwaser-Klimovsky},
  \citenamefont {Bylinskii}, \citenamefont {Gangloff}, \citenamefont {Islam},
  \citenamefont {Aspuru-Guzik},\ and\ \citenamefont
  {Vuletic}}]{gelbwaser2018single}%
  \BibitemOpen
  \bibfield  {author} {\bibinfo {author} {\bibfnamefont {David}\ \bibnamefont
  {Gelbwaser-Klimovsky}}, \bibinfo {author} {\bibfnamefont {Alexei}\
  \bibnamefont {Bylinskii}}, \bibinfo {author} {\bibfnamefont {Dorian}\
  \bibnamefont {Gangloff}}, \bibinfo {author} {\bibfnamefont {Rajibul}\
  \bibnamefont {Islam}}, \bibinfo {author} {\bibfnamefont {Al\'an}\
  \bibnamefont {Aspuru-Guzik}}, \ and\ \bibinfo {author} {\bibfnamefont
  {Vladan}\ \bibnamefont {Vuletic}},\ }\bibfield  {title} {\enquote {\bibinfo
  {title} {Single-atom heat machines enabled by energy quantization},}\ }\href
  {\doibase 10.1103/PhysRevLett.120.170601} {\bibfield  {journal} {\bibinfo
  {journal} {Phys. Rev. Lett.}\ }\textbf {\bibinfo {volume} {120}},\ \bibinfo
  {pages} {170601} (\bibinfo {year} {2018}{\natexlab{b}})}\BibitemShut
  {NoStop}%
\bibitem [{\citenamefont {Levy}\ and\ \citenamefont
  {Gelbwaser-Klimovsky}(2018)}]{levy2018}%
  \BibitemOpen
  \bibfield  {author} {\bibinfo {author} {\bibfnamefont {Amikam}\ \bibnamefont
  {Levy}}\ and\ \bibinfo {author} {\bibfnamefont {David}\ \bibnamefont
  {Gelbwaser-Klimovsky}},\ }\enquote {\bibinfo {title} {Quantum features and
  signatures of quantum thermal machines},}\ in\ \href {\doibase
  10.1007/978-3-319-99046-0_4} {\emph {\bibinfo {booktitle} {Thermodynamics in
  the Quantum Regime: Fundamental Aspects and New Directions}}},\ \bibinfo
  {editor} {edited by\ \bibinfo {editor} {\bibfnamefont {Felix}\ \bibnamefont
  {Binder}}, \bibinfo {editor} {\bibfnamefont {Luis~A.}\ \bibnamefont
  {Correa}}, \bibinfo {editor} {\bibfnamefont {Christian}\ \bibnamefont
  {Gogolin}}, \bibinfo {editor} {\bibfnamefont {Janet}\ \bibnamefont {Anders}},
  \ and\ \bibinfo {editor} {\bibfnamefont {Gerardo}\ \bibnamefont {Adesso}}}\
  (\bibinfo  {publisher} {Springer International Publishing},\ \bibinfo
  {address} {Cham},\ \bibinfo {year} {2018})\ pp.\ \bibinfo {pages}
  {87--126}\BibitemShut {NoStop}%
\bibitem [{\citenamefont {Kosloff}\ and\ \citenamefont
  {Rezek}(2017)}]{kosloff2017quantum}%
  \BibitemOpen
  \bibfield  {author} {\bibinfo {author} {\bibfnamefont {Ronnie}\ \bibnamefont
  {Kosloff}}\ and\ \bibinfo {author} {\bibfnamefont {Yair}\ \bibnamefont
  {Rezek}},\ }\bibfield  {title} {\enquote {\bibinfo {title} {The quantum
  harmonic otto cycle},}\ }\href {\doibase 10.3390/e19040136} {\bibfield
  {journal} {\bibinfo  {journal} {Entropy}\ }\textbf {\bibinfo {volume} {19}},\
  \bibinfo {pages} {136} (\bibinfo {year} {2017})}\BibitemShut {NoStop}%
\bibitem [{\citenamefont {Huang}\ \emph {et~al.}(2013)\citenamefont {Huang},
  \citenamefont {Xu}, \citenamefont {Niu},\ and\ \citenamefont
  {Fu}}]{huang_2013}%
  \BibitemOpen
  \bibfield  {author} {\bibinfo {author} {\bibfnamefont {X~L}\ \bibnamefont
  {Huang}}, \bibinfo {author} {\bibfnamefont {Huan}\ \bibnamefont {Xu}},
  \bibinfo {author} {\bibfnamefont {X~Y}\ \bibnamefont {Niu}}, \ and\ \bibinfo
  {author} {\bibfnamefont {Y~D}\ \bibnamefont {Fu}},\ }\bibfield  {title}
  {\enquote {\bibinfo {title} {A special entangled quantum heat engine based on
  the two-qubit heisenberg xx model},}\ }\href {\doibase
  10.1088/0031-8949/88/06/065008} {\bibfield  {journal} {\bibinfo  {journal}
  {Physica Scripta}\ }\textbf {\bibinfo {volume} {88}},\ \bibinfo {pages}
  {065008} (\bibinfo {year} {2013})}\BibitemShut {NoStop}%
\bibitem [{\citenamefont {Huang}\ \emph {et~al.}(2014)\citenamefont {Huang},
  \citenamefont {Liu}, \citenamefont {Wang},\ and\ \citenamefont
  {Niu}}]{huang_2014}%
  \BibitemOpen
  \bibfield  {author} {\bibinfo {author} {\bibfnamefont {X.~L.}\ \bibnamefont
  {Huang}}, \bibinfo {author} {\bibfnamefont {Yang}\ \bibnamefont {Liu}},
  \bibinfo {author} {\bibfnamefont {Zhen}\ \bibnamefont {Wang}}, \ and\
  \bibinfo {author} {\bibfnamefont {X.~Y.}\ \bibnamefont {Niu}},\ }\bibfield
  {title} {\enquote {\bibinfo {title} {Special coupled quantum otto cycles},}\
  }\href {\doibase 10.1140/epjp/i2014-14004-8} {\bibfield  {journal} {\bibinfo
  {journal} {The European Physical Journal Plus}\ }\textbf {\bibinfo {volume}
  {129}},\ \bibinfo {pages} {4} (\bibinfo {year} {2014})}\BibitemShut {NoStop}%
\bibitem [{\citenamefont {Abah}\ and\ \citenamefont {Lutz}(2016)}]{abah_2016}%
  \BibitemOpen
  \bibfield  {author} {\bibinfo {author} {\bibfnamefont {Obinna}\ \bibnamefont
  {Abah}}\ and\ \bibinfo {author} {\bibfnamefont {Eric}\ \bibnamefont {Lutz}},\
  }\bibfield  {title} {\enquote {\bibinfo {title} {Optimal performance of a
  quantum otto refrigerator},}\ }\href {\doibase 10.1209/0295-5075/113/60002}
  {\bibfield  {journal} {\bibinfo  {journal} {{EPL} (Europhysics Letters)}\
  }\textbf {\bibinfo {volume} {113}},\ \bibinfo {pages} {60002} (\bibinfo
  {year} {2016})}\BibitemShut {NoStop}%
\bibitem [{\citenamefont {Stefanatos}(2014)}]{stefanatos_2014}%
  \BibitemOpen
  \bibfield  {author} {\bibinfo {author} {\bibfnamefont {Dionisis}\
  \bibnamefont {Stefanatos}},\ }\bibfield  {title} {\enquote {\bibinfo {title}
  {Optimal efficiency of a noisy quantum heat engine},}\ }\href {\doibase
  10.1103/PhysRevE.90.012119} {\bibfield  {journal} {\bibinfo  {journal} {Phys.
  Rev. E}\ }\textbf {\bibinfo {volume} {90}},\ \bibinfo {pages} {012119}
  (\bibinfo {year} {2014})}\BibitemShut {NoStop}%
\bibitem [{\citenamefont {Stefanatos}(2017)}]{stefanatos_2017}%
  \BibitemOpen
  \bibfield  {author} {\bibinfo {author} {\bibfnamefont {Dionisis}\
  \bibnamefont {Stefanatos}},\ }\bibfield  {title} {\enquote {\bibinfo {title}
  {Exponential bound in the quest for absolute zero},}\ }\href {\doibase
  10.1103/PhysRevE.96.042103} {\bibfield  {journal} {\bibinfo  {journal} {Phys.
  Rev. E}\ }\textbf {\bibinfo {volume} {96}},\ \bibinfo {pages} {042103}
  (\bibinfo {year} {2017})}\BibitemShut {NoStop}%
\bibitem [{\citenamefont {Chatterjee}\ \emph {et~al.}(2021)\citenamefont
  {Chatterjee}, \citenamefont {Koner}, \citenamefont {Chatterjee},\ and\
  \citenamefont {Kumar}}]{chatterjee_2020}%
  \BibitemOpen
  \bibfield  {author} {\bibinfo {author} {\bibfnamefont {Sarbani}\ \bibnamefont
  {Chatterjee}}, \bibinfo {author} {\bibfnamefont {Arghadip}\ \bibnamefont
  {Koner}}, \bibinfo {author} {\bibfnamefont {Sohini}\ \bibnamefont
  {Chatterjee}}, \ and\ \bibinfo {author} {\bibfnamefont {Chandan}\
  \bibnamefont {Kumar}},\ }\bibfield  {title} {\enquote {\bibinfo {title}
  {Temperature-dependent maximization of work and efficiency in a
  degeneracy-assisted quantum stirling heat engine},}\ }\href {\doibase
  10.1103/PhysRevE.103.062109} {\bibfield  {journal} {\bibinfo  {journal}
  {Phys. Rev. E}\ }\textbf {\bibinfo {volume} {103}},\ \bibinfo {pages}
  {062109} (\bibinfo {year} {2021})}\BibitemShut {NoStop}%
\bibitem [{\citenamefont {Chattopadhyay}\ \emph {et~al.}(2021)\citenamefont
  {Chattopadhyay}, \citenamefont {Mitra}, \citenamefont {Paul},\ and\
  \citenamefont {Zarikas}}]{Chattopadhyay_2019}%
  \BibitemOpen
  \bibfield  {author} {\bibinfo {author} {\bibfnamefont {Pritam}\ \bibnamefont
  {Chattopadhyay}}, \bibinfo {author} {\bibfnamefont {Ayan}\ \bibnamefont
  {Mitra}}, \bibinfo {author} {\bibfnamefont {Goutam}\ \bibnamefont {Paul}}, \
  and\ \bibinfo {author} {\bibfnamefont {Vasilios}\ \bibnamefont {Zarikas}},\
  }\bibfield  {title} {\enquote {\bibinfo {title} {Bound on efficiency of heat
  engine from uncertainty relation viewpoint},}\ }\href {\doibase
  10.3390/e23040439} {\bibfield  {journal} {\bibinfo  {journal} {Entropy}\
  }\textbf {\bibinfo {volume} {23}} (\bibinfo {year} {2021}),\
  10.3390/e23040439}\BibitemShut {NoStop}%
\bibitem [{\citenamefont {Chowdhury}\ and\ \citenamefont
  {Stauffer}(2000)}]{chowdhury_2000}%
  \BibitemOpen
  \bibfield  {author} {\bibinfo {author} {\bibfnamefont {Debasish}\
  \bibnamefont {Chowdhury}}\ and\ \bibinfo {author} {\bibfnamefont {Dietrich}\
  \bibnamefont {Stauffer}},\ }\href@noop {} {\emph {\bibinfo {title}
  {Principles of Equilibrium Statistical Mechanics}}}\ (\bibinfo  {publisher}
  {WILEY-VCH},\ \bibinfo {year} {2000})\BibitemShut {NoStop}%
\bibitem [{\citenamefont {Pathria}\ and\ \citenamefont
  {Beale}(2011{\natexlab{a}})}]{PATHRIA2011179}%
  \BibitemOpen
  \bibfield  {author} {\bibinfo {author} {\bibfnamefont {R.K.}\ \bibnamefont
  {Pathria}}\ and\ \bibinfo {author} {\bibfnamefont {Paul~D.}\ \bibnamefont
  {Beale}},\ }\bibfield  {title} {\enquote {\bibinfo {title} {7 - ideal bose
  systems},}\ }in\ \href {\doibase
  https://doi.org/10.1016/B978-0-12-382188-1.00007-4} {\emph {\bibinfo
  {booktitle} {Statistical Mechanics (Third Edition)}}},\ \bibinfo {editor}
  {edited by\ \bibinfo {editor} {\bibfnamefont {R.K.}\ \bibnamefont {Pathria}}\
  and\ \bibinfo {editor} {\bibfnamefont {Paul~D.}\ \bibnamefont {Beale}}}\
  (\bibinfo  {publisher} {Academic Press},\ \bibinfo {address} {Boston},\
  \bibinfo {year} {2011})\ \bibinfo {edition} {third edition}\ ed.,\ pp.\
  \bibinfo {pages} {179--229}\BibitemShut {NoStop}%
\bibitem [{\citenamefont {Pathria}\ and\ \citenamefont
  {Beale}(2011{\natexlab{b}})}]{PATHRIA2011231}%
  \BibitemOpen
  \bibfield  {author} {\bibinfo {author} {\bibfnamefont {R.K.}\ \bibnamefont
  {Pathria}}\ and\ \bibinfo {author} {\bibfnamefont {Paul~D.}\ \bibnamefont
  {Beale}},\ }\bibfield  {title} {\enquote {\bibinfo {title} {8 - ideal fermi
  systems},}\ }in\ \href {\doibase
  https://doi.org/10.1016/B978-0-12-382188-1.00008-6} {\emph {\bibinfo
  {booktitle} {Statistical Mechanics (Third Edition)}}},\ \bibinfo {editor}
  {edited by\ \bibinfo {editor} {\bibfnamefont {R.K.}\ \bibnamefont {Pathria}}\
  and\ \bibinfo {editor} {\bibfnamefont {Paul~D.}\ \bibnamefont {Beale}}}\
  (\bibinfo  {publisher} {Academic Press},\ \bibinfo {address} {Boston},\
  \bibinfo {year} {2011})\ \bibinfo {edition} {third edition}\ ed.,\ pp.\
  \bibinfo {pages} {231--273}\BibitemShut {NoStop}%
\bibitem [{\citenamefont {Jaramillo}\ \emph {et~al.}(2016)\citenamefont
  {Jaramillo}, \citenamefont {Beau},\ and\ \citenamefont {del
  Campo}}]{jaramillo2016quantum}%
  \BibitemOpen
  \bibfield  {author} {\bibinfo {author} {\bibfnamefont {Juan}\ \bibnamefont
  {Jaramillo}}, \bibinfo {author} {\bibfnamefont {Mathieu}\ \bibnamefont
  {Beau}}, \ and\ \bibinfo {author} {\bibfnamefont {Adolfo}\ \bibnamefont {del
  Campo}},\ }\bibfield  {title} {\enquote {\bibinfo {title} {Quantum supremacy
  of many-particle thermal machines},}\ }\href {\doibase
  10.1088/1367-2630/18/7/075019} {\bibfield  {journal} {\bibinfo  {journal}
  {New J. Phys.}\ }\textbf {\bibinfo {volume} {18}},\ \bibinfo {pages} {075019}
  (\bibinfo {year} {2016})}\BibitemShut {NoStop}%
\bibitem [{\citenamefont {Chen}\ \emph {et~al.}(2019)\citenamefont {Chen},
  \citenamefont {Watanabe}, \citenamefont {Yu}, \citenamefont {Guan},\ and\
  \citenamefont {del Campo}}]{chen2019}%
  \BibitemOpen
  \bibfield  {author} {\bibinfo {author} {\bibfnamefont {Yang-Yang}\
  \bibnamefont {Chen}}, \bibinfo {author} {\bibfnamefont {Gentaro}\
  \bibnamefont {Watanabe}}, \bibinfo {author} {\bibfnamefont {Yi-Cong}\
  \bibnamefont {Yu}}, \bibinfo {author} {\bibfnamefont {Xi-Wen}\ \bibnamefont
  {Guan}}, \ and\ \bibinfo {author} {\bibfnamefont {Adolfo}\ \bibnamefont {del
  Campo}},\ }\bibfield  {title} {\enquote {\bibinfo {title} {An
  interaction-driven many-particle quantum heat engine and its universal
  behavior},}\ }\href {\doibase 10.1038/s41534-019-0204-5} {\bibfield
  {journal} {\bibinfo  {journal} {npj Quantum Information}\ }\textbf {\bibinfo
  {volume} {5}},\ \bibinfo {pages} {88} (\bibinfo {year} {2019})}\BibitemShut
  {NoStop}%
\bibitem [{\citenamefont {Watanabe}\ \emph {et~al.}(2020)\citenamefont
  {Watanabe}, \citenamefont {Venkatesh}, \citenamefont {Talkner}, \citenamefont
  {Hwang},\ and\ \citenamefont {del Campo}}]{watanabe2020}%
  \BibitemOpen
  \bibfield  {author} {\bibinfo {author} {\bibfnamefont {Gentaro}\ \bibnamefont
  {Watanabe}}, \bibinfo {author} {\bibfnamefont {B.~Prasanna}\ \bibnamefont
  {Venkatesh}}, \bibinfo {author} {\bibfnamefont {Peter}\ \bibnamefont
  {Talkner}}, \bibinfo {author} {\bibfnamefont {Myung-Joong}\ \bibnamefont
  {Hwang}}, \ and\ \bibinfo {author} {\bibfnamefont {Adolfo}\ \bibnamefont {del
  Campo}},\ }\bibfield  {title} {\enquote {\bibinfo {title} {Quantum
  statistical enhancement of the collective performance of multiple bosonic
  engines},}\ }\href {\doibase 10.1103/PhysRevLett.124.210603} {\bibfield
  {journal} {\bibinfo  {journal} {Phys. Rev. Lett.}\ }\textbf {\bibinfo
  {volume} {124}},\ \bibinfo {pages} {210603} (\bibinfo {year}
  {2020})}\BibitemShut {NoStop}%
\bibitem [{\citenamefont {Griffiths}(2005)}]{griffithsbookquantummechanics}%
  \BibitemOpen
  \bibfield  {author} {\bibinfo {author} {\bibfnamefont {David}\ \bibnamefont
  {Griffiths}},\ }\href@noop {} {\emph {\bibinfo {title} {Introduction to
  Quantum Mechanics}}}\ (\bibinfo  {publisher} {Pearson Prentice Hall},\
  \bibinfo {year} {2005})\BibitemShut {NoStop}%
\bibitem [{\citenamefont {Belloni}\ and\ \citenamefont
  {Robinett}(2014)}]{belloni_2014}%
  \BibitemOpen
  \bibfield  {author} {\bibinfo {author} {\bibfnamefont {M.}~\bibnamefont
  {Belloni}}\ and\ \bibinfo {author} {\bibfnamefont {R.W.}\ \bibnamefont
  {Robinett}},\ }\bibfield  {title} {\enquote {\bibinfo {title} {The infinite
  well and dirac delta function potentials as pedagogical, mathematical and
  physical models in quantum mechanics},}\ }\href {\doibase
  https://doi.org/10.1016/j.physrep.2014.02.005} {\bibfield  {journal}
  {\bibinfo  {journal} {Physics Reports}\ }\textbf {\bibinfo {volume} {540}},\
  \bibinfo {pages} {25 -- 122} (\bibinfo {year} {2014})}\BibitemShut {NoStop}%
\bibitem [{\citenamefont {Gurtin}\ \emph {et~al.}(2010)\citenamefont {Gurtin},
  \citenamefont {Fried},\ and\ \citenamefont
  {Anand}}]{gurtin_fried_anand_2010}%
  \BibitemOpen
  \bibfield  {author} {\bibinfo {author} {\bibfnamefont {Morton~E.}\
  \bibnamefont {Gurtin}}, \bibinfo {author} {\bibfnamefont {Eliot}\
  \bibnamefont {Fried}}, \ and\ \bibinfo {author} {\bibfnamefont {Lallit}\
  \bibnamefont {Anand}},\ }\href {\doibase 10.1017/CBO9780511762956} {\emph
  {\bibinfo {title} {The Mechanics and Thermodynamics of Continua}}}\ (\bibinfo
   {publisher} {Cambridge University Press},\ \bibinfo {year}
  {2010})\BibitemShut {NoStop}%
\bibitem [{\citenamefont {Paolucci}(2016)}]{paolucci_2016}%
  \BibitemOpen
  \bibfield  {author} {\bibinfo {author} {\bibfnamefont {S.}~\bibnamefont
  {Paolucci}},\ }\href {\doibase 10.1017/CBO9781316106167} {\emph {\bibinfo
  {title} {Continuum Mechanics and Thermodynamics of Matter}}}\ (\bibinfo
  {publisher} {Cambridge University Press},\ \bibinfo {year}
  {2016})\BibitemShut {NoStop}%
\bibitem [{\citenamefont {Ro{\ss}nagel}\ \emph {et~al.}(2016)\citenamefont
  {Ro{\ss}nagel}, \citenamefont {Dawkins}, \citenamefont {Tolazzi},
  \citenamefont {Abah}, \citenamefont {Lutz}, \citenamefont {Schmidt-Kaler},\
  and\ \citenamefont {Singer}}]{rossnagel2016single}%
  \BibitemOpen
  \bibfield  {author} {\bibinfo {author} {\bibfnamefont {Johannes}\
  \bibnamefont {Ro{\ss}nagel}}, \bibinfo {author} {\bibfnamefont {Samuel~T}\
  \bibnamefont {Dawkins}}, \bibinfo {author} {\bibfnamefont {Karl~N}\
  \bibnamefont {Tolazzi}}, \bibinfo {author} {\bibfnamefont {Obinna}\
  \bibnamefont {Abah}}, \bibinfo {author} {\bibfnamefont {Eric}\ \bibnamefont
  {Lutz}}, \bibinfo {author} {\bibfnamefont {Ferdinand}\ \bibnamefont
  {Schmidt-Kaler}}, \ and\ \bibinfo {author} {\bibfnamefont {Kilian}\
  \bibnamefont {Singer}},\ }\bibfield  {title} {\enquote {\bibinfo {title} {A
  single-atom heat engine},}\ }\href {\doibase 10.1126/science.aad6320}
  {\bibfield  {journal} {\bibinfo  {journal} {Science}\ }\textbf {\bibinfo
  {volume} {352}},\ \bibinfo {pages} {325--329} (\bibinfo {year}
  {2016})}\BibitemShut {NoStop}%
\bibitem [{\citenamefont {von Lindenfels}\ \emph {et~al.}(2019)\citenamefont
  {von Lindenfels}, \citenamefont {Gr\"ab}, \citenamefont {Schmiegelow},
  \citenamefont {Kaushal}, \citenamefont {Schulz}, \citenamefont {Mitchison},
  \citenamefont {Goold}, \citenamefont {Schmidt-Kaler},\ and\ \citenamefont
  {Poschinger}}]{lindenfels2019spin}%
  \BibitemOpen
  \bibfield  {author} {\bibinfo {author} {\bibfnamefont {D.}~\bibnamefont {von
  Lindenfels}}, \bibinfo {author} {\bibfnamefont {O.}~\bibnamefont {Gr\"ab}},
  \bibinfo {author} {\bibfnamefont {C.~T.}\ \bibnamefont {Schmiegelow}},
  \bibinfo {author} {\bibfnamefont {V.}~\bibnamefont {Kaushal}}, \bibinfo
  {author} {\bibfnamefont {J.}~\bibnamefont {Schulz}}, \bibinfo {author}
  {\bibfnamefont {Mark~T.}\ \bibnamefont {Mitchison}}, \bibinfo {author}
  {\bibfnamefont {John}\ \bibnamefont {Goold}}, \bibinfo {author}
  {\bibfnamefont {F.}~\bibnamefont {Schmidt-Kaler}}, \ and\ \bibinfo {author}
  {\bibfnamefont {U.~G.}\ \bibnamefont {Poschinger}},\ }\bibfield  {title}
  {\enquote {\bibinfo {title} {Spin heat engine coupled to a
  harmonic-oscillator flywheel},}\ }\href {\doibase
  10.1103/PhysRevLett.123.080602} {\bibfield  {journal} {\bibinfo  {journal}
  {Phys. Rev. Lett.}\ }\textbf {\bibinfo {volume} {123}},\ \bibinfo {pages}
  {080602} (\bibinfo {year} {2019})}\BibitemShut {NoStop}%
\bibitem [{\citenamefont {Deng}\ \emph {et~al.}(2018)\citenamefont {Deng},
  \citenamefont {Chenu}, \citenamefont {Diao}, \citenamefont {Li},
  \citenamefont {Yu}, \citenamefont {Coulamy}, \citenamefont {del Campo},\ and\
  \citenamefont {Wu}}]{deng2018}%
  \BibitemOpen
  \bibfield  {author} {\bibinfo {author} {\bibfnamefont {Shujin}\ \bibnamefont
  {Deng}}, \bibinfo {author} {\bibfnamefont {Aurélia}\ \bibnamefont {Chenu}},
  \bibinfo {author} {\bibfnamefont {Pengpeng}\ \bibnamefont {Diao}}, \bibinfo
  {author} {\bibfnamefont {Fang}\ \bibnamefont {Li}}, \bibinfo {author}
  {\bibfnamefont {Shi}\ \bibnamefont {Yu}}, \bibinfo {author} {\bibfnamefont
  {Ivan}\ \bibnamefont {Coulamy}}, \bibinfo {author} {\bibfnamefont {Adolfo}\
  \bibnamefont {del Campo}}, \ and\ \bibinfo {author} {\bibfnamefont {Haibin}\
  \bibnamefont {Wu}},\ }\bibfield  {title} {\enquote {\bibinfo {title}
  {Superadiabatic quantum friction suppression in finite-time
  thermodynamics},}\ }\href {\doibase 10.1126/sciadv.aar5909} {\bibfield
  {journal} {\bibinfo  {journal} {Science Advances}\ }\textbf {\bibinfo
  {volume} {4}},\ \bibinfo {pages} {eaar5909} (\bibinfo {year} {2018})},\
  \Eprint
  {http://arxiv.org/abs/https://www.science.org/doi/pdf/10.1126/sciadv.aar5909}
  {https://www.science.org/doi/pdf/10.1126/sciadv.aar5909} \BibitemShut
  {NoStop}%
\bibitem [{\citenamefont {Batalh\~ao}\ \emph {et~al.}(2014)\citenamefont
  {Batalh\~ao}, \citenamefont {Souza}, \citenamefont {Mazzola}, \citenamefont
  {Auccaise}, \citenamefont {Sarthour}, \citenamefont {Oliveira}, \citenamefont
  {Goold}, \citenamefont {De~Chiara}, \citenamefont {Paternostro},\ and\
  \citenamefont {Serra}}]{batalho_2014}%
  \BibitemOpen
  \bibfield  {author} {\bibinfo {author} {\bibfnamefont {Tiago~B.}\
  \bibnamefont {Batalh\~ao}}, \bibinfo {author} {\bibfnamefont {Alexandre~M.}\
  \bibnamefont {Souza}}, \bibinfo {author} {\bibfnamefont {Laura}\ \bibnamefont
  {Mazzola}}, \bibinfo {author} {\bibfnamefont {Ruben}\ \bibnamefont
  {Auccaise}}, \bibinfo {author} {\bibfnamefont {Roberto~S.}\ \bibnamefont
  {Sarthour}}, \bibinfo {author} {\bibfnamefont {Ivan~S.}\ \bibnamefont
  {Oliveira}}, \bibinfo {author} {\bibfnamefont {John}\ \bibnamefont {Goold}},
  \bibinfo {author} {\bibfnamefont {Gabriele}\ \bibnamefont {De~Chiara}},
  \bibinfo {author} {\bibfnamefont {Mauro}\ \bibnamefont {Paternostro}}, \ and\
  \bibinfo {author} {\bibfnamefont {Roberto~M.}\ \bibnamefont {Serra}},\
  }\bibfield  {title} {\enquote {\bibinfo {title} {Experimental reconstruction
  of work distribution and study of fluctuation relations in a closed quantum
  system},}\ }\href {\doibase 10.1103/PhysRevLett.113.140601} {\bibfield
  {journal} {\bibinfo  {journal} {Phys. Rev. Lett.}\ }\textbf {\bibinfo
  {volume} {113}},\ \bibinfo {pages} {140601} (\bibinfo {year}
  {2014})}\BibitemShut {NoStop}%
\bibitem [{\citenamefont {Josefsson}\ \emph {et~al.}(2018)\citenamefont
  {Josefsson}, \citenamefont {Svilans}, \citenamefont {Burke}, \citenamefont
  {Hoffmann}, \citenamefont {Fahlvik}, \citenamefont {Thelander}, \citenamefont
  {Leijnse},\ and\ \citenamefont {Linke}}]{josefsson_2018}%
  \BibitemOpen
  \bibfield  {author} {\bibinfo {author} {\bibfnamefont {Martin}\ \bibnamefont
  {Josefsson}}, \bibinfo {author} {\bibfnamefont {Artis}\ \bibnamefont
  {Svilans}}, \bibinfo {author} {\bibfnamefont {Adam~M.}\ \bibnamefont
  {Burke}}, \bibinfo {author} {\bibfnamefont {Eric~A.}\ \bibnamefont
  {Hoffmann}}, \bibinfo {author} {\bibfnamefont {Sofia}\ \bibnamefont
  {Fahlvik}}, \bibinfo {author} {\bibfnamefont {Claes}\ \bibnamefont
  {Thelander}}, \bibinfo {author} {\bibfnamefont {Martin}\ \bibnamefont
  {Leijnse}}, \ and\ \bibinfo {author} {\bibfnamefont {Heiner}\ \bibnamefont
  {Linke}},\ }\bibfield  {title} {\enquote {\bibinfo {title} {A quantum-dot
  heat engine operating close to the thermodynamic efficiency limits},}\ }\href
  {\doibase 10.1038/s41565-018-0200-5} {\bibfield  {journal} {\bibinfo
  {journal} {Nature Nanotechnology}\ }\textbf {\bibinfo {volume} {13}},\
  \bibinfo {pages} {920--924} (\bibinfo {year} {2018})}\BibitemShut {NoStop}%
\bibitem [{\citenamefont {Sattler}(2010)}]{sattler_2010}%
  \BibitemOpen
  \bibfield  {author} {\bibinfo {author} {\bibfnamefont {K.~D.}\ \bibnamefont
  {Sattler}},\ }\href@noop {} {\emph {\bibinfo {title} {Handbook of
  nanophysics: Nanotubes and Nanowires}}}\ (\bibinfo  {publisher} {CRC Press},\
  \bibinfo {year} {2010})\BibitemShut {NoStop}%
\bibitem [{\citenamefont {Liu}\ \emph {et~al.}(2013)\citenamefont {Liu},
  \citenamefont {Yang}, \citenamefont {Hong}, \citenamefont {Si}, \citenamefont
  {Chi},\ and\ \citenamefont {Guo}}]{liu_2013}%
  \BibitemOpen
  \bibfield  {author} {\bibinfo {author} {\bibfnamefont {Y.~S.}\ \bibnamefont
  {Liu}}, \bibinfo {author} {\bibfnamefont {X.~F.}\ \bibnamefont {Yang}},
  \bibinfo {author} {\bibfnamefont {X.~K.}\ \bibnamefont {Hong}}, \bibinfo
  {author} {\bibfnamefont {M.~S.}\ \bibnamefont {Si}}, \bibinfo {author}
  {\bibfnamefont {F.}~\bibnamefont {Chi}}, \ and\ \bibinfo {author}
  {\bibfnamefont {Y.}~\bibnamefont {Guo}},\ }\bibfield  {title} {\enquote
  {\bibinfo {title} {A high-efficiency double quantum dot heat engine},}\
  }\href {\doibase 10.1063/1.4819852} {\bibfield  {journal} {\bibinfo
  {journal} {Applied Physics Letters}\ }\textbf {\bibinfo {volume} {103}},\
  \bibinfo {pages} {093901} (\bibinfo {year} {2013})},\ \Eprint
  {http://arxiv.org/abs/https://doi.org/10.1063/1.4819852}
  {https://doi.org/10.1063/1.4819852} \BibitemShut {NoStop}%
\bibitem [{\citenamefont {Du}\ \emph {et~al.}(2020)\citenamefont {Du},
  \citenamefont {Shen}, \citenamefont {Zhang}, \citenamefont {Su},\ and\
  \citenamefont {Chen}}]{du_2020}%
  \BibitemOpen
  \bibfield  {author} {\bibinfo {author} {\bibfnamefont {Jianying}\
  \bibnamefont {Du}}, \bibinfo {author} {\bibfnamefont {Wei}\ \bibnamefont
  {Shen}}, \bibinfo {author} {\bibfnamefont {Xin}\ \bibnamefont {Zhang}},
  \bibinfo {author} {\bibfnamefont {Shanhe}\ \bibnamefont {Su}}, \ and\
  \bibinfo {author} {\bibfnamefont {Jincan}\ \bibnamefont {Chen}},\ }\bibfield
  {title} {\enquote {\bibinfo {title} {Quantum-dot heat engines with
  irreversible heat transfer},}\ }\href {\doibase
  10.1103/PhysRevResearch.2.013259} {\bibfield  {journal} {\bibinfo  {journal}
  {Phys. Rev. Research}\ }\textbf {\bibinfo {volume} {2}},\ \bibinfo {pages}
  {013259} (\bibinfo {year} {2020})}\BibitemShut {NoStop}%
\bibitem [{\citenamefont {Stern}\ \emph {et~al.}(2014)\citenamefont {Stern},
  \citenamefont {Umansky},\ and\ \citenamefont {Bar-Joseph}}]{stern_2014}%
  \BibitemOpen
  \bibfield  {author} {\bibinfo {author} {\bibfnamefont {Michael}\ \bibnamefont
  {Stern}}, \bibinfo {author} {\bibfnamefont {Vladimir}\ \bibnamefont
  {Umansky}}, \ and\ \bibinfo {author} {\bibfnamefont {Israel}\ \bibnamefont
  {Bar-Joseph}},\ }\bibfield  {title} {\enquote {\bibinfo {title} {Exciton
  liquid in coupled quantum wells},}\ }\href {\doibase 10.1126/science.1243409}
  {\bibfield  {journal} {\bibinfo  {journal} {Science}\ }\textbf {\bibinfo
  {volume} {343}},\ \bibinfo {pages} {55--57} (\bibinfo {year} {2014})},\
  \Eprint
  {http://arxiv.org/abs/https://www.science.org/doi/pdf/10.1126/science.1243409}
  {https://www.science.org/doi/pdf/10.1126/science.1243409} \BibitemShut
  {NoStop}%
\bibitem [{\citenamefont {Gangloff}\ \emph {et~al.}(2015)\citenamefont
  {Gangloff}, \citenamefont {Bylinskii}, \citenamefont {Counts}, \citenamefont
  {Jhe},\ and\ \citenamefont {Vuleti{\'{c}}}}]{gangloff2015}%
  \BibitemOpen
  \bibfield  {author} {\bibinfo {author} {\bibfnamefont {Dorian}\ \bibnamefont
  {Gangloff}}, \bibinfo {author} {\bibfnamefont {Alexei}\ \bibnamefont
  {Bylinskii}}, \bibinfo {author} {\bibfnamefont {Ian}\ \bibnamefont {Counts}},
  \bibinfo {author} {\bibfnamefont {Wonho}\ \bibnamefont {Jhe}}, \ and\
  \bibinfo {author} {\bibfnamefont {Vladan}\ \bibnamefont {Vuleti{\'{c}}}},\
  }\bibfield  {title} {\enquote {\bibinfo {title} {Velocity tuning of friction
  with two trapped atoms},}\ }\href {\doibase 10.1038/nphys3459} {\bibfield
  {journal} {\bibinfo  {journal} {Nature Physics}\ }\textbf {\bibinfo {volume}
  {11}},\ \bibinfo {pages} {915--919} (\bibinfo {year} {2015})}\BibitemShut
  {NoStop}%
\bibitem [{\citenamefont {Bylinskii}\ \emph {et~al.}(2015)\citenamefont
  {Bylinskii}, \citenamefont {Gangloff},\ and\ \citenamefont
  {Vuletic}}]{bylinskii_2015}%
  \BibitemOpen
  \bibfield  {author} {\bibinfo {author} {\bibfnamefont {Alexei}\ \bibnamefont
  {Bylinskii}}, \bibinfo {author} {\bibfnamefont {Dorian}\ \bibnamefont
  {Gangloff}}, \ and\ \bibinfo {author} {\bibfnamefont {Vladan}\ \bibnamefont
  {Vuletic}},\ }\bibfield  {title} {\enquote {\bibinfo {title} {Tuning friction
  atom-by-atom in an ion-crystal simulator},}\ }\href {\doibase
  10.1126/science.1261422} {\bibfield  {journal} {\bibinfo  {journal}
  {Science}\ }\textbf {\bibinfo {volume} {348}},\ \bibinfo {pages} {1115--1118}
  (\bibinfo {year} {2015})}\BibitemShut {NoStop}%
\bibitem [{\citenamefont {Karpa}\ \emph {et~al.}(2013)\citenamefont {Karpa},
  \citenamefont {Bylinskii}, \citenamefont {Gangloff}, \citenamefont {Cetina},\
  and\ \citenamefont {Vuleti\ifmmode~\acute{c}\else \'{c}\fi{}}}]{karpa_2013}%
  \BibitemOpen
  \bibfield  {author} {\bibinfo {author} {\bibfnamefont {Leon}\ \bibnamefont
  {Karpa}}, \bibinfo {author} {\bibfnamefont {Alexei}\ \bibnamefont
  {Bylinskii}}, \bibinfo {author} {\bibfnamefont {Dorian}\ \bibnamefont
  {Gangloff}}, \bibinfo {author} {\bibfnamefont {Marko}\ \bibnamefont
  {Cetina}}, \ and\ \bibinfo {author} {\bibfnamefont {Vladan}\ \bibnamefont
  {Vuleti\ifmmode~\acute{c}\else \'{c}\fi{}}},\ }\bibfield  {title} {\enquote
  {\bibinfo {title} {Suppression of ion transport due to long-lived
  subwavelength localization by an optical lattice},}\ }\href {\doibase
  10.1103/PhysRevLett.111.163002} {\bibfield  {journal} {\bibinfo  {journal}
  {Phys. Rev. Lett.}\ }\textbf {\bibinfo {volume} {111}},\ \bibinfo {pages}
  {163002} (\bibinfo {year} {2013})}\BibitemShut {NoStop}%
\bibitem [{\citenamefont {Carnot}(1897)}]{carnot1897}%
  \BibitemOpen
  \bibfield  {author} {\bibinfo {author} {\bibfnamefont {Sadi}\ \bibnamefont
  {Carnot}},\ }\bibfield  {title} {\enquote {\bibinfo {title} {R\'eflexions sur
  la puissance motrice du feu et sur les machines propres \`a d\'evelopper atte
  puissance},}\ }\href@noop {} {\  (\bibinfo {year} {1897})}\BibitemShut
  {NoStop}%
\bibitem [{\citenamefont {Schroeder}(2000)}]{schroeder}%
  \BibitemOpen
  \bibfield  {author} {\bibinfo {author} {\bibfnamefont {Daniel~V.}\
  \bibnamefont {Schroeder}},\ }\href@noop {} {\emph {\bibinfo {title} {An
  introduction to thermal physics}}}\ (\bibinfo  {publisher} {San Francisco, CA
  : Addison Wesley},\ \bibinfo {year} {2000})\ \bibinfo {note} {includes
  bibliographical references (pages 397-405) and index.}\BibitemShut {Stop}%
\bibitem [{\citenamefont {Arovas}(2013)}]{arovas}%
  \BibitemOpen
  \bibfield  {author} {\bibinfo {author} {\bibfnamefont {Daniel}\ \bibnamefont
  {Arovas}},\ }\href@noop {} {\emph {\bibinfo {title} {Arovas Lecture Notes On
  Thermodynamics And Statistical Mechanics}}}\ (\bibinfo  {publisher}
  {University of California, San Diego},\ \bibinfo {year} {2013})\BibitemShut
  {NoStop}%
\bibitem [{\citenamefont {Acharyya}(2010)}]{ma_2010}%
  \BibitemOpen
  \bibfield  {author} {\bibinfo {author} {\bibfnamefont {Muktish}\ \bibnamefont
  {Acharyya}},\ }\bibfield  {title} {\enquote {\bibinfo {title} {Noninteracting
  fermions in infinite dimensions},}\ }\href {\doibase
  10.1088/0143-0807/31/6/l01} {\bibfield  {journal} {\bibinfo  {journal}
  {European Journal of Physics}\ }\textbf {\bibinfo {volume} {31}},\ \bibinfo
  {pages} {L89--L91} (\bibinfo {year} {2010})}\BibitemShut {NoStop}%
\bibitem [{\citenamefont {Blumenson}(1960)}]{blumenson}%
  \BibitemOpen
  \bibfield  {author} {\bibinfo {author} {\bibfnamefont {L.~E.}\ \bibnamefont
  {Blumenson}},\ }\bibfield  {title} {\enquote {\bibinfo {title} {A derivation
  of n-dimensional spherical coordinates},}\ }\href
  {http://www.jstor.org/stable/2308932} {\bibfield  {journal} {\bibinfo
  {journal} {The American Mathematical Monthly}\ }\textbf {\bibinfo {volume}
  {67}},\ \bibinfo {pages} {63--66} (\bibinfo {year} {1960})}\BibitemShut
  {NoStop}%
\bibitem [{Note1()}]{Note1}%
  \BibitemOpen
  \bibinfo {note} {$[\alpha ] = [\hbar ]^2 [k_B]^{-1} M^{-1} L^{-2}\\ =
  (ML^2T^{-1})^2(ML^2T^{-2}\Theta ^{-1})^{-1} M^{-1} L^{-2} = \Theta
  $}\BibitemShut {NoStop}%
\bibitem [{\citenamefont {Cowan}(2019)}]{cowan_2019}%
  \BibitemOpen
  \bibfield  {author} {\bibinfo {author} {\bibfnamefont {Brian}\ \bibnamefont
  {Cowan}},\ }\bibfield  {title} {\enquote {\bibinfo {title} {On the chemical
  potential of ideal fermi and bose gases},}\ }\href {\doibase
  10.1007/s10909-019-02228-0} {\bibfield  {journal} {\bibinfo  {journal}
  {Journal of Low Temperature Physics}\ }\textbf {\bibinfo {volume} {197}},\
  \bibinfo {pages} {412--444} (\bibinfo {year} {2019})}\BibitemShut {NoStop}%
\bibitem [{\citenamefont {Myers}\ \emph {et~al.}(2022)\citenamefont {Myers},
  \citenamefont {Peña}, \citenamefont {Negrete}, \citenamefont {Vargas},
  \citenamefont {Chiara},\ and\ \citenamefont {Deffner}}]{myers_bec_2022}%
  \BibitemOpen
  \bibfield  {author} {\bibinfo {author} {\bibfnamefont {Nathan~M}\
  \bibnamefont {Myers}}, \bibinfo {author} {\bibfnamefont {Francisco~J}\
  \bibnamefont {Peña}}, \bibinfo {author} {\bibfnamefont {Oscar}\ \bibnamefont
  {Negrete}}, \bibinfo {author} {\bibfnamefont {Patricio}\ \bibnamefont
  {Vargas}}, \bibinfo {author} {\bibfnamefont {Gabriele~De}\ \bibnamefont
  {Chiara}}, \ and\ \bibinfo {author} {\bibfnamefont {Sebastian}\ \bibnamefont
  {Deffner}},\ }\bibfield  {title} {\enquote {\bibinfo {title} {Boosting engine
  performance with bose–einstein condensation},}\ }\href {\doibase
  10.1088/1367-2630/ac47cc} {\bibfield  {journal} {\bibinfo  {journal} {New
  Journal of Physics}\ }\textbf {\bibinfo {volume} {24}},\ \bibinfo {pages}
  {025001} (\bibinfo {year} {2022})}\BibitemShut {NoStop}%
\bibitem [{\citenamefont {Niedenzu}\ \emph
  {et~al.}(2019{\natexlab{b}})\citenamefont {Niedenzu}, \citenamefont {Mazets},
  \citenamefont {Kurizki},\ and\ \citenamefont
  {Jendrzejewski}}]{niedenzu_bec_2019}%
  \BibitemOpen
  \bibfield  {author} {\bibinfo {author} {\bibfnamefont {Wolfgang}\
  \bibnamefont {Niedenzu}}, \bibinfo {author} {\bibfnamefont {Igor}\
  \bibnamefont {Mazets}}, \bibinfo {author} {\bibfnamefont {Gershon}\
  \bibnamefont {Kurizki}}, \ and\ \bibinfo {author} {\bibfnamefont {Fred}\
  \bibnamefont {Jendrzejewski}},\ }\bibfield  {title} {\enquote {\bibinfo
  {title} {Quantized refrigerator for an atomic cloud},}\ }\href {\doibase
  10.22331/q-2019-06-28-155} {\bibfield  {journal} {\bibinfo  {journal}
  {{Quantum}}\ }\textbf {\bibinfo {volume} {3}},\ \bibinfo {pages} {155}
  (\bibinfo {year} {2019}{\natexlab{b}})}\BibitemShut {NoStop}%
\end{thebibliography}%

\end{document}